\PassOptionsToPackage{prologue,table}{xcolor}
\documentclass[acmsmall]{acmart}
\AtBeginDocument{%
  }

\setcopyright{acmlicensed}
\acmJournal{PACMHCI}
\acmYear{2025} \acmVolume{9} \acmNumber{2} \acmArticle{CSCW146} \acmMonth{4}\acmDOI{10.1145/3711044}

\usepackage{graphicx}
\usepackage{booktabs}
\usepackage{layouts}  % to print textwidth in inches
\usepackage{subcaption} % to add subcaptions to multi-figures
\usepackage{makecell} % to make cells with linebreaks
\usepackage{multirow} % for multirow cells in tables
\usepackage[normalem]{ulem} % from https://www.tablesgenerator.com/
\useunder{\uline}{\ul}{} % from https://www.tablesgenerator.com/
\usepackage{tabularx}

\definecolor{offwhite}{HTML}{ebebeb}

\newcommand{\rev}[1]{#1}

\begin{document}

%%
%% The "title" command has an optional parameter,
%% allowing the author to define a "short title" to be used in page headers.
\title[Peer Recommendation Interventions for Health-related Social Support: a Feasibility Assessment]{Peer Recommendation Interventions for Health-related Social Support: a Feasibility Assessment}

%%
%% The "author" command and its associated commands are used to define
%% the authors and their affiliations.
%% Of note is the shared affiliation of the first two authors, and the
%% "authornote" and "authornotemark" commands
%% used to denote shared contribution to the research.
\author{Zachary Levonian}
\email{zacharylevonian@gmail.com}
\orcid{0000-0002-8932-1489}
\affiliation{%
  \institution{GroupLens, University of Minnesota}
  \city{Minneapolis}
  \state{Minnesota}
  \country{US}
}

\author{Matthew Zent}
\email{zentx005@umn.edu}
\orcid{0000-0003-4555-8764}
\affiliation{%
  \institution{GroupLens, University of Minnesota}
  \city{Minneapolis}
  \state{Minnesota}
  \country{US}
}

\author{Ngan Nguyen}
\email{nguy4068@umn.edu}
\orcid{0009-0008-4752-7674}
\affiliation{%
  \institution{GroupLens, University of Minnesota}
  \city{Minneapolis}
  \state{Minnesota}
  \country{US}
}

\author{Matthew McNamara}
\email{mcnam385@umn.edu}
\orcid{0009-0000-5579-4838}
\affiliation{%
  \institution{GroupLens, University of Minnesota}
  \city{Minneapolis}
  \state{Minnesota}
  \country{US}
}

\author{Loren Terveen}
\email{terveen@umn.edu}
\orcid{0000-0002-8843-4035}
\affiliation{%
  \institution{GroupLens, University of Minnesota}
  \city{Minneapolis}
  \state{Minnesota}
  \country{US}
}

\author{Svetlana Yarosh}
\email{lana@umn.edu}
\orcid{0000-0001-8389-2064}
\affiliation{%
  \institution{GroupLens, University of Minnesota}
  \city{Minneapolis}
  \state{Minnesota}
  \country{US}
}

%%
%% By default, the full list of authors will be used in the page
%% headers. Often, this list is too long, and will overlap
%% other information printed in the page headers. This command allows
%% the author to define a more concise list
%% of authors' names for this purpose.
\renewcommand{\shortauthors}{Zachary Levonian et al.}

%%
%% The abstract is a short summary of the work to be presented in the
%% article.
\begin{abstract}
Online health communities (OHCs) offer the promise of connecting with supportive peers.
Forming these connections first requires \textit{finding} relevant peers---a process that can be time-consuming.
Peer recommendation systems are a computational approach to make finding peers easier during a health journey.
By encouraging OHC users to alter their online social networks, peer recommendations could increase available support.
But these benefits are hypothetical and based on mixed, observational evidence.
To experimentally evaluate the effect of peer recommendations, we conceptualize these systems as health interventions designed to increase specific beneficial connection behaviors.
In this paper, we designed a peer recommendation intervention to increase two behaviors: reading about peer experiences and interacting with peers.
We conducted an initial feasibility assessment of this intervention by conducting a 12-week field study in which 79 users of CaringBridge.org received weekly peer recommendations via email.
Our results support the usefulness and demand for peer recommendation and suggest benefits to evaluating larger peer recommendation interventions.
Our contributions include practical guidance on the development and evaluation of peer recommendation interventions for OHCs.
\end{abstract}

%%
%% The code below is generated by the tool at http://dl.acm.org/ccs.cfm.
%% Please copy and paste the code instead of the example below.
%%


%%
%% Keywords. The author(s) should pick words that accurately describe
%% the work being presented. Separate the keywords with commas.

\received{January 2024}
\received[revised]{July 2024}
\received[accepted]{October 2024}

%%
%% This command processes the author and affiliation and title
%% information and builds the first part of the formatted document.
\maketitle

\section{Introduction}
Social support helps people in health crises cope with stressful circumstances.  
Access to emotional, informational, and instrumental support is associated with increased quality of life~\cite{ashbury_one--one_1998}, improved psychosocial health~\cite{umberson_social_2010}, and physical health~\cite{holt-lunstad_social_2010}.
Support from \textit{peers}---people who have had similar health experiences---is particularly useful~\cite{simoni_peer_2011,allison_logging_2021}.
However, a person's existing offline support community may lack peers~\cite{thoits_mechanisms_2011,turner_developing_2001}. 
Online communities provide a place for people to support each other in ways that their existing offline support communities cannot by offering opportunities to connect with peers.
While health support is exchanged online on diverse platforms~\cite{chancellor_recovery_2016,sharma_mental_2018,bender_seeking_2011}, online health communities (OHCs) are specifically designed for health-related discussion and support~\cite{allison_logging_2021,young_this_2019}.

Even in OHCs, however, finding supportive peers can be time-consuming~\cite{gatos_how_2021,eschler_im_2017}.
Algorithmic matching systems could enable design features that help OHC users to find peers, but existing approaches are limited or have remained entirely theoretical~\cite{hartzler_leveraging_2016,levonian_patterns_2021}.
For example, a common approach to peer finding requires a user to \textit{explicitly} search or filter for people or topics of interest, a process that support seekers and providers may find labor-intensive and discouraging in health-related contexts~\cite{gatos_how_2021,pretorius_searching_2020}.
%However, in health-related contexts, support seekers and providers may not be able to articulate what they're looking for in a way the system can understand.
\textit{Recommendation systems} for peer finding can incorporate the user's past behaviors to \textit{implicitly} identify potential matches---people with valuable similar experiences~\cite{yang_seekers_2019}.
However, while recommendation systems are commonly seen in real-world OHCs, no rigorous experimental evidence exists linking peer recommendations with increases in beneficial peer connection behaviors.
Given the hypothesized benefits, why so little evidence?
%As we will see, evaluating recommendation efficacy requires reckoning with many practical complexities. 
%no existing recommendation system has been evaluated for peer matching in OHCs.
%Specifically, no experimental evidence links availability of peer recommendations with increases in beneficial peer connection behaviors.
%However, no existing recommendation system has been evaluated for peer matching in OHCs: no experimental evidence links the availability of peer recommendations with hypothesized increases in beneficial peer connection behavior.
%demonstrates if or how much the availability of peer recommendations will increase beneficial peer connection behavior.
%the feasibility of using recommendation systems to increase peer support behaviors for facilitating peer support remains unclear.
%However, to our knowledge, no existing recommendation system has been created, used, or evaluated for peer matching in OHCs. Despite much research motivated by the goal of designing recommendation systems to match peers for mutual support~\cite{cohen_social_2004,oleary_design_2017,eschler_im_2017,newman_its_2011,chen_make_2009,hartzler_leveraging_2016,yang_seekers_2019,levonian_patterns_2021}, the feasibility of recommendation systems for peer matching remains unclear.

%The goal of this paper is to explore the idea of connecting peers in online health communities for mutual support. 
The goal of this paper is to identify and address the conceptual and practical barriers to researching the impact of peer recommendations on connection behavior.
%hypothetical support benefits of connecting peers in online health communites.
We accomplish this by designing, developing, and evaluating an email-based peer recommendation system for users of the online community CaringBridge.
Despite recent research arguing for the potential utility of peer recommendation for providing social support~\cite{oleary_design_2017,eschler_im_2017,newman_its_2011,chen_make_2009,hartzler_leveraging_2016,yang_seekers_2019,macleod_be_2017,levonian_patterns_2021,levonian_bridging_2020,gatos_how_2021,barta_similar_2023,poon_computer-mediated_2023,ceron-guzman_its_2025}, no study yet describes use of peer recommendation in practice.
% online peer-matching recommendation system, we conduct a \textit{feasibility study}.
We argue that peer recommendation should be conceptualized as a health intervention that can change specific user behaviors---behaviors that are linked to psychosocial or physical health.
Inspired by public health research approaches, we conduct this initial study as a \textit{feasibility assessment} to identify barriers to the implementation of a recommendation intervention for increasing social supportiveness and to determine how the intervention should be evaluated in a larger, more comprehensive study~\cite{bowen_how_2009}.
%We conducted a \textit{feasibility assessment} to determine the usefulness of and demand for peer recommendation. %the lack of usage data.
%Feasibility assessments are designed to determine if an intervention should be evaluated in a larger or more comprehensive study~\cite{bowen_how_2009}.
%We argue that peer recommendation should be conceptualized as a health intervention that can change specific user behaviors---behaviors that are linked to psychosocial or physical health.
%Our proposed intervention is designed to increase two behaviors: reading about peers' health experiences and increasing interaction among peers.
In the rest of this introduction, we summarize the proposed intervention, its evaluation in a 12-week field study, and the encouraging findings for our core contribution: evidence for the feasibility of using recommendations to connect peers in OHCs. % based on feedback and behavior changes driven by actual system use

\subsection{What is the proposed intervention?}
\label{sec:proposed_intervention}

The proposed intervention is displaying recommended peers to current OHC users.
This design intervention has two aspects: an \textit{interface} that displays details about recommendations and an \textit{algorithm} that selects peers to recommend in the interface.
The interface shows ``profiles''---summaries of each individual peer being recommended~\cite{hartzler_design_2016}. 
By including previews of and links to a user's recent activity within the OHC, we enable prospective peers to evaluate the relevance of the recommended user~\cite{andalibi_considerations_2021}.
The algorithm identifies ``relevant'' peers for a specific user by incorporating their prior activity in the OHC and ranking potential peers based on their mutual activity.
%More than just similarity between users, relevance is a multi-faceted notion that we discuss in the Related Work.  

We ground our feasibility assessment in the context of a large existing OHC---CaringBridge.org. CaringBridge users write blogs to describe their health experiences to their broader support networks.  
In the CaringBridge context, we will be recommending blogs written by peers to the authors of existing blogs.
A recommendation intervention offers the potential for CaringBridge blog authors to form connections with peers that they may not be explicitly seeking but for whom they can give or receive meaningful support.
We use a weekly email as the interface to display recommendations, copying an email alert design used by CaringBridge authors.
%, producing textual profiles that contain previews of recent blog activity.
We use a machine learning-based recommendation system to rank potential peers based on historical interaction behavior on CaringBridge.

As an intervention, recommending the blogs of peers to OHC users aims to directly manipulate those users' natural social networks.
Such manipulations are necessarily complex~\cite{su_experimental_2020}, so a focus on specific user behaviors that the manipulation will induce is important.
%Our proposed intervention is designed to increase two behaviors: reading about peers' health experiences and increasing interaction among peers.
The intervention is designed to increase two behaviors: reading about the experiences of peers and interacting with peers.
Both of these behaviors are associated with benefits, such as reduced stress, useful coping information, and a sense of community~\cite{rains_coping_2018}.
%---discussed further in sec.~\ref{sec:benefits}.
However, the \textit{experimental} evidence linking peer connection with benefits is mixed~\cite{allison_logging_2021}, and includes risks such as increased distress~\cite{rains_coping_2018}.
These mixed outcomes motivate us to carefully evaluate the feasibility of an intervention to increase these behaviors and produce benefits for participants.
%motivating us to carefully evaluate the feasibility of an intervention to increase these behaviors.
%We discuss the benefits and risks associated with increasing these user behaviors in the next section.
% thought: peer recommendation is benefical because it can in principle function for both support providers and support seekers 

\subsection{How did we assess feasibility of the intervention?}
%What are the feasibility outcomes for this study?
\label{sec:feasibility_outcomes}

%We conducted a field study to assess the feasibility of a peer recommendation intervention and to identify requirements for running a larger randomized controlled trial.

\textit{Feasibility} refers to the ability to use an intervention to achieve desired behavior changes in reality.
Content recommendations are in use on many social networking sites---including some OHCs\footnote{Online health communities include both dedicated sites and sub-communities on larger social networking sites e.g.\ Reddit and Facebook~\cite{young_this_2019}. While all sub-communities on larger sites will include post or possibly follow recommendations, these recommenders are not specifically designed or evaluated for increasing social support.}---with the goal of increasing user engagement~\cite{jannach_measuring_2019,van_couvering_is_2007,zhao_recommending_2019,freyne_increasing_2009}.
Most evidence for the causal effect of content recommenders is not public~\cite{jannach_measuring_2019}, and we’re aware of no evidence for the causal effect of recommenders on OHC users specifically. 
However online health communication is a sensitive context and people use OHCs differently than they use other social networking sites~\cite{newman_its_2011}, so we need to be particularly careful about potential negative outcomes from the deployment of recommenders.
Potential harms---including increased user distress---are a serious risk not only for users who receive recommendations but also the users recommended by these systems i.e.\ negative second-order effects.
So the feasibility assessment we conduct is focused both on designing for the specifics of the OHC context and on evaluating the possibility of harms.
One could proceed directly to a randomized controlled trial (RCT) focused directly on a construct like perceived social support rather than on-platform behavior, but without a reasonable power analysis or other minimal information about the acceptability or efficacy of peer recommendations the success of such a trial is unlikely.
The feasibility assessment we conduct is designed to provide the evidence needed to run a successful RCT focused on increasing social support via peer recommendations.
%The intervention we describe here is focused on increasing the provision of peer social support.
As feasibility is multifaceted, we adapt Bowen et al.'s useful framework for feasibility studies~\cite{bowen_how_2009}, collecting evidence of feasibility in five focus areas summarized in Table~\ref{tab:feasibility_outcomes}: Demand, Implementation, Practicality, Acceptability, and Efficacy.
We designed a system and ran a field study to get converging lines of evidence for each of the feasibility focus areas.

%\textit{Feasibility} refers to the ability to use an intervention in reality: a broad and necessarily multifaceted construct~\cite{bowen_how_2009}. 
%Thus, we collected evidence of feasibility in five focus areas, summarized in Table~\ref{tab:feasibility_outcomes}: Demand, Implementation, Practicality, Acceptability, and Efficacy.
%We designed a system and ran a field study to get converging lines of evidence for each of the feasibility focus areas.

\begin{figure}
\centering
\includegraphics[width=\textwidth]{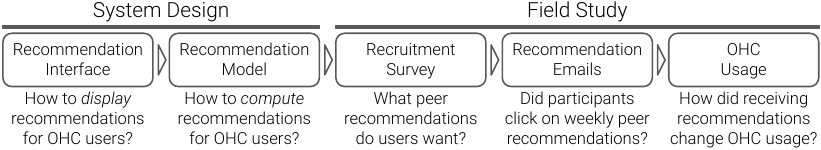}
\caption{Study outline: assessing the feasibility of a peer recommendation intervention in an OHC.}
\label{fig:paper_outline}
\Description{Two phases: System Design and Field Study. Recommendation Interface: ``How to display recommendations for OHC users?'' Recommendation Model: ``How to compute recommendations for OHC users?'' Recruitment Survey: ``What peer recommendations do users want?'' Recommendation Emails: ``Did participants click on weekly peer recommendations?'' OHC Usage: ``How did receiving recommendations change OHC usage?''}
\end{figure}

The primary contribution of this paper is our determination that peer recommendation interventions designed to facilitate OHC user connections are ethically and practically feasible to evaluate in a larger RCT.
%We made this assessment by designing a system and conducting a field study that provides converging lines of evidence for each of the feasibility focus areas.
Figure~\ref{fig:paper_outline} outlines the paper and the primary questions we address by analyzing participants' survey responses and behaviors on CaringBridge.
%We offer this feasibility assessment in terms of demand for the intervention, implementation challenges, practicality of administration, acceptability to participants, and efficacy.
%We present our system design as a model for future peer recommender systems.
During the 12-week field study, 79 participants received weekly peer recommendations via email, leading to hundreds of repeat visits to blogs and hundreds of extra peer interactions.
Participants clicked 5\% of recommendations, although less than half of the participants clicked any recommendation and fewer still chose to visibly interact with recommended blogs.
We find no evidence of significant second-order harms or benefits, and overall find an interest in and willingness to engage with blogs written by peer strangers.
These results provide tangible guidance for the future development of peer recommendation systems for OHCs.
We identify design trade-offs and implications for successful future deployments, discussing each of the five feasibility focus areas and arguing that peer recommendations can facilitate connections that authors may not be explicitly seeking---and that those connections can facilitate meaningful support.

\begin{table}[]
\caption{Feasibility assessment of a peer recommendation intervention.  We collected evidence in five focus areas (originally described by Bowen et al.~\cite{bowen_how_2009}).}
\label{tab:feasibility_outcomes}
\centering
\rowcolors{1}{offwhite}{}
% for a 3-column table, appear to have exactly 92.6% of the \textwidth to work with
% why is \arraybackslash needed in the rightmost column? see: https://tex.stackexchange.com/a/387741/142583
\begin{tabular}{>{\raggedright} p{0.15\textwidth} >{\raggedright} p{0.34\textwidth} >{\raggedright\arraybackslash} p{0.436\textwidth} }
\toprule
\rowcolor{white}
Feasibility Area  & Description & Evidence (Section) \\ \midrule
Demand & Interest in the intervention & Expressed interest (\ref{sec:survey_motivations}), actual use (\ref{sec:click_rate}) \\
Implementation & Tangible design and engineering to implement the intervention in a particular context & The system design: both interface (\ref{sec:sse_design}) and model (\ref{sec:model_development}) \\
Practicality & Requirements for administering the intervention & Data and compute requirements (App.~\ref{app:sec:hyperparameter_search}) \\
Acceptability & How participants react to the intervention & Explicit participant preferences (\ref{sec:survey_characteristics}) and feedback (\ref{sec:explicit_rec_feedback}) \\
Efficacy & How much the intervention affects the desired behaviors & Reading behavior (\ref{sec:reading_behavior}), interaction behavior (\ref{sec:interaction_behavior}), second-order effects~(\ref{sec:retention_outcomes}) \\
\bottomrule
\end{tabular}
\end{table}

\section{Related Work}

To design a peer recommendation system for users of online health communities (OHCs), we drew from existing work on OHCs, on peer matching, and on algorithmic recommendation.
%In this paper, we study a peer recommendation system for users of online health communities.
%In designing this system, we draw from existing work on online health communities, on peer matching, and on algorithmic recommendation.
%The most direct precursor to our current work is Hartzler et al.'s study of peer mentor recommendations for cancer patients and caregivers~\cite{hartzler_leveraging_2016}, which we discuss as an algorithmic approach to peer matching.

%Much research is motivated by the goal of designing recommendation systems to form new relationships~\cite{cohen_social_2004,oleary_design_2017,eschler_im_2017,newman_its_2011,chen_make_2009,hartzler_leveraging_2016,yang_seekers_2019,macleod_be_2017,levonian_patterns_2021,levonian_bridging_2020}.

\subsection{Online health community use}
\label{sec:benefits}

People use OHCs in the hope that they will gain useful support. 
Both those experiencing a health condition themselves and the caregivers of others are motivated to use OHCs to overcome isolation by offsetting deficits in existing relationships and identifying people who have had similar experiences~\cite{rains_coping_2018}.
While the effects of using OHCs are somewhat unclear, feeling socially supported is a key determinant of health linked to OHC use~\cite{allison_logging_2021,holt-lunstad_social_2010}.
Meta-reviews reveal consistent associations between social support and a variety of health outcomes e.g. mortality~\cite{holt-lunstad_social_2010,uchino_social_2006,uchino_social_2018}.
OHCs have diverse interfaces and affordances, including forums, listservs, blogs, chatrooms, Q\&A sites, and update feeds~\cite{rains_coping_2018}.
The social support available from OHCs is also diverse, including informational support from discussions of treatments and symptoms, emotional support from sympathetic others and being a part of a community, and other forms of support including the instrumental and the spiritual~\cite{allison_logging_2021,smith_what_2021}.
Support can come from many types of users~\cite{yang_seekers_2019}, but one particular benefit of OHCs is that they expose you to many \textit{peers}.

Peers are people who have similar experiences~\cite{simoni_peer_2011}.
Peer relationships differ from professional/patient and mentor/mentee relationships in that no formal role or expectation structures the relationship~\cite{simoni_peer_2011}.
Compared to a person's existing offline support networks, peers can provide more useful support~\cite{thoits_mechanisms_2011,simoni_peer_2011}.
%Peers share similar experiences, enabling them to provide information, timely emotional support, and a sense of shared community.
%An intervention to connect with peers links to those supports either through the valuable information captured in a peer's writing about their own experiences or through the exchange of supportive messages.
A variety of theoretical models support the potential benefits of peer connection~\cite{allison_logging_2021,thoits_mechanisms_2011,simoni_peer_2011,turner_developing_2001}.
We avoid adopting a specific theoretical basis for our current study, as we do not operationalize or measure social support directly, instead focusing on behaviors---reading and interaction---that are compatible with multiple theoretical models. 
If interaction is the primary focus of online social support research~\cite{allison_logging_2021}, we view ``reading'' as an important but distinct way to experience online support~\cite{tixier_counting_2016,han_social_2012}.
We motivate our focus on these two behaviors by describing prior work linking these behaviors to benefits.
%We do this for two reasons: firstly, we do not attempt measure peer social support as a construct, and secondly we argue for diverse potential benefits of peer recommendation. 
%Specifically, benefits occur related to two behaviors: reading about the experiences of peers and interacting with peers.
%In the next two paragraphs, we motivate our focus on these two behaviors by describing prior work linking these behaviors to benefits. 
%If interaction is the primary focus of online social support research~\cite{allison_logging_2021}, we view ``reading'' as an important but distinct way to experience online support~\cite{tixier_counting_2016,han_social_2012}.

\subsubsection{Reading about peer experiences}

Reading about the experiences of peers can be beneficial even in the absence of interaction~\cite{allison_logging_2021}.
In addition to learning from the valuable information contained in peers' writing e.g.\ coping strategies~\cite{yli-uotila_online_2014}, reading peer experiences can build a sense of community~\cite{saksono_algorithmic_2021}.
Further, reading can reduce loneliness~\cite{tixier_counting_2016}, contribute to feelings of normalcy and hope~\cite{tixier_counting_2016,holbrey_qualitative_2013}, reduce uncertainty and anxiety~\cite{yli-uotila_online_2014}, and enable collective sensemaking about one's journey~\cite{gatos_how_2021}.
In general, reading the experiences of others can benefit readers by enabling positive and normalizing social comparisons to the experiences of others~\cite{malik_computer-mediated_2008,skea_enabling_2011}.
But, making social comparisons is not without risk: the negative experiences of others can produce a sense of helplessness or increase distress~\cite{malik_computer-mediated_2008,karusala_that_2021,barta_similar_2023}.

\subsubsection{Interacting with peers}
Interacting with peers offers many potential benefits---among them are membership in a community, acquisition of new information, normalization of one's experiences, and relief from distress~\cite{rains_coping_2018}.
Online peer interaction can take three general forms: providing support to others, receiving support from others, and forming reciprocal relationships.
While receiving support from others has obvious appeal, not all support is perceived as wanted or useful, and in general there is mixed causal evidence for the benefits of online interaction-based social support~\cite{allison_logging_2021}.
A gap between received and perceived support bedevils designers of social support interventions: increased received support is only weakly correlated with perceptions of that support~\cite{rains_social_2011,thoits_mechanisms_2011}.
\textit{Providing} support to peers, on the other hand, may be more beneficial to the provider than the receiver~\cite{riessman_helper_1965}.
Support needs differ over the course of a health journey, and providing support presents an opportunity to ``give back'' and enact self-efficacy~\cite{introne_narrative_2021,jacobs_cancer_2016}.
Reciprocal peer relationships can offer the best of both worlds, but also present significant risks in health contexts:
stress can increase if online contacts are doing poorly \textit{or} doing well due to social comparisons~\cite{malik_computer-mediated_2008,rains_coping_2018}. 
The sudden drop-out of a connection, due to churn or patient death, can also increase distress~\cite{attard_thematic_2012}.
Further, peers might be unintentionally unsupportive due to differences in communication style~\cite{peng_exploring_2020}.
Due to the risks of interacting with peers, interventions designed to increase interaction cannot be deployed without careful evaluation of the risks and benefits---which motivates us to conduct an initial feasibility assessment of peer recommendation specifically.

%Good for stress, specifically for caregivers \cite{faw_supporting_2018}.

\subsection{Social support interventions}

The archetypal peer support intervention is the support group~\cite{cohen_social_2004}.
Online support groups offer similar approaches using a different medium, although generally still designed for and managed in a clinical setting~\cite{kowitt_peer_2019,walshe_peer_2018}.
Other clinical approaches bridge the gap to OHCs: Haldar et al.\ designed an OHC for people in the same hospital~\cite{haldar_patient_2020}.
Peer support interventions can have many goals: providing emotional support, providing health information or education, developing self-efficacy (e.g. by vicarious viewing of peer behavior~\cite{saksono_algorithmic_2021}), adjusting social norms (e.g. encouraging treatment compliance), or even facilitating social movements for patient advocacy~\cite{simoni_peer_2011}.
In 2004, Cohen expressed skepticism of existing peer support interventions, arguing that peer support groups produce minimal effects and that interventions should focus instead on forming weak ties and propping up existing support networks~\cite{cohen_social_2004}.
Twenty years later, the challenges associated with designing effective peer support interventions remain~\cite{allison_logging_2021}.
As an intervention into people's online social networks, we examine recommender systems as a mechanism for encouraging initial interactions that can blossom into weak tie relationships.
Recommendation aims to improve the quality of received peer support by matching users according to some measure of ``fit''.
%We might try to improve the quality of the support receiving by matching peers in some way, a line of work we will discuss next.
%We think social support is the most plausible immediate effect of recommendation, in part because instrumental advice and the sorts of detailed treatment and symptom management behaviors are uncommon on CaringBridge~\cite{smith_i_2020}.

\subsection{Health peer matching}
\label{sec:related_work_peer_matching}
Health peer matching has occurred largely in the context of hospital-attached programs where mentors and mentees are matched by a 3rd-party broker, usually a nurse or program manager~\cite{taylor_peer_2016,moulton_woman_2013,andalibi_considerations_2021}. Consider ``woman-to-woman'', a peer support program for women with gynecologic cancer: when a new participant expresses interest in the program, the program manager selects a match ``of similar diagnosis and age'' from a pool of volunteer mentors~\cite{moulton_woman_2016}. In contrast, online peer recommendation is not constrained to formal mentor/mentee pairings and can draw from a much larger pool of prospective ``volunteers'' at the cost of the clear expectations that come with structure and a human coordinator. 
In this study, we aim to seriously consider non-coordinated peer matching as a health-related social support intervention.

Little explicit guidance exists for peer matching~\cite{anderson_peer_2021,simoni_peer_2011}.
Prior work identifies a variety of peer characteristics that may be important for effective peer matching: we list all of the relevant papers we're aware of in Appendix~\ref{app:sec:prior_work_characteristics}.
A general finding from HCI research is that a shared health condition is not a requirement for peer communication; people perceive value to learning from and communicating with others based on many factors that shift throughout a health journey as needs and communication goals change~\cite{eschler_im_2017,barta_similar_2023}.
While prior work identifies some relevant peer characteristics, minimal \textit{comparative} work exists to identify the most important characteristics~\cite{simoni_peer_2011}.
% "little research has examined precisely what qualities of perceived or actual similarity might be most important." \cite{simoni_peer_2011}
Hartzler et al.\ are a notable exception, running scenario-based sessions in which participants explicitly evaluated five potential peer mentors based on provided health information~\cite{hartzler_leveraging_2016}.
%We surveyed participants to identify their preferences for specific peer characteristics so they can be incorporated in future systems.
For this feasibility assessment, we make recommendations using machine learning approaches that \textit{learn} valued peer characteristics from prior peer interactions.
%\footnote{We discuss approaches to peer matching other than recommendation in App.~\ref{app:sec:recommendation_alternatives}.}

\subsection{Algorithmic recommendation for peer matching}
Few published works explicitly discuss computational recommendation systems for online health communities.
Hartzler et al.\ matched peer mentors on the basis of shared health interests, language style, and demographics---as extracted from prior posts made in the CancerConnect OHC~\cite{hartzler_leveraging_2016}.
The other notable example is described only in a PhD thesis: Yang developed and deployed a recommendation system in the American Cancer Society's Cancer Survivor Network (CSN) forums ``to direct participants to useful and informative threads that they might be interested in''~\cite{yang_computational_2019}. 
They evaluated a model based on implicit feedback from prior commenting behavior by presenting recommended threads and users within the CSN interface, reporting greater thread click-through rate compared to a baseline model recommending recently popular threads.
%In contrast to the CSN forums, CaringBridge is a blogging platform without existing interface recommendation features for discovering other blogs.
We evaluate with both click rate and more explicit metrics to understand the influence of recommendations on reading and interaction behavior.
%We present an evaluation for a new CaringBridge recommender that extends beyond click rate into explicit feedback and the effect of the recommendations on a variety of user behaviors.

Outside of health, a variety of problem formulations and modeling methods have been used for the problem of recommending people.
Recommendation models are generally supervised machine learning models that optimize a loss function comparing the model's output to ``ground truth'' labels: explicit or implicit feedback provided by users.  Xu et al.\ present a useful review~\cite{xu_deep_2020}.
Use of implicit behavioral feedback is based on relevance assumptions, e.g. that clicked items are relevant while non-clicked items are not relevant~\cite{joachims_optimizing_2002}, which may not hold true in practice~\cite{lu_between_2018,wen_leveraging_2019}. 
%We assume that interactions between OHC users indicates relevance, discussed further in sec. \ref{sec:implicit_feedback}.
Given historical user/user interactions, one can then optimize a pointwise loss that rewards high scores for assumed-relevant user/user pairs.
%and low scores for non-interacting user/user pairs.
Input features vary from IDs for the users---which gives the classic matrix factorization approach to collaborative filtering~\cite{rendle_neural_2020}---to side information about the context where the recommendation was generated (e.g. the time and place) or the two users' prior content (e.g. their comments)~\cite{steck_deep_2021,goldschmitt_shaping_2019}.  

%\subsubsection{Machine learning for matching people}
Person-to-person recommendation is typically modeled as similar to the conventional user/item recommendation problem.
Facebook's deep learning recommendation model (DLRM) represents a common approach, using embeddings for categorical features (including user IDs) and multilayer perceptrons for creating dense representations of other features, then combining all representations with additional linear layers~\cite{naumov_deep_2019}.
We use a simplified form of DLRM to design a recommendation system appropriate for the peer support context.

\section{System Design}

\begin{figure}
\centering
\begin{subfigure}[t]{0.33\textwidth}
  \includegraphics[width=\textwidth]{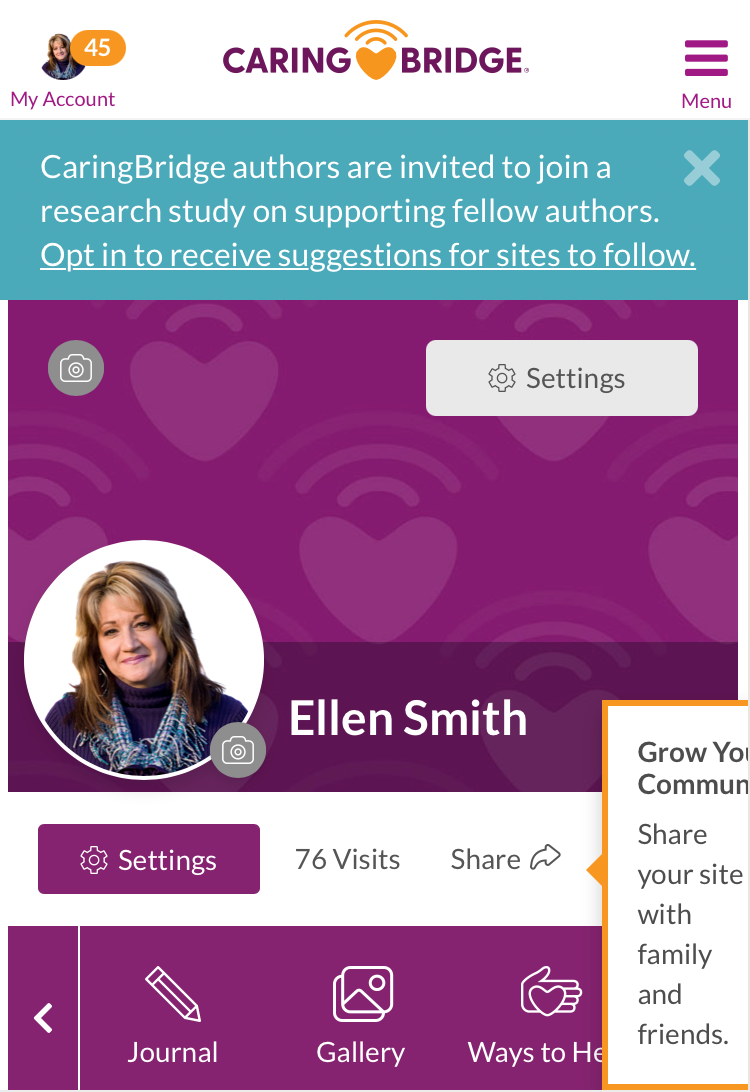}
\caption{Site home page, as viewed by\\a logged-in author of that site.\\The study recruitment banner is\\visible at the top of the page.}
\label{fig:sub:home_page}
\end{subfigure}%
\begin{subfigure}[t]{0.33\textwidth}
  \includegraphics[width=\textwidth]{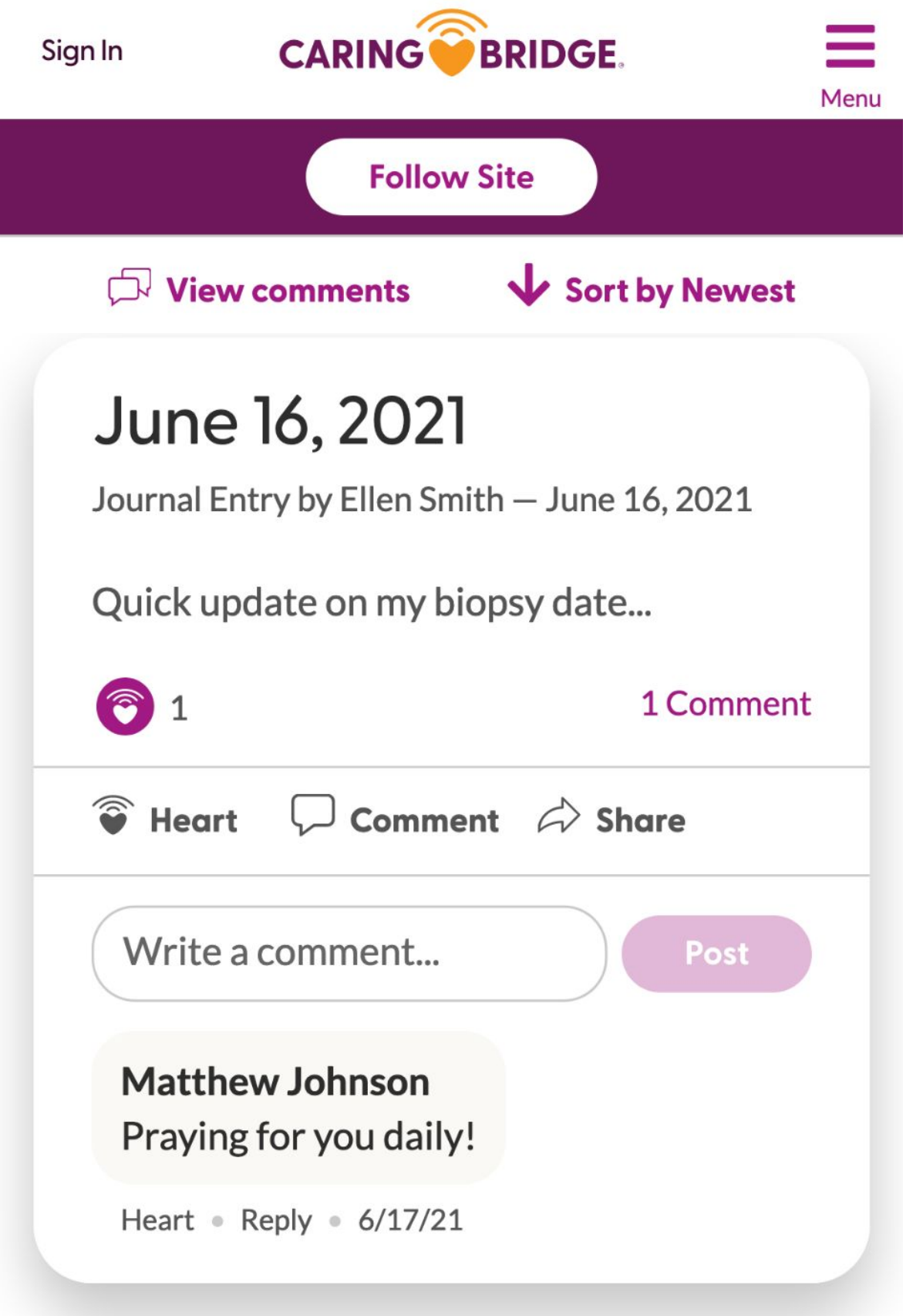}
\caption{Journal page and entry,\\as viewed by a logged-out visitor.}
\label{fig:sub:journal_page}
\end{subfigure}
\begin{subfigure}[t]{0.33\textwidth}
  \includegraphics[width=\textwidth]{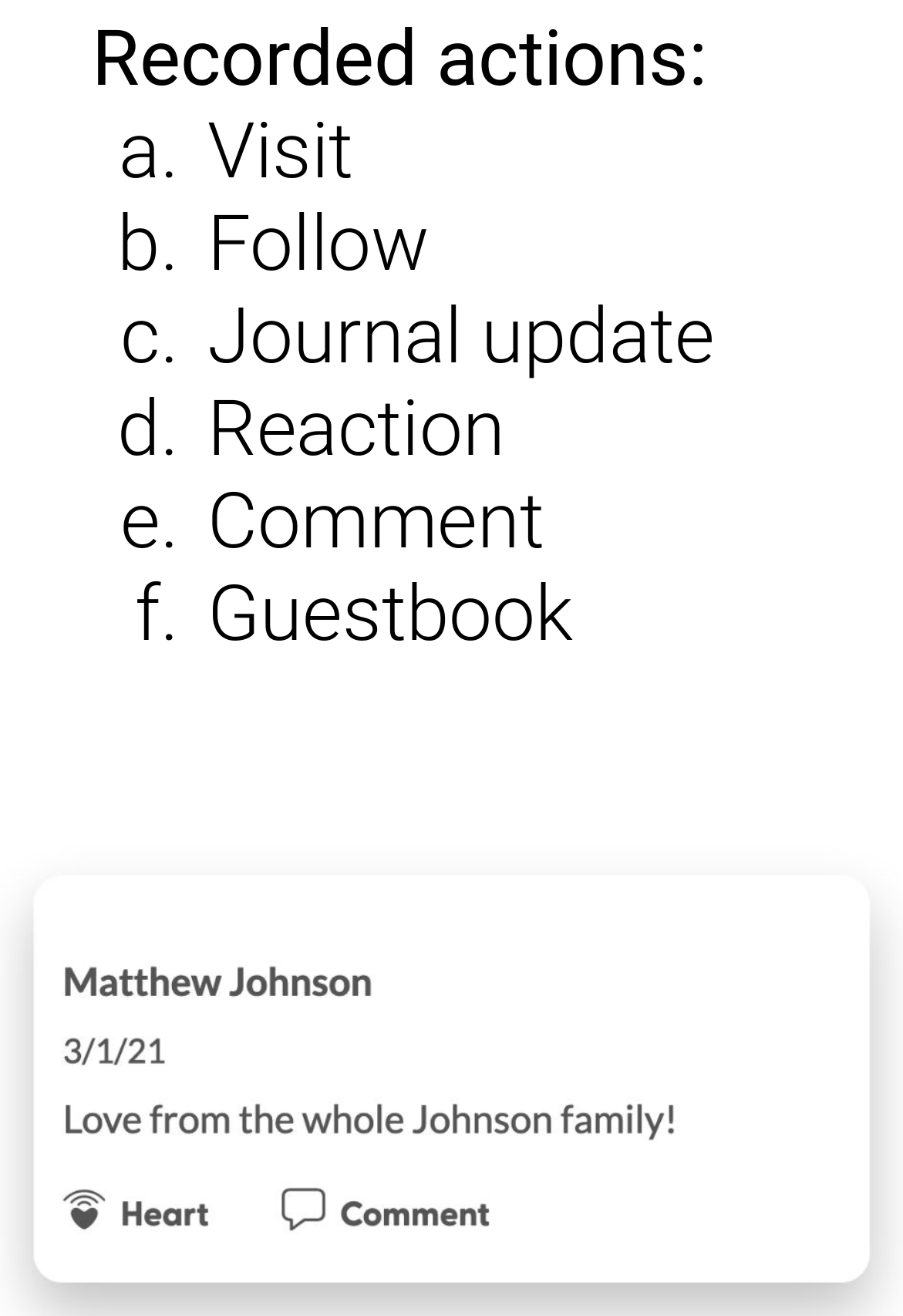}
\caption{The six logged-in user actions, above a guestbook on the Well Wishes page.}
\label{fig:sub:guestbook_page}
\end{subfigure}
\caption{The CaringBridge interface. We record six logged-in user actions: visits (to any of these site pages), follows (clicks on ``Follow Site'' and others, see sec. \ref{sec:prior_use}), Journal updates, reactions (on Journal updates, comments, or guestbooks), comments (on Journal updates or guestbooks), and guestbooks.}
\label{fig:cb_interface}
\Description{The CaringBridge web interface. Recorded actions: Visit, Follow, Journal update, Reaction, Comment, Guestbook.}
\end{figure}

\subsection{Observed prior use of CaringBridge for peer connection}\label{sec:prior_use}

CaringBridge.org is an online health community and blogging platform that has been the focus of prior HCI research e.g.~\cite{levonian_patterns_2021,smith_i_2020,ma_write_2017}.
In collaboration with the CaringBridge organization, we were given access to usage data from the CaringBridge website.
Registered users on CaringBridge can create blogs called \textit{sites}, on which they can publish blog posts called \textit{Journal updates}.
An \textit{author} is a user who has published at least one Journal update.
Much usage of CaringBridge is based around notification emails (an example is shown in Figure \ref{fig:email_notifications}).
Visitors can follow a site to be notified via email when an author of that site publishes a new Journal update.\footnote{Follows are somewhat complex on CaringBridge. Clicking the Follow Site button (depicted in Fig. \ref{fig:sub:journal_page}) has the effect of subscribing the visitor to one of several types of notification emails, but users can accomplish the equivalent by turning on notifications for that site under the list of ``Sites You Visit'' contained on the user's Notifications page.  Simply \textit{visiting} a site will add it to the Sites You Visit list. Further complicating matters, removing a site from the Sites You Visit list is unrelated to Follows, and a visitor will automatically Follow a site if they visit \textit{and} interact with that site.}
The majority of visitors to a site will be a patient's existing support network~\cite{levonian_patterns_2021}.
Registered visitors can leave \textit{reactions} to Journal updates (a reaction labeled ``Heart'' is the default) to show their support~\cite{smith_thoughts_2023}.
Alternately, they can leave text-based \textit{comments} on individual Journal updates or write \textit{guestbooks} which are comments that appear on a special Well Wishes page.  
These interface components and the associated actions we track for this study are shown in Figure~\ref{fig:cb_interface}.

We are interested here in \textit{interactions} (via reactions, comments, and guestbooks) between registered CaringBridge authors.  
An \textit{initiation} is the first interaction between an author and a site on which they are not an author.
In assessing demand for a peer recommendation intervention, we consider the volume of inter-author communication on CaringBridge, studied in detail by Levonian et al.~\cite{levonian_patterns_2021}.
Between 2010 and 2021, 275K authors initiated with other authors, more than 32.3\% of all 852K authors active during that period;
% note: Levonian et al. estimate that 12\% of text interactions are identifiably from authors who the receiving author met after the health event that precipitated the creation of a CaringBridge site.
this interaction occurs despite minimal existing discovery features to facilitate finding peers.\footnote{While CaringBridge offers a search feature, authors use this tool to find \textit{specific} authors, not for general searches for e.g. particular health conditions (see Appendix \ref{app:sec:search}).}
Further, there is associational evidence---detailed in Appendix~\ref{app:sec:assocational_peer_behavior_change}---that interactions from peer authors change author behavior.
Sites that receive at least one interaction from a peer author will publish on CaringBridge for a median of 3.2 additional months (with 6 additional updates) compared to sites that receive only interactions from non-author visitors.
Staying longer on CaringBridge is associated with benefits to authors~\cite{ma_write_2017}; a peer recommendation intervention aims to induce these positive author behavior changes by causing more peer interaction.

\subsection{Adapting a recommendation system to CaringBridge}
\label{sec:sse_design}

\begin{figure}
\centering
\begin{subfigure}[t]{0.33\textwidth}
  \includegraphics[width=\textwidth]{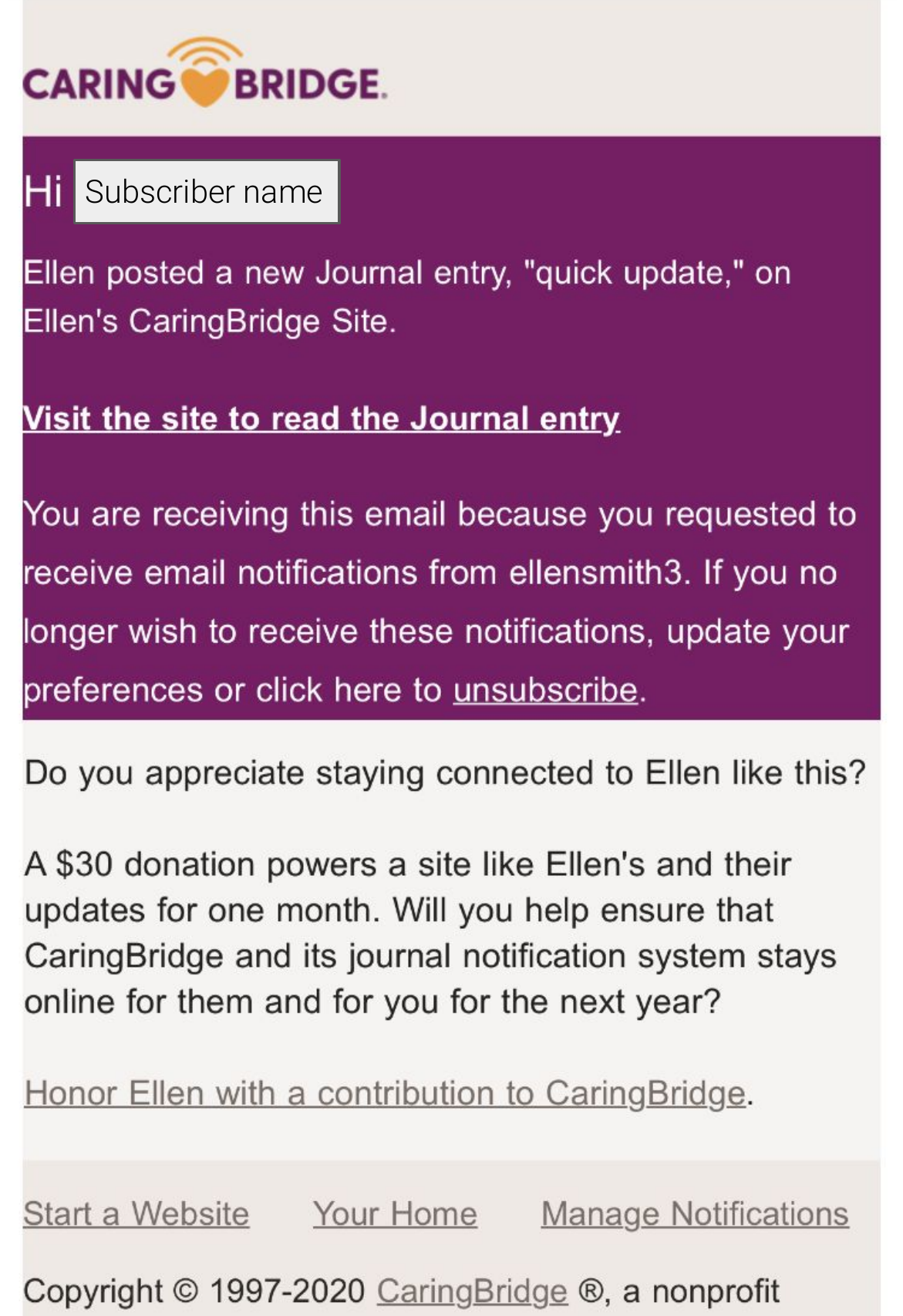}
\caption{Existing design:\\Visitor notification email}
\label{fig:sub:email_visitor}
\end{subfigure}%
\begin{subfigure}[t]{0.33\textwidth}
  \includegraphics[width=\textwidth]{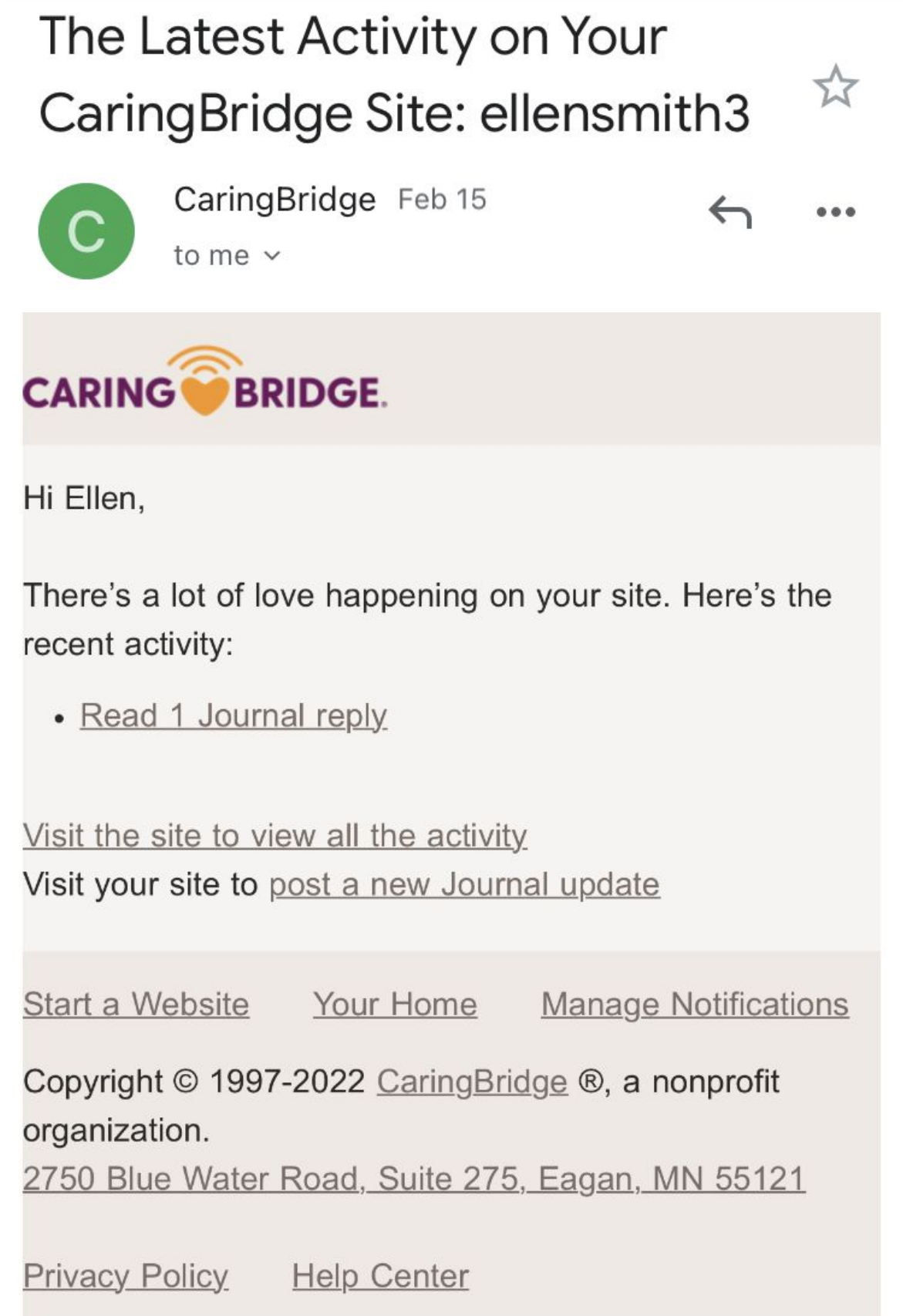}
\caption{Existing design:\\Author notification email}
\label{fig:sub:email_author}
\end{subfigure}
\begin{subfigure}[t]{0.33\textwidth}
  \includegraphics[width=\textwidth]{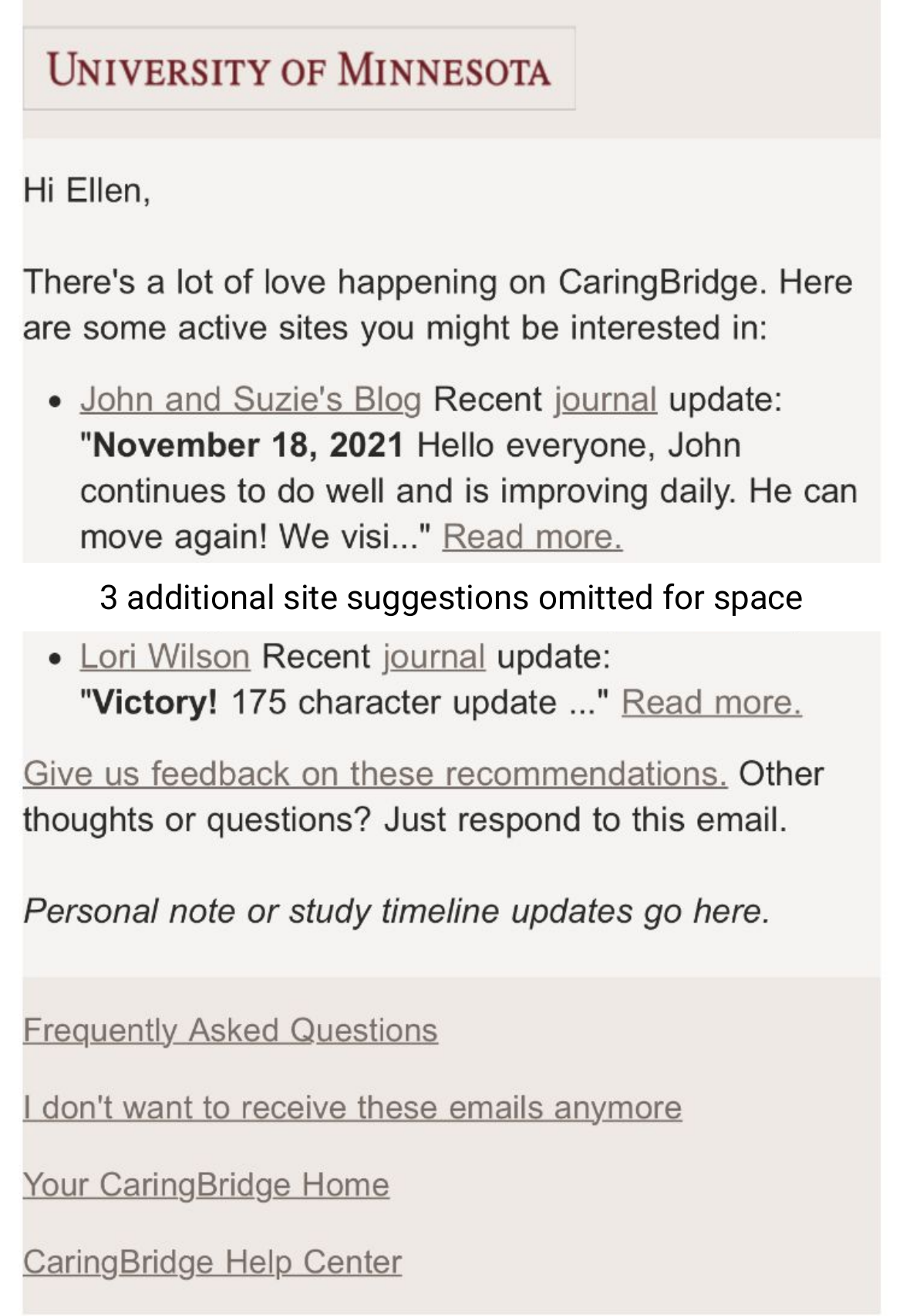}
\caption{\textbf{New design:}\\Site Suggestion email}
\label{fig:sub:email_sse}
\end{subfigure}
\caption{Existing design of email notifications on CaringBridge and the Site Suggestion email interface designed for this study. Data shown is a representative fabrication.}
\label{fig:email_notifications}
\Description{CaringBridge notification email design. The new design sample reads: Hi Ellen, There's a lot of love happening on CaringBridge. Here are some active sites you might be interested in: John and Suzie's Blog: Recent journal update: ``November 18, 2021 Hello everyone, John continues to do well and is improving daily. He can move again! We visi...'' Read more. Lori  Wilson: Recent journal update: ``Victory! 175 character update ...'' Read more. Give us feedback on these recommendations. Other thoughts or questions? Just respond to this email. Personal note or study timeline updates go here. Footer links: Frequently Asked Questions, I don't want to receive these emails anymore, Your CaringBridge Home, CaringBridge Help Center.}
\end{figure}

For a peer recommender intervention to be successful, the system must be adapted to the specifics of the context.  
First, CaringBridge offers only a limited public profile view, so only the content of the sites themselves provides a meaningful view into who an author is. 
For this reason, we optimize recommendations at the author level, but present those recommendations as site summaries.
In other words, recommendations appear to users to be recommendations for \textit{sites} but are actually recommendations for that site's \textit{author}.
Second, most CaringBridge usage is motivated by email notifications, so we chose to deliver recommendations via email as well.  
Author notification emails (depicted in Figure \ref{fig:sub:email_author}) are sent to site authors when a visitor interacts on a site or a co-author publishes a Journal update.
As authors are familiar with this design, we use the same layout and style for our Site Suggestion emails (depicted in Figure \ref{fig:sub:email_sse}).
Site Suggestion emails contain 5 bullet-pointed site recommendations, a request for feedback, a link to a feedback survey, a link to the study FAQ, and an unsubscribe link.
Each site recommendation is presented with the site's title (usually the patient's name), a link to the site's Journal page, and a preview of the most recent Journal update---inspired by Hartzler et al.'s finding that the most useful feature for evaluating a peer mentor is sample text~\cite{hartzler_leveraging_2016}.

\subsection{Model development}\label{sec:model_development}

\begin{figure}
\centering
\includegraphics[width=\textwidth]{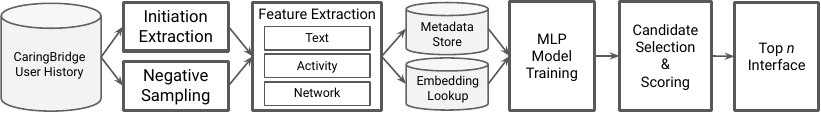}
\caption{CaringBridge peer recommendation system overview. Full details in Appendix~\ref{app:sec:hyperparameter_search}.}
\label{fig:system_overview}
\Description{Flowchart: CaringBridge User History, Initiation Extraction, Negative Sampling, Feature Extraction (Text, Activity, Network), Metadata Store, Embedding Lookup, MLP Model Training, Candidate Selection and Scoring, Top n interface.}
\end{figure}

To present recommended sites to a recommendation-seeking author, we include the top 5 sites as scored by a recommendation model.
We adhered to two key design requirements: First, recommendations should be \textit{personalized}, focusing on the right connection rather than a popular connection. We should not recommend any one site to a lot of people as that could create a negative experience for that site. Second, recommendations should be available even for authors who have never visited or interacted with another CaringBridge site. Because support-based reading and interaction is our goal, an author's initial ``cold start'' recommendations should be of similar quality to recommendations for long-time authors.

To provide recommendations that meet these two criteria, we implemented a content-based recommendation system. 
%Social matching systems require users to disclose sensitive personal information~\cite{terveen_social_2005}, and CaringBridge is a context where authors are already making those sensitive disclosures in the content of the Journal updates they publish~\cite{levonian_bridging_2020}.
We use implicit feedback from user activity and content-based (text) features to train a multilayer perceptron model that predicts historical peer connections and, ultimately, new site recommendations. 
To support the future development of similar systems, we provide a full accounting of the specific recommendation models, features, and computational resources we used in Appendix~\ref{app:sec:hyperparameter_search}.
The offline evaluation we conducted used a train/validation/test split containing 7 years, 6 months, and 6 months of data respectively.
We used standard recommendation system metrics appropriate for this context (discussed in detail in App.~\ref{sec:offline_evaluation_methods}), achieving a hit rate (HR@5) of 19\% and a mean reciprocal rank (MRR) of 0.16, which we deemed more than acceptable.
Recommendations were based on authors' recent activity on CaringBridge, their prior author interactions, and their recent Journal updates.
To preserve user privacy, future recommendation systems could make use of less data; we discuss the trade-offs involved in App.~\ref{sec:feature_ablations}.
%We discuss and compare several recommendation models (sec. \ref{sec:offline_evaluation}), as well as what features (sec. \ref{sec:feature_ablations}) and resources (sec. \ref{sec:required_resources}) are required to generate peer recommendations using our system.

%We adhered to two design requirements:
%-Personalized: The system should not require previous inter-author interaction or explicit rating in order to provide personalized recommendations. 
%-Integrated: The system should not replace the primary function of CaringBridge, i.e. blogging to communicate with one's existing support network.

%A key design requirement is that the system be capable of producing appropriate recommendations when the author has no prior interactions... which suggests a content-based approach. (Similarly: one implication is that it is NOT important to produce good recommendations for a brand new user... we require them to already be using the platform as intended, i.e. maybe a design requirement is that we do not supplant the current primary use of CB? The system should gel)
%(Terveen and McDonald point out that social matching systems require users to disclose sensitive personal information~\cite{terveen_social_2005}; CaringBridge is a context where sensitive disclosure was already occurring.)

\section{Methods}

\begin{figure}
    \centering
    \includegraphics[width=\textwidth]{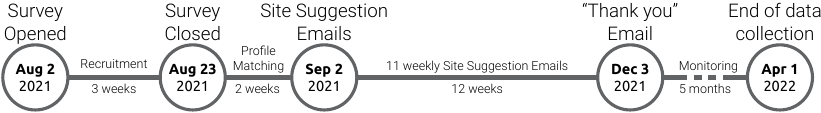}
    \caption{Field study timeline.}
    \label{fig:study_timeline}
    \Description{Survey opened August 2nd, 2021. Recruitment lasted 3 weeks. Survey closed August 23rd, 2021. Profile matching lasted 2 weeks. Site Suggestion Emails were first sent on September 2nd, 2021. 11 weekly Site Suggestion Emails were sent over the next 12 weeks. A ``Thank you'' email was sent December 3rd, 2021. Monitoring lasted 5 months. The end of data collection was April 1, 2022.}
\end{figure}

To evaluate our recommendation system, we conducted a field study.
%The previous section summarized aspects of feasibility related to the system design, while the next section summarizes aspects of feasibility related to the field study. 
%Table \ref{tab:study_phases} summarizes these two phases and the associated components, alongside the relevant feasibility aspect.
%We discuss the assembled evidence by feasibility area in the Discussion (sec. \ref{sec:discussion_feasibility}). 
The field study consisted of recruiting CaringBridge authors and sending them 11 weekly Site Suggestion emails.
The full analysis timeline is shown in Figure \ref{fig:study_timeline}.
%We analyzed field study data using both qualitative and quantitative methods, discussed in subsequent sections.
System implementation and analysis code are available on GitHub.\footnote{\url{https://github.com/levon003/HealthBlogRec}}
%Analysis code makes primary use of Python's scikit-learn~\cite{pedregosa_scikit-learn_2011}, statsmodels~\cite{seabold_statsmodels_2010}, transformers~\cite{wolf_transformers_2020}, NumPy~\cite{harris_array_2020}, pandas~\cite{mckinney_data_2010}, and Matplotlib \cite{hunter_matplotlib_2007} packages.
%The recommendation model was trained using PyTorch~\cite{paszke_pytorch_2019}.

\subsection{Recruitment Survey}

We recruited active CaringBridge authors by displaying a banner ad with a link to an opt-in survey. 
The banner appeared only to logged-in CaringBridge users on the home page of sites on which they are an author, as shown in Figure \ref{fig:sub:home_page}.\footnote{The recruitment banner being visible only on the home page of an authored site means that it will not be seen by users who authored a site in the past but are not regularly visiting their site(s). However, this recruitment method also excludes active authors who (a) visit a site sub-page like the Journal page (Fig. \ref{fig:sub:journal_page}) directly  e.g. via a bookmark or (b) access CaringBridge using the mobile app.}
The recruitment banner and opt-in survey were active for three weeks in August 2021.
The survey asked authors to opt-in to the study and included three optional questions on peer connection: on prior use of CaringBridge for connecting with strangers, on motivations for peer connection, and on characteristics that make a peer connection appealing. 
All survey texts are available in Appendix~\ref{app:survey}.
%Our results summarize responses to those questions: on prior use of CaringBridge for connecting with strangers (sec.~\ref{sec:survey_prior_use}), motivations for peer connection (sec.~\ref{sec:survey_motivations}), and potential characteristics that make peer connection appealing (sec.~\ref{sec:survey_characteristics}). 
%Then, we discuss the eligibility criteria and process for matching survey respondents to CaringBridge user accounts  (sec. \ref{sec:participant_matching}).
%Finally, we characterize the on-CaringBridge activity of the 79 authors who enrolled and were eligible to receive peer recommendations (sec. \ref{sec:participant_prior_use}).

\subsubsection{Participant matching}
\label{sec:participant_matching}

\begin{figure}
    \centering
    \includegraphics[width=\textwidth]{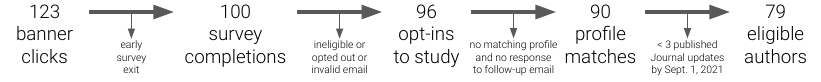}
    \caption{Field study recruitment pipeline.}
    \label{fig:participant_matching}
    \Description{123 banner clicks. 23 early survey exits. 100 survey completions. 4 ineligible or opted out or invalid email. 96 opt-ins to study. 6 had no matching profile and no response to follow-up email. 90 profile matches. 11 had fewer than 3 published Journal updates by September 1, 2021. Finally, 79 eligible authors remained.}
\end{figure}

Figure \ref{fig:participant_matching} shows the recruitment pipeline.
Of the 100 survey completions, 96 opted-in and met the three study consent criteria: being at least 18 years old, being a current CaringBridge author, and consenting to provide the email address associated with their CaringBridge account in order to receive Site Suggestion emails. Electronic consent was stored in Qualtrics.

We matched survey responses to a specific CaringBridge account based on their provided email. For the 8 cases where we couldn't find an associated account, we sent a follow-up email asking for profile information, matching 2 additional profiles.
Finally, we excluded 11 participants who had published less than 3 Journal updates by September 1, 2021.\footnote{See App.~\ref{sec:implicit_feedback} for a discussion of why we require three updates to produce recommendations.}
Ultimately, 79 participants were sent Site Suggestion emails.

\subsubsection{Observed prior use}
\label{sec:participant_prior_use}

\begin{table}[htb]
\caption{Pre-study CaringBridge usage by participants enrolled in the field study and by a pseudo-control group of eligible non-enrolled authors. Author tenure is the number of days between a user's first published Journal update and Sept.~1, 2021. *Indicates a significant difference at the 99.5\% threshold for a Welch's $t$-test on the mean difference and for a Mann-Whitney $U$ test (reported as the common language effect size).}
\label{tab:prestudy_difference}
% on CLES and ROC AUC from (stats.stackexchange.com/a/71950): "the probability that a randomly sampled positive (or case) will have a higher marker value than a negative (or control)"
\begin{tabular}{@{}lrlrlll@{}}
\toprule
 & \multicolumn{2}{l}{Participants ($n_\text{P}$=79)} & \multicolumn{2}{l}{Pseudo-Control ($n_\text{C}$=1759)} & & \\
 & Med. & M (SD) & Med. & M (SD) & $\text{M}_\text{P}$ - $\text{M}_\text{C}$ & $U_\text{P} / (n_\text{P}n_\text{C})$     \\ \midrule
%%%% copy cell output below
     Author tenure (days) & 179 & 709.4 (1146.6) & 3894 & 4078.5 (983.1) & -3369.1* & 3.2\%* \\
          Journal updates & 28 & 98.6 (262.4) & 77 & 167.1 (307.7) & -68.5 & 31.1\%* \\
         Peer site visits & 3 & 6.5 (12.3) & 10 & 32.0 (84.0) & -25.5* & 23.9\%* \\
    Peer site initiations & 1 & 2.5 (4.4) & 5 & 11.5 (35.2) & -9.0* & 23.8\%* \\
   Peer site interactions & 2 & 39.7 (94.4) & 30 & 201.2 (1384.5) & -161.5* & 29.0\%* \\
%\includegraphics[width=0.8in]{figures/n_interactions_hist_small.pdf} & & & & & & \\
%%%% copy cell output above
\bottomrule
\end{tabular}
\end{table}

The median participant had been writing Journal updates on CaringBridge for fewer than 6 months at the time of enrollment.
But, consistent with observations by Levonian et al.~\cite{levonian_patterns_2021}, a majority of the 79 participants both visited and interacted with at least one fellow author's site.
That suggests at least some demand for peer connection, although the field study extends beyond known contacts to peer strangers.
To quantify the differences between authors who chose to enroll in the study and other CaringBridge authors, we identified 30K authors who visited their own site while the banner survey was live and thus could have seen the banner, filtering down to 1,759 who had at least 3 Journal updates and thus could have received Site Suggestion emails.
Table \ref{tab:prestudy_difference} compares participants to this set of non-enrolled authors, which we term a \textit{pseudo-control} group, revealing that participants are generally newer to CaringBridge than other eligible authors.
While the field study uses an uncontrolled design, we will use this pseudo-control group to estimate the potential impact of the peer recommendation intervention on author behavior.

\subsection{Recommendation Emails}
\label{sec:rec_email_methods}

After we sent an initial Site Suggestion email on September 2, 2021, we conducted an initial assessment of interest and determined the study could proceed; we resumed sending emails on September 17, 2021 at a weekly pace. After 11 Site Suggestion emails, a ``thank you'' email was sent with a final request for feedback.
To evaluate the effectiveness of the Site Suggestion emails as a recommendation interface, we analyzed the rate of clicks on recommendations and the explicit feedback we received from participants.
%To evaluate the effectiveness of the Site Suggestion emails as a recommendation interface, we analyzed the rate of clicks on recommendations (sec. \ref{sec:click_rate}), the explicit feedback we received from participants (sec. \ref{sec:explicit_rec_feedback}), and the characteristics of the site representations we included in the emails (sec. \ref{sec:recommendation_characteristics}).

Click estimates are based on multiple data sources, although the primary source was via UTM tags embedded in each email's links. Appendix \ref{app:sec:click_data} presents additional details.
Explicit participant feedback was collected from four sources: direct responses to the emails, responses to a feedback survey linked in each email, an unsubscription survey linked in each email, and responses to the final ``thank you'' email sent on December 3, 2021.
The Site Suggestion email was previously described in sec.~\ref{sec:sse_design}.
%The survey texts are presented in Appendix~\ref{app:sec:feedback_survey} and~\ref{app:sec:unsubscribe_survey}.
The ``thank you'' email was sent in two versions. 
To participants who clicked on none of the recommendations, we asked for feedback on whether they had seen the emails and why they chose not to visit any of the recommended sites.  
To participants who clicked on at least one recommendation, we listed up to three random sites they clicked and asked for reflections.

%The primary information available to participants while they were deciding to click on a recommendation was the text preview of a recent Journal update. 
%To capture the preview characteristics that our participants could see and respond to, we conducted a thematic content analysis of these previews.
%Two researchers generated open and axial codes independently, then used an affinity mapping process based on the Grounded Theory method to identify three high-level themes~\cite{charmaz_constructing_2006}.
%\red{To produce quantitative prevalence estimates, we integrated our three themes with taxonomic descriptions from prior work to identify a set of four categories~\cite{ma_write_2017,smith_i_2020}.
%We conducted a regression analysis to identify which categories were associated with recommendation clicks.
%Further qualitative and quantitative method details are presented in Appendix~\ref{app:sec:thematic_analysis}.}
%Based on our thematic analysis and on existing work with CaringBridge Journal updates~\cite{ma_write_2017,smith_i_2020}, we isolated a set of four categories in order to produce quantitative prevalence estimates for each of the categories and to identify which categories were associated with clicks---further details in Appendix~\ref{app:sec:thematic_analysis}.

\subsection{Observed behavior on CaringBridge}

\subsubsection{Reading and interaction behavior}
To analyze the two primary behavioral outcomes---reading and interaction behavior---we identified measured proxies.
We used repeated visits and site Follows as proxies for interest in reading a site. 
While a single site visit may still indicate value for the reader, we would need either explicit feedback or some measure of dwell time on the site. 
Thus, we focus on instances where a participant returns to a site at least once. 
Follow actions are harder to interpret (see \ref{sec:prior_use}), but at a minimum indicate an interest in continuing to read updates on the followed site. 
We use reactions, comments, and guestbooks left on a recommended site as evidence of interaction.

We analyzed the textual content of the comments and guestbooks created by participants.
Three researchers conducted a qualitative analysis of the participant/site dyads where interaction occurred. Using Grounded Theory methods, we wrote axial codes and memos based on the Journal updates and comments in which participants and authors of recommended sites interacted, identifying emergent themes through code comparison~\cite{charmaz_constructing_2006}.  % $\approx$2000 visitor

\subsubsection{Second-order effects of recommendations on behavior}

In addition to the reading and interaction behavior of participants, receiving Site Suggestion emails and visiting strangers' CaringBridge sites might have second-order effects: harms or benefits that accrue to both participants and the authors of the sites they visit.  
%Our primary objective is to check for potential harms via quantitative modeling of CaringBridge user behavior.
%We now consider \textit{implicit} evidence of any harm or benefit induced by the recommendation intervention. 
%We check for harms using quantitative modeling of observed CaringBridge author behavior for both the participants (who received the emails) and for recommended sites visited by at least one participant (who experienced a change in their visitors).
%We consider potential impacts on (a) publishing Journal updates---the ``primary'' use of CaringBridge by authors---and (b) visits to and interactions with fellow authors' sites---a hypothesized outcome of peer interaction.
While we cannot draw firm causal conclusions from our non-experimental study, we can check for potential harms by estimating the effect of recommendations on two secondary behavior outcomes: (a) publishing Journal updates---the ``primary'' use of CaringBridge by authors---and (b) visits to and interactions with fellow authors' sites---a hypothesized outcome of peer interaction.
We estimate the effects of two ``treatments'': 
for authors, we estimate the \textit{participation effect} of receiving 11 weeks of Site Suggestion emails by comparing participants to the pseudo-control group of eligible unenrolled authors (introduced in sec. \ref{sec:participant_prior_use});
for sites, we estimate the \textit{visit effect} of receiving site visits from peer strangers by comparing visited recommended sites to both \textit{non-visited} recommended sites and a pseudo-control group of \textit{non-recommended} sites.\footnote{The non-recommended pseudo-control sites are the five highest-scoring sites for each participant and each batch after removing the sites we actually recommended---additional details and comparison to recommended sites in Appendix~\ref{app:sec:site_rank_analysis}.}
%We use these group comparisons to estimate the effect of two ``treatments'': receiving 11 weeks of Site Suggestion emails and receiving site visits from peer strangers respectively.
%The non-recommended pseudo-control sites are the five highest-scoring sites for each participant and each batch after removing the sites we actually recommended---additional details and comparison to recommended sites in Appendix~\ref{app:sec:site_rank_analysis}.
%Given the differences between the ``treated'' groups and the pseudo-control group, we can make untestable assumptions to explore the behavioral differences between these groups.

We quantify the difference between the ``treated'' authors/sites and the untreated authors/sites, producing three estimates: associational, model-adjusted, and causal.
The associational difference for a behavioral outcome $Y$ is $\text{E}[Y|T=1] - \text{E}[Y|T=0]$; the raw observed difference in the mean between treated and untreated authors/sites, where $\text{E}[Y|T=1]$ is the mean outcome for the treated group i.e.\ participant authors or visited sites. 
% (T=1 is treated, T=0 is untreated)
The model-adjusted estimate uses linear regression (OLS) to adjust for activity variables $A$, adding assumptions about model misspecification to compute the quantity $\text{E}[Y|T=1,A] - \text{E}[Y|T=0,A]$.
The causal estimate requires us to make untestable assumptions about the modeled relationship between the treated and untreated groups~\cite{hernan_causal_2020}---see Appendix~\ref{app:sec:retention_outcomes} for a detailed discussion of these assumptions.
%five assumptions: exchangeability, positivity, consistency, no measurement error, and no model misspecification.
%We defer a full discussion of these assumptions to Appendix~\ref{app:sec:retention_outcomes} but recommend using these estimates---which differ only mildly from the OLS estimates---as soft evidence towards triangulating on the true effect.
Using potential outcomes notation, we define $\text{E}[Y^{t=1}]$ as the mean behavioral outcome that would have been observed if all authors in the participant and pseudo-control group had been participants in the study.
The true causal effect $\text{E}[Y^{t=1}] - \text{E}[Y^{t=0}]$ is the influence on $Y$ that would occur if Site Suggestion emails were sent to all eligible authors.
We use the Bang-Robins doubly robust estimator to compute the causal effect, a modeling approach which combines inverse probability weighting and standardization~\cite{bang_doubly_2005} (see also~\cite{hernan_causal_2020}, Ch. 13).\footnote{A common alternative approach is propensity score matching, but PSM is sensitive to misspecifications of the propensity model~\cite{king_why_2019,kang_demystifying_2007}. We leave application of more sophisticated modeling approaches to future work.}

The author and site outcomes are measured as number of actions in the 13 weeks post-study and post-visit respectively. 
%We compare sites that were visited to both sites that were not visited and to the pseudo-control group of non-recommended sites. 
As the non-visited comparison sites were not visited by participants, we fabricate a visit time by sampling a random visit time from among the sites that \textit{were} visited in that same batch.
The activity variables $A$ are computed based on an equivalent time window before the event of interest---13 weeks before the study for participants and 5 weeks before the site visit respectively.\footnote{We include only 5 weeks of activity context due to a lack of available repeat visit data before August 2021.}
The analysis is not sensitive to the time window over which the pre-study features and post-study outcomes were measured (see Appendix \ref{app:sec:retention_outcomes}).
The associational and causal estimates require specification of a model that includes all relevant confounds.
For example, as we saw in Table~\ref{tab:prestudy_difference}, pseudo-control authors have been active on CaringBridge longer than participants.
Activity variables used are shown in Appendix Table~\ref{app:tab:covariates}, although we cannot measure important confounds such as ``interest in receiving Site Suggestion emails'', which is unobserved and potentially unexplained by the activity variables we do observe.

\subsubsection{Sample sizes needed for a powered RCT}
\label{sec:sample_size_method}
We use observed participant behavior during our field study to estimate the effect sizes of the peer recommendation intervention's impact on reading and interaction behavior.
%on both the authors who are exposed to recommendations and the recommended sites that receive visits from strangers. 
The structure of our data let us consider two potential future interventions: a one-time recommendation email and a recurring, weekly recommendation email.
We estimate the effects of a one-time recommendation email by including only visits and interactions resulting from the first batch of recommendation emails.
We compute sample sizes for both a replication (no control group) and an RCT (with control group) from the estimated standardized effect sizes at 80\% power with \(\alpha\)=0.05 using G*Power's one-tailed point biserial model~\cite{faul_g*power_2007}.  % version 3.1.9.7
Additional details are available in Appendix~\ref{app:sec:sample_size_calculations}.

\subsection{Ethical Considerations}
Peer support during health journeys is a sensitive research context.
To mitigate risk, we manually reviewed all recommendations before they were sent (sec.~\ref{sec:required_resources}), monitored for negative experiences during the study, and restricted the pool of recommended sites to include only sites with the lowest privacy settings i.e.\ those that are indexable by search engines and visible to all visitors.
%To mitigate potential negative impacts from unwanted contact, we 
The University of Minnesota Institutional Review Board determined that this research was exempt from full review.
%This research was deemed exempt from full review by the [Omitted for Review] Institutional Review Board.
All usage data was collected in compliance with the CaringBridge terms of service and privacy policy, and data was anonymized with placeholder IDs before transfer to an encrypted data repository accessible only by the research team.
Throughout this paper, quotes from ``public'' Journal updates and comments are representative but not literal to reduce discoverability~\cite{markham_fabrication_2012}.

\section{Results}

We recruited authors via a survey (sec. \ref{sec:recruitment_survey}).
Then, we sent 79 participants weekly emails (sec. \ref{sec:rec_email}).
Finally, we observed the subsequent usage of CaringBridge by both the participants and the authors of recommended sites (sec. \ref{sec:observed_study_behavior}).

\subsection{Recruitment Survey}
\label{sec:recruitment_survey}

\begin{table}[]
\caption{Recruitment survey responses.}
\label{tab:optin_survey_responses}
\begin{tabular}{@{}llrr@{}}
& & \# checked & \% \\ \cmidrule{2-4} 
\multirow{17}{*}{\includegraphics[]{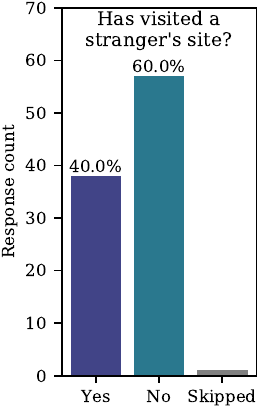}} & \textbf{Motivations}     &               &    \\
& Learn from others & 75 & 79.8\% \\
& Communicate with peers & 43 & 45.7\% \\
& Receive experienced support & 41 & 43.6\% \\
& Mentor newer authors & 27 & 28.7\% \\
& Not interested, but maybe in future & 6 & 6.4\% \\
& Never interested & 2 & 2.1\% \\
& Not interested, but maybe in past & 0 & 0.0\% \\
& Something else & 8 & 8.5\% \\
\cmidrule{2-4} 
& \textbf{Characteristics} &               &    \\
& Similar diagnosis or symptoms & 79 & 84.0\% \\
& Similar treatment & 51 & 54.3\% \\
& High-quality writing & 48 & 51.1\% \\
& Same caregiver relationship & 30 & 31.9\% \\
& Lives near me & 23 & 24.5\% \\
& Similar cultural background & 13 & 13.8\% \\
& Something else & 8 & 8.5\% \\
\cmidrule{2-4}
\end{tabular}
\end{table}

Quantitative survey responses are summarized in Table \ref{tab:optin_survey_responses}.

\subsubsection{Self-reported prior use}
\label{sec:survey_prior_use}

40\% of participants said they visited the CaringBridge site of an author who they did not know personally.\footnote{If we assume that participants visited CaringBridge sites with the account they used to fill the survey, we can compare self-reported use to actual use: 76\% of participants had made at least one logged-in visit to a CaringBridge they do not author---considering only the 79 participants we could link to an existing CaringBridge account (see sec.~\ref{sec:participant_matching}). So, among authors who have visited one or more CaringBridge sites, 65\% report visiting the site of a stranger.}
This prior usage provides evidence that the visits and interactions among peers includes strangers, although most interactions are likely between people who already know each other offline~\cite{levonian_patterns_2021}.

\subsubsection{Interest \& motivations}
\label{sec:survey_motivations}

We asked participants why they might visit a fellow author's CaringBridge site. 
The top motivation was to learn from others (80\%), then to communicate with peers (46\%) and to receive experienced support (44\%). A smaller percentage (29\%) were motivated by mentoring newer authors.
8 respondents (9\%) indicated they were not interested in visiting stranger's sites, although 6 suggested they could be interested in the future.  
Free response motivations centered on learning from others, including finding inspiration, hearing how others handled similar issues, and learning ``ideas on how to engage readers'' of their own site.\footnote{We report and discuss a variety of participant quotes in Appendix~\ref{app:sec:participant_quotes}.}

%Motivations:
%To see what a site looks like from a non-author viewer perspective
%To find inspiration
%I care about people and I am truly interested in their journeys.
%To hear about how someone else battled a similar issue as mine. 
%I don't know in a way this is difficult, not as bad as it could have been so far, and it's hard to think about joining in someone else's journey.  I can focus on writing this blog discussing our journey because it is letting friends and family know what is happening so it is narrow enough that it doesn't take away from the other things needed to be accomplished during the rest of the day. 
%I have been busy on my own journey and not taken time to view other's sites
%Ideas on how to engage readers - the people who love the friend I write for.

\subsubsection{Peer characteristics}
\label{sec:survey_characteristics}

We asked participants what characteristics of peer sites would make them want to read and engage with that site.
The most selected characteristic was a similar diagnosis or symptoms (84\%), a theme echoed in more specific free responses e.g. ``neurological conditions'' and a participant calling shared diagnosis the \textit{most} important characteristic. 
A majority of respondents indicated that similar treatments (54\%) and high-quality writing (51\%) were important. Free responses indicated a desire for ``inspirational'', ``honest'' writing with specific details that show ``positive and negative aspects'': a ``glimpse into the future''.
%One participant wanted to see \textit{``multiple posts all the way through death. I wanted to see what I would probably be writing as time progressed. A glimpse into the future if you will.''}
Fewer respondents (32\%) indicated that having the same caregiver relationship with the patient was important, although that checkbox was only relevant for non-patient respondents.
25\% of respondents indicated that geography was important; one specified sharing the same hospital as a relevant characteristic.
Few selected similar cultural background (14\%): a divergence from prior work~\cite{embuldeniya_experience_2013}, although the question may have been phrased too euphemistically to elicit specific preferences about e.g. age, gender, and ethnicity. 
%Three participants specifically identified age of the author or patient as a relevant detail.
%Two participants listed shared social context as important characteristics, such as already knowing the person or having ``common friendships''.
%One participant said they sought Spanish-language updates, while another said they were looking for sites authored by healthcare professionals.
%These responses provide evidence for the types of peer recommendations that are perceived as most useful and acceptable by CaringBridge users---as well as indicating that preferences are diverse, necessitating personalization.

%Specific characteristics:
%Someone I know
%People we have common friendships with
%Inspirational writing
%Probably medical issues since my twins are affected and have been for twenty two years - and they are special need.
%Receives care at same hospital
%Health care professional credentials
%Honesty and specific details
%Someone who can use extra support  Young adult rare disease especially neurological and even un undiagnosed complex conditions
%Age of patient (young adult), multiple posts all the way through death. I wanted to see what I would probably be writing as time progressed. A glimpse into the future if you will. 
%If it were in Spanish
%Most important would be shared diagnosis. Also same age (caregiver in their 20's, 30's caring for a parent in their 50's or 60's.)
%CaringBridge is unique because it allows people to discuss the positive and negative aspects of life. Doesn't censor the writings of people who feel victimized.

\subsection{Recommendation Emails}
\label{sec:rec_email}

In total, we sent 4,190 recommendations to 79 participants, with 526 unique sites recommended.
To evaluate the effectiveness of the Site Suggestion emails as a recommendation interface, we analyzed the rate of clicks on recommendations (sec. \ref{sec:click_rate}) and the explicit feedback we received from participants (sec. \ref{sec:explicit_rec_feedback}).
%, and the characteristics of the site previews we included in the emails (sec.~\ref{sec:recommendation_characteristics}).

\subsubsection{Click rate}
\label{sec:click_rate}

\begin{figure}
\centering
\begin{subfigure}[t]{0.81\textwidth}
  \includegraphics[width=\textwidth]{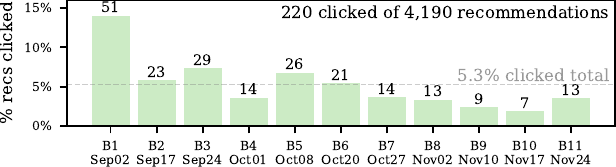}
\caption{Recommendation clicks by email batch}
\label{fig:sub:batch_clicks}
\end{subfigure}
\begin{subfigure}[t]{0.18\textwidth}
  \includegraphics[width=\textwidth]{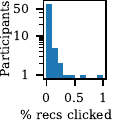}
\caption{By participant}
\label{fig:sub:participant_clicks}
\end{subfigure}
\caption{Recommendation click counts and percentages by email batch and participant. Participants are shown grouped by decile; 49 participants never clicked a recommendation, while 2 clicked more than half.}
\label{fig:click_counts}
\Description{220 clicked of 4190 recommendations. Click counts for the 11 weeks are: 51, 23, 29, 14, 26, 21, 14, 13, 9, 7, 13. 5.3\% of recommendations were clicked in total.}
\end{figure}

Over the 11 weekly batches of Site Suggestion emails, participants clicked 220 (5.3\%) of the 4,190 recommendations.\footnote{We would like to estimate click rate conditional on opening the email (i.e.~the click-through rate) independently from the base click rate, but for logistical and ethical reasons we did not use email trackers in the email HTML so we assume every email was received and opened.}
Figure \ref{fig:click_counts} breaks down observed clicks by email batch and by participant.  Clicks are approximately log-normal; only 30 of 79 (38\%) participants clicked any recommendations, and the most active participant, who we designate P1, was responsible for 54 (24.5\%) of the recommendation clicks---and the majority of the interactions, as we will see in sec.~\ref{sec:observed_study_behavior}.

\subsubsection{Explicit feedback}
\label{sec:explicit_rec_feedback}

\begin{table}[]
\caption{Explicit quantitative feedback to 11 Site Suggestion emails from 8 total participants.}
\label{tab:explicit_feedback}
\begin{tabular}{@{}rlrl@{}}
\toprule
\multicolumn{2}{l}{Recommendations interesting in general?} & \multicolumn{2}{l}{Specific recommendations relevant?} \\
\midrule
 & & Very Relevant & 20 \\
Yes & 4 & Somewhat Relevant & 6 \\
Unsure/Neutral & 3 & Unsure/Neutral & 8 \\
No & 3 & Somewhat Irrelevant & 17 \\
& & Very Irrelevant/Offensive & 5 \\
\bottomrule
\end{tabular}
\end{table}

%No authors of recommended sites contacted us or the CaringBridge support team with any questions or feedback related to our study.
%We discuss the observed interactions between our participants and authors of recommended sites in sec. \ref{sec:interaction_behavior}.
Only 8 participants provided explicit feedback on the recommendations, across 13 responses. 
No participants provided recommendation feedback after batch 7, so responses reflect initial impressions.
Table \ref{tab:explicit_feedback} summarizes the quantitative feedback received: only 4 of 10 responses found the recommendations interesting in general, while 46\% of responses to specific recommendations were deemed relevant compared to 39\% irrelevant.\footnote{As the sample is small and the textual feedback is more informative, we conducted no further statistical analysis on recommendation relevance feedback.} 
Five of the recommendations were deemed ``very irrelevant or offensive'', and associated text feedback can tell us why.
Participants objected to sites due to having poor writing, describing the death of a patient (\textit{``don't want to read about ppl dying of cancer''}), being ``too religious'', and describing patients of a very different age than the participant. 
On the positive side, participants valued the ``wide range of experiences'' present in the recommended sites. One participant was surprised by recommendations for different medical conditions but describes realizing that \textit{``what is important to me is to see and be inspired by how others deal with} any \textit{difficult situation.''}  % emphasis added

6 participants unsubscribed during the course of the study. 
5 provided unsubscription motivations:
2 participants indicated a disinterest in receiving recommendations (\textit{``I thought it was something else''}).
1 indicated that they were no longer using CaringBridge due to the death of their loved one.
2 indicated a lack of time to engage with the recommendations, with one adding in addition: \textit{``it's kind of depressing. Since we're already going through cancer treatments, it's hard to look at what other people are going through.''}
We received only four replies to our final ``thank you'' email.
The one non-clicking participant who replied indicated they had seen a few of the Site Suggestion emails but focused on using CaringBridge to get support from their existing network. 
The remaining three replies emphasized the importance of identifying some kind of common ground when reading a stranger's site. 

%Based on the recommendations sent in the first batch, 85.2\% of previews reported the patient's health status in some way. 
%These disclosures were more likely to be positive (31.5\%) than negative (27.9\%).

\subsection{Observed behavior on CaringBridge}
\label{sec:observed_study_behavior}

We describe the observed impact of the recommendation intervention on behaviors: reading (sec.~\ref{sec:reading_behavior}), interaction (sec.~\ref{sec:interaction_behavior}), and second-order effects on other behaviors (sec.~\ref{sec:retention_outcomes}).  
Participant reading and interaction behaviors are the primary outcomes and are summarized in Table \ref{tab:behavior_outcomes}.

\begin{table}[]
\caption{Observed participant behavior in response to Site Suggestion emails, from the start of the study (September 2021) to the end of data collection (April 2022).}
\label{tab:behavior_outcomes}
\begin{tabular}{@{}lrlrlrl@{}}
\toprule
 & \multicolumn{2}{c}{Recommendations} & \multicolumn{2}{c}{Participants} & \multicolumn{2}{c}{Recced sites} \\
 Behavior & $n$          & \% (of 4190 total)        & $n$               & \% (of 79 total)            & $n$               & \% (of 526 total)            \\ \midrule
First Visits/Clicks & 220 & 5.5\% & 30 & 38.0\% & 158 & 30.0\% \\
%Repeat Clicks? \\ % we could include repeat clicks, but they're poorly defined; do we only count Cloudfront clicks? If yes, count is wrong. If no, hard to map site_profile updates to cloudfront clicks
Second Visits & 86 & 2.1\% & 17 & 21.5\% & 76 & 14.4\% \\
\quad Repeat Visits & 589 & - & 17 & 21.5\% & 76 & 14.4\% \\
Follows & 24 & 0.1\% & 5 & 6.3\% & 23 & 4.4\% \\
Initiations & 36 & 0.9\% & 9 & 11.4\% & 33 & 6.3\% \\
\quad Interactions & 948 & - & 9 & 11.4\% & 33 & 6.3\% \\
\quad Text Interactions & 268 & - & 4 & 5.1\% & 20 & 3.8\% \\
Relationships & 1 & 0.0\% & 1 & 1.3\% & 1 & 0.2\% \\
\bottomrule
\end{tabular}
\end{table}

\subsubsection{Reading behavior}
\label{sec:reading_behavior}
%Table \ref{tab:behavior_outcomes} presents counts for both repeat visits and site Follow actions. 
Follows were rare; only five participants ever followed a site. Repeat visits were more common: 86 (39.1\%) of 220 clicked recommendations were visited a second time and collectively accrued 589 repeat visits during and after the study.

\subsubsection{Interaction behavior}
\label{sec:interaction_behavior}
Nine participants interacted with recommended sites, which is 30\% of those who visited at least one recommended site. 
Collectively, 33 sites received 948 additional interactions as a result of this study, although the majority of these accrue to just a few sites: median interactions with a single site was 6 (M=28.7; SD=56.9).
Participants initiated with 38 non-recommended sites during the course of the study as well---likely with sites authored by people they already knew~\cite{levonian_patterns_2021}. 
Thus, participation was associated with a 94.7\% increase in total initiations during the study period.\footnote{At an average of 0.37 initiations per author during the study period, participants also initiated more often than the 0.05 initiation average for non-participating active authors. For context, participant initiations comprised 0.2\% of all the author initiations that occurred on CaringBridge during the study period, from among 39.5K active authors.}
71.7\% of the interactions were reactions: we discuss text-based interactions next.\footnote{This proportion is similar to the percentage of reaction interactions for all users (71.6\%) and all participant interactions with non-recommended sites (71.2\%) during the same period.}

Only four participants interacted using comments or guestbooks, resulting in 20 participant/site dyads in which interaction occurred.
Our qualitative analysis identified two areas of interest: \textit{first-contact strategies} and \textit{potential norm violations}. 
We observed a diversity of \textit{first-contact strategies}, varying from formally introducing the self (\textit{``I am managing a CaringBridge site for my sister. ... Your site was mentioned in an email from CaringBridge, I took interest in your story.''}\footnote{Three participants described the Site Suggestion emails as coming ``from CaringBridge''.}), to general expressions of support (\textit{``Hope you get some rest soon''}), to establishing common ground by sharing their personal health experiences (\textit{``I kind of know the road you are traveling. [personal health history]''}).
A larger study could investigate first-contact strategies that are particularly effective, either at encouraging a response or at providing useful support.
In addition, we observed \textit{potential norm violations}: situations where participant comments diverge from other visitor comments.
These divergences include specific behaviors: first-time sympathetic comments on posts announcing the death of the patient, being the first and only commenter on a Journal update, bringing in potentially-unwanted religious messaging (\textit{``God bless you''}), and asking explicit questions of the update author.
The commenting norms perceived by `power user' peers may diverge from most visitors, enabling them to provide support in situations others may not but running the risk of producing uniquely negative experiences. 
P1 provided the only comment on a death announcement update: \textit{``I am so very, very sorry for the passing of your loved one.''}
We have no evidence of how these comments were received by authors of recced sites; we observed no objections as comment replies or indirectly via CaringBridge help/support lines.

The author of a recced site responded to a participant comment in only a single instance,\footnote{The observed 5\% reciprocation rate to participant text initiations diverges from the 12\% reciprocation rate observed by Levonian et al.~\cite{levonian_patterns_2021}.} and their subsequent interactions expanded into a relationship.
While one relationship is insufficient for generalization, we sketch it here as a rich example. P1's first contact with R1 shared personal experiences and offered support.
\textit{``You don’t know me but you sound like you’re handling this all very well. .... I know what it’s like to get chemo and I know what it’s like after.''}
Within weeks, R1 left supportive comments on P1's recent Journal updates (\textit{``I can't even begin to imagine how strong you are''}).
Support exchange continued, with evidence that off-CaringBridge communication had been established.
Two months into their relationship, R1 explicitly referenced P1 in a Journal update (\textit{``I have a friend, I think she'd agree with that title.  She somehow found me through Caring Bridge, and has been a constant support to me and my family since.''})
As a frequent commenter, P1 occasionally generated further discussion (Visitor: \textit{``I agree with [P1].''}) and on one occasion was thanked for their support by R1's relative.
While we cannot quantify the expected number of relationships per recommendation from this single example, we take the creation of a new relationship as an important `existence proof' for the expected effects of peer recommendation.
%benefits to peer recommendation, although it is hard to produce estimates for the expected number of supportive relationships if more peer recommendations were given. 
P1 is much more active than other authors, being on the 97th percentile by number of Journal updates and the 99th percentile by number of initiations and interactions. Will authors that are less active still form peer relationships given the opportunity?
We estimate the sample sizes needed for a larger trial in sec.~\ref{sec:effect_size_estimation}.

\subsubsection{Second-order effects on participant and recommended site behavior}  % consider: Secondary behaviors or Peripheral behaviors
\label{sec:retention_outcomes}

\begin{figure}
    \centering
    \includegraphics{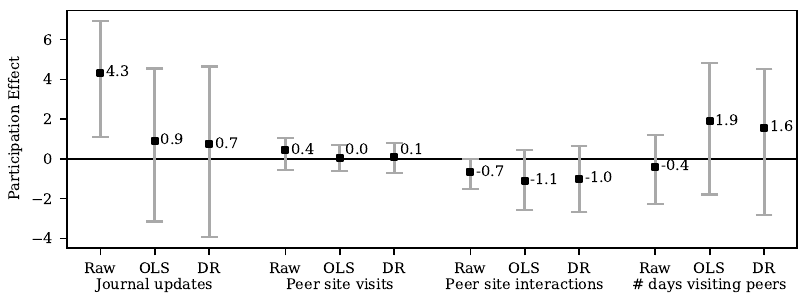}
    \caption{Participation effect: Estimated impacts of receiving Site Suggestion emails on author behavior in the 90 days after the study. Raw estimate is the mean difference between the participants and the pseudo-control group of unenrolled but eligible authors. The OLS estimate adjusts for pre-study activity on CaringBridge while the doubly robust (DR) estimate adds additional causal identification assumptions. 95\% confidence intervals are computed via bootstrapping (1000 iterations).}
    \label{fig:participant_outcomes}
    \Description{A significant Participation Effect is visible only for the raw Journal updates outcome; OLS and DR estimates have CIs that contain 0. All other outcomes and models have CIs that contain 0.}
\end{figure}

\begin{figure}
    \centering
    \includegraphics{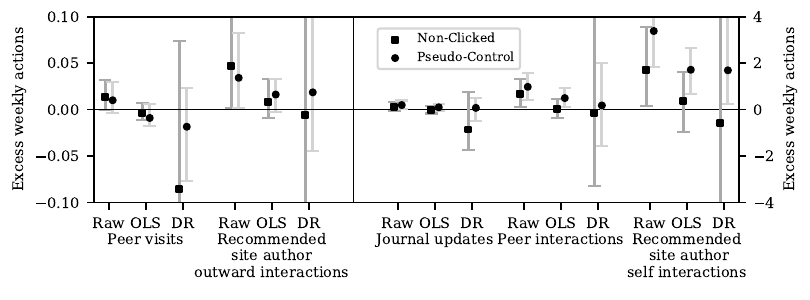}
    \caption{Visit effect: Estimated impacts of receiving a stranger visit on recommended site behavior in the 90 days after the visit when compared to non-clicked recommendations (square, dark-gray) and pseudo-control recommendations (circle, light-gray). Peer visits and interactions count \textit{non}-participant author behavior directed at the site. Recommended site author interactions count behavior of authors of the site. The OLS estimate adjusts for pre-study activity on CaringBridge while the doubly robust (DR) estimate adds additional causal identification assumptions. 95\% confidence intervals are computed via bootstrapping (1000 iterations).}
    \label{fig:recommendation_outcomes}
    \Description{Excess weekly actions for non-clicked recommendations and pseudo-control recommendations, both raw and modeled (OLS and DR). Actions include: Peer visits, Recommended site author outward interactions, Journal updates, Peer interactions, Recommended site author self interactions.}
\end{figure}

Estimates of the participation effect are shown in Figure \ref{fig:participant_outcomes}.\footnote{Additional outcomes and analysis for both participants and clicked sites are provided in Appendix~\ref{app:sec:retention_outcomes}.}
While the associational effect of recommendation indicates that participants published more Journal updates than non-participants, we suspect this difference is primarily due to participants' shorter tenure, and the effect disappears after adjusting for author tenure. We present a selection of outcomes related to visiting and interacting with others' sites (specifically excluding any sites recommended during the study); none of these estimates suggest a positive or negative effect at the 95\% significance level.
We conclude that receiving recommendation emails had a small or high-variance effect on participant behavior---and no clear harms.

Estimates of the visit effect are shown in Figure \ref{fig:recommendation_outcomes}---derived by comparing clicked sites to both non-clicked recommendations and to lower-scoring sites that were not recommended (see Appendix~\ref{app:sec:retention_outcomes}).
The impact of an individual visit should be small, so we expect and observe small, statistically indistinguishable differences.
We observe that peer visits from clicking participants are associated with additional author interactions on their own site and on peer sites, although these effects disappear after adjustment.\footnote{A notable exception in Figure \ref{fig:recommendation_outcomes} is that visited sites have on average ~2 additional self-interactions per week compared to the pseudo-control sites, although pseudo-control sites already had fewer self-interactions pre-study.}
Absent clear evidence of a harmful effect on associated behaviors and given the successful primary behavior manipulation, we recommend proceeding to a randomized controlled trial (RCT) for the peer recommendation intervention.
%, and discuss effect and sample size estimates in sec.~\ref{sec:effect_size_estimation}.

%Matching produces causal estimates of the recommendation intervention for the \textit{participant} population~\cite{hernan_causal_2020}, i.e.~authors who click on the banner for a research study.  We assume that participant authors are more responsive to peer recommendation than the average author, so an effect size estimate derived from matching will plausibly exaggerate the true effect size in the general author population.

\subsubsection{Sample sizes needed for a powered RCT}
\label{sec:effect_size_estimation}

\begin{figure}
\centering
\includegraphics[width=\textwidth]{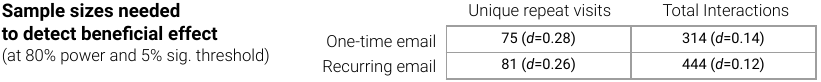}
\caption{Sample sizes needed to detect effects of the magnitude we observed during the field study. Field study effect sizes are shown parenthetically as Cohen's $d$.}
\label{fig:power_analysis}
\Description{Sample sizes needed to detect beneficial effect (at 80\% power and 5\% significance threshold). For a one-time email focused on unique repeat visits or total interactions a sample size of 75 (d=0.28) or 314 (d=0.14) respectively is needed. For a recurring email, a sample size of 81 (d=0.26) or 444 (d=0.12) is needed.}
\end{figure}

Figure~\ref{fig:power_analysis} shows the sample sizes needed for an uncontrolled replication of our field study, based on the observed visit and interaction behavior.  If sending a one-time recommendation email, at least 314 authors should be included in order to obtain a reliable estimate of total peer interactions with (and visits to) recommended sites.
Designing a trial to estimate the effect of recommendations on relationships is more challenging, as no relationships formed due to the first batch of Site Suggestion emails and only one relationship formed during the entire study. Based on that one relationship, a recurring email study would need 478 participants to detect at least 1 relationship per 79 participants ($d$=0.11, $\alpha=0.05$, $\beta$=0.2), although 10 times that number of participants would be needed if the proportion of participants that form relationships is closer to 1 in 1000 than 1 in 100---and tens of thousands are needed if relationship formation is uniformly probable across participant/recommendation pairs (as only 1 in 4190 recommendations led to a relationship).
We also estimated effect sizes relative to the pseudo-control group (see full results in Appendix~\ref{app:sec:sample_size_calculations}).
Based on unadjusted effect size differences between those groups, a 300-participant recurring-email RCT---with 50\% not receiving recommendation emails---should be sufficient to estimate the relative impact of recommendation on total peer visits and interactions. Assuming the same click rate we observed in this study (5.3\%), such an RCT would also be sufficiently powered to investigate any potential negative impact of peer visits on recommended site update frequency.

% \ref{fig:click_counts} - batch 1 has higher click counts

\section{Discussion: Feasibility of a peer recommendation intervention}
\label{sec:discussion_feasibility}

In this study, we implemented a peer recommendation system in order to assess its feasibility as a behavior-change intervention. 
We collected evidence for feasibility in five areas (Table \ref{tab:feasibility_outcomes}): Demand, Implementation, Practicality, Acceptability, and Efficacy.
Here, we summarize our results as evidence for each area.
A top-line summary of results is shown in Table~\ref{tab:quantitative_result_summary}.

\begin{table}[htb]
\caption{Summary of quantitative results.}
\begin{tabularx}{\textwidth}{lXl}
\toprule
Feasibility Area & Result & Section \\ \midrule
Demand & 1 in 3 authors interacted with at least one peer author. & s\ref{sec:prior_use} \\
Demand & Interactions from peer authors increase retention relative to interactions from non-author visitors. & s\ref{sec:prior_use} \\
Demand & 40\% of surveyed authors reported visiting a stranger’s site. & s\ref{sec:survey_prior_use} \\
Demand & 80\% of surveyed authors reported an interest in learning from the experiences of others. & s\ref{sec:survey_motivations} \\
Demand & Surveyed authors were less motivated to interact (46\% interested in communicating with peers, 44\% interested in receiving experienced support.) & s\ref{sec:survey_motivations} \\
Demand & 14\% click rate on the first batch of recommendations, 5\% overall. & s\ref{sec:click_rate} \\
Demand & 62\% (49) participants never clicked a recommendation. & s\ref{sec:click_rate}  \\
Acceptability & A majority of surveyed authors identified a similar diagnosis (84\%), treatment (54\%), and engaging writing (51\%) as important characteristics for potential peer connections. & s\ref{sec:survey_characteristics} \\
Acceptability & Only 10\% of participants (8) gave explicit feedback. In that limited feedback, only 4 of 11 batches and 46\% (26 of 56) of specific recommended sites were deemed interesting. & s\ref{sec:explicit_rec_feedback}  \\
Efficacy & 57\% of participants visited at least one recommended site twice. 39\% of clicked sites were visited at least twice. & s\ref{sec:reading_behavior} \\
Efficacy & 30\% of participants interacted with at least one recommended site. Participants made 948 interactions on recommended sites. 72\% of interactions were non-textual post reactions. & s\ref{sec:interaction_behavior} \\
Efficacy & 4 participants (5\%) left textual comments, and 1 relationship formed. & s\ref{sec:interaction_behavior} \\
Efficacy & Study participation had non-significant impacts on author engagement behavior 3 months post-study. & s\ref{sec:retention_outcomes} \\
Efficacy & Visits from study participants had a nonsignificant impact on the behavior of visited-site authors. & s\ref{sec:retention_outcomes} \\
Efficacy & An RCT aiming to increase peer interaction via a one-time recommendation intervention ought to recruit at least 314 OHC users. & s\ref{sec:effect_size_estimation} \\ \bottomrule
\end{tabularx}
\label{tab:quantitative_result_summary}
\end{table}

\subsection{Demand}

\textbf{CaringBridge users are motivated to read and interact with peer authors---and do so in practice.}
Demand refers to interest in the intervention.
%Specifically, we collected evidence around demand via prior use (sec. \ref{sec:prior_use}), expressed interest (sec. \ref{sec:survey_motivations}), and actual use (sec. \ref{sec:click_rate}).
Prior use indicates both a large number of peer author interactions and a positive correlation between interaction with peer authors and retention (sec. \ref{sec:prior_use}).
Peer interaction includes interactions with peer strangers as well: 40\% of participants reported previously visiting the site of a stranger (sec. \ref{sec:survey_motivations}).
Participants indicated a motivation to connect with peers, although less interest in interaction specifically; 80\% reported an interest in learning from the experiences of others, compared to 46\% reporting an interest in communication ``with a peer who understands''.
These motivations suggest that participants primarily approach peer recommendations as an opportunity to receive social support by reading the experiences of others rather than as an opportunity to form reciprocal relationships or to provide support to others.
Expressed demand was matched in practice with reasonable recommendation click rates---5\% overall, 14\% in the first batch (sec. \ref{sec:click_rate}).\footnote{While comparisons are tricky, these click rates are broadly consistent with other academic studies of email recommendations e.g.~\cite{huang_toward_2016,beel_persistence_2013}.}

\subsection{Implementation}

\textbf{A text-based email interface with recommendations from an implicit feedback model was effective.}
Implementation refers to the tangible design and engineering required to implement the intervention.
%Specifically, we collected evidence around implementation requirements for both the recommendation interface (sec. \ref{sec:sse_design}) and the recommendation model (sec. \ref{sec:model_development}).
Choosing an appropriate recommendation interface is important, and the text-centric email interface we chose may be inappropriate on other platforms (sec. \ref{sec:sse_design}).
%We used email as the recommendation medium to adapt peer recommendation to the CaringBridge context, due to the ubiquity of email notifications on the platform (sec. \ref{sec:sse_design}).
%However, a text-centric email interface may be inappropriate on platforms that emphasize non-text content or want to provide recommendations in ``always-available'' recommendation interfaces that may be more familiar to users.
%Further, email content is static and differs from ``always-available'' recommendation interfaces that may be more familiar to users.
Future studies should adapt the interface design to the platform, integrating recommendations into OHC interfaces as appropriate.
%---although we found that email recommendations were reasonable and accepted by participants.  % one-time and recurring
Representing peer profiles in an interface remains an important open question for future work~\cite{hartzler_design_2016}; text previews were effective, but their focus on recency is a trade-off relative to curated profiles or previews that attempt to highlight the expected utility of the recommended site to the viewer e.g.~via explanations~\cite{yang_seekers_2019}.
The model underlying the interface was based on implicit feedback from historical peer interactions (sec. \ref{sec:model_development}).
%This implicit feedback choice biases  
%The model was developed based on implicit feedback from historical peer interactions.
Future peer recommender systems should carefully consider the available implicit feedback and the implications of optimizing for historical patterns when new interaction patterns are desired.
For example, learning from the experiences of others was the most common participant motivation, but learning was likely not the primary motivation for authors' historical interactions---collecting explicit feedback could supplement or replace this implicit feedback to better align the training objective and user motivations.
%For example, optimizing for first interactions among authors biases recommendations toward moments when first interactions are more likely
% one option: collect explicit feedback
% historical patterns may not align with motivations for viewing recommendations

%Method reflection: multiple recommendation approaches; always-available interface, one-time email, recurring emails. Our choice was reasonable, future studies could choose differently

%Choice of implicit feedback method.

\subsection{Practicality}

\textbf{Prior user interaction data are useful for training recommenders that recreate historical interactions.}
Practicality refers to requirements for administering the intervention in practice. 
%Specifically, we collected evidence of practicality by investigating model quality (sec. \ref{sec:offline_evaluation}), required data (sec. \ref{sec:feature_ablations}), and compute time (sec. \ref{sec:required_resources}).
%We presented an offline evaluation based on historical data that enables model comparison.
During model development, none of the non-personalized approaches produced recommendations that effectively captured historical peer interaction dynamics, but such approaches may still be appropriate in other contexts that want to avoid using sensitive text-based disclosures as the basis for recommendations (Appendix~\ref{app:sec:hyperparameter_search}).
For example, if only activity data is available, it may be reasonable to form peer cohorts based on sign-up time and activity level~\cite{harper_supporting_2007}.
%If interaction data is available, we found that traditional interaction-based recommendation models may be effective at recreating historical patterns without the need for elaborate disclosures.
%The RoBERTa-based similarity approach we considered was generally ineffective; the utility of sensitive health disclosures to peer recommendations needs further consideration, such as using feature extraction approaches targeted to the domain e.g.\ activity or health role classifiers~\cite{yang_seekers_2019,levonian_patterns_2021}.
We did not attempt to compare multiple models during our field study; future investigations will need to link the specific peer connection benefits sought with the type of modeling approach used and conduct appropriate online evaluations.
Given our interest in interaction, we recommended only recent sites in order to make candidate scoring more practical---$\approx$13K authors active within a week vs 1 million total authors---but a system focused on recommending historical information or completed journeys might consider other candidate-generation approaches~\cite{lin_pretrained_2020}.

\subsection{Acceptability}

% to include: These responses provide evidence for the types of peer recommendations that are perceived as most useful and acceptable by CaringBridge users---as well as indicating that preferences are diverse, necessitating personalization.

% todo, thought: how important is health condition?  maybe not so much, echoing prior work
\textbf{Recommendations were inflexible---and often rejected by participants.}
Acceptability refers to how participants react to the intervention.
%Specifically, we collected evidence of acceptability via explicit participant preferences (sec.~\ref{sec:survey_characteristics}) and feedback (sec.~\ref{sec:explicit_rec_feedback}).
Pre-study, the characteristics identified as most important for potential peer connections were a similar diagnosis, treatment, and engaging writing (sec.~\ref{sec:survey_characteristics}). These were the most frequently mentioned characteristics in recommendation feedback as well, with the addition of similar age.
%These characteristics suggest that participants approached peer recommendations as an opportunity for receiving passive support, without a specific desire to function as a
Only 8 participants gave explicit feedback (sec.~\ref{sec:explicit_rec_feedback}).
To those participants, recommendations were only modestly acceptable; 39\% of specific recommendations were deemed irrelevant, frequently due to a lack of common ground between the participant and the recommended site.
%We chose a modeling approach that optimizes for interactions and does not explicitly reward similarity.
Future peer recommenders might consider including model features that capture the specific aspects of similarity deemed most important to users. For example, users might volunteer information about health condition, or it might be inferred from existing disclosures~\cite{li_condition_2018}.

We received three explicit requests from users to alter the types of recommendations they were receiving, and one request from a user for the ability to filter the recommendations.
Functionally, these requests speak to the coordination role played by a recommendation system.
In clinical contexts, a human coordinator can incorporate this feedback to alter their peer matching approach on an individual level~\cite{taylor_peer_2016}.
An algorithmic recommendation system faces a much greater challenge soliciting useful feedback~\cite{terveen_social_2005}.
By designing feedback interfaces, recommendation users could \textit{self}-coordinate, at the risk of trade-offs between user learning and ease of use~\cite{gajos_people_2022}.
%Designing for self-coordination introduces trade-offs between user learning, system explanations, and ease of use~\cite{gajos_people_2022}.
%An alternative might be designing peer matching systems with a human coordinator ``in the loop''; the recommendation system could function as an assistance tool to facilitate matching thousands of peers simultaneously.

\subsection{Efficacy}

\textbf{Receiving recommendations increased peers' reading and interaction behavior---and did not reduce other usage.}
Efficacy refers to how much the intervention affects the desired behaviors.
%We collected evidence of efficacy by examining both reading and interaction behavior effects (secs. \ref{sec:reading_behavior} and \ref{sec:interaction_behavior}) as well as second-order behavior effects (sec. \ref{sec:retention_outcomes}).
Given the small sample and the uncontrolled design, we can only evaluate efficacy in a limited way, although our results are promising. 
39\% of visited sites were visited a second time and a majority (57\%) of participants who clicked at least once went on to visit at least one recommended site twice, suggesting that participants were interested in reading about ongoing health journeys (sec. \ref{sec:reading_behavior}).
A smaller percentage (30\%) of participants interacted with at least one recommended site, and only one reciprocal relationship formed as a result of the recommendations (sec. \ref{sec:interaction_behavior}).
These low percentages indicate likely barriers to forming deep connections that the current design is ineffective at overcoming.
Unclear social norms around interaction with strangers may present a barrier to greater interaction, such as whether an author is open to receiving unsolicited support~\cite{levonian_patterns_2021}.
Future design work could explore soliciting indicators of openness to interactions from strangers and providing first-contact writing assistance for peer supporters who are not sure what to say~\cite{smith_what_2021,peng_exploring_2020}.
We observed no evidence of second-order engagement harms due to recommendations.
These results suggest that efficacy should be evaluated in a larger, controlled trial.

All of the interactions we observed in this study---excepting the one relationship---saw our participants providing support to others.
Being a peer supporter may by more attractive and accessible than being a support recipient~\cite{taylor_peer_2016}.
On one hand, a recommender system that encourages a supporter role runs the risk of generating self-fulfilling prophecies and preventing exploration of other roles~\cite{gatos_how_2021}---leading users to deprioritise their own needs~\cite{allison_logging_2021}---as well as contributing to exclusion of new users and reinforcement of existing cliques~\cite{nakikj_park_2017}.
On the other hand, exposure to peer recommendations might lead to \textit{more} exploration than what users choose without this scaffolding~\cite{su_effect_2016}, and supporting others might serve as the endpoint for a personal transition to working in broader service of a health community~\cite{introne_narrative_2021,levonian_bridging_2020}.
In general, the impacts of recommendation on community social dynamics are hard to predict~\cite{nakikj_lost_2018}---and it is for that reason that future trials are necessary. 
%Our feasibility assessment increases our confidence that peer recommendation will have positive benefits, including for equity. 
%While only 1 in 40 of pre-recommender initiations are with sites without prior peer interactions, 1 in 10 of the model recommendations are for these siloed sites.
%1 in 10 of the model's recommendations are for sites without peer initiations, compared to 1 in 40 of the 
%a higher rate than the 1 in 40 historical initiations from peers.
%compared to 1 in 40 actual, pre-recommender initiations.
In future studies, recommendation and visit dynamics should be carefully monitored for harmful social dynamics or intervention-generated inequalities~\cite{veinot_good_2018}.
% what percentage of initiations are with siloed people?  2.8\% during the test period

%Being a supporter can be the culmination of a transition from someone needing social support to someone working in broader service of the community~\cite{introne_narrative_2021,levonian_bridging_2020}.

%To the discussion point on reinforcing existing patterns of communication or siloing people: \cite{nakikj_park_2017}. Particularly relevant if we talk about/do an analysis on new members specifically. (i.e. does our recommendation model seem to exclude new users?)
%Further, to build on that point, there’s a more general problem: unintended negative impact on the social dynamics within a community~\cite{nakikj_lost_2018}. It was for that reason that we did a feasibility study in the first place, and while we are more confident now that the types of changes induced are likely to be positive it’s still something to monitor for in larger trials (and indiscriminately deploying on the basis of the current study’s findings would be irresponsible).

%``Members in need of support may deprioritise their own psychosocial needs if they feel expected to help others'' \cite{allison_logging_2021}.
%Risk of becoming a self-fulfilling property; recommendations could take away from people's exploration of other patterns of use/roles ~\cite{gatos_how_2021}. (But, incredibly hard to reason about this problem: recommendations might lead to \textit{more} exploration than the behaviors users choose on their own \cite{su_effect_2016}, e.g. word-of-mouth and off-platform link sharing.)

\section{Discussion: Designing for peer recommendations}
\label{sec:discussion_facilitation}

\subsection{Design implications}
% design implications for peer support

%implications for system design for health-related social support? 
%did the authors observe any effect of the intervention on community building?

\textbf{Give recommendations to OHC users.}
Despite negative reception in the limited explicit feedback we received, we observed no quantifiable harms to user engagement.
Taken together with evidence of positive benefits---specifically, an increase in reading and peer interaction---our results suggest that designers should try giving recommendations to OHC users.
In other words: recommendations probably won't hurt and might help.
Some users will ignore or dislike recommendations, as with the participant who found reading other sites ``depressing''; others will use recommendations to discover sites they find valuable. 
%It is feasible to deliver recommendations to OHC users, and giving these recommendations will enable deeper study of the benefits and risks.
We also see opportunities for disrupting ``rich get richer'' social network effects, using recommendations to promote equitable support exchange e.g. by recommending new sites that might derive more benefit from incoming support~\cite{stray_building_2022}.
On CaringBridge, while only 1 in 40 of non-recommended initiations are on sites without any prior peer interactions, 1 in 10 of the recommendations we gave during the field study are for those ``siloed'' sites.
%In most OHCs, cold-start recommendations based on prior implicit interaction data will result in a wide diversity in the types of recommended peer content.
%Diverse recommendations were well-received in some participants' explicit feedback and represent a good trade-off between usefulness and equitable support exchange e.g. by recommending active sites that are more likely to be new and might derive more benefit from the support~\cite{stray_building_2022}.
%Deploying recommendations to OHC users is also an opportunity to design new metrics based on implicit feedback.
%The recent history of recommender systems development involves a shift from explicit to implicit feedback~\cite{zhao_explicit_2018}, echoed in our ease building a recommendation system from existing implicit usage data. But creating a recommender system that embodies values like well-being and safety requires a process to operationalize those values as metrics from implicit feedback~\cite{stray_building_2022}.
%Designers can use small-scale feasibility studies like this one to encode the assumptions around implicit feedback metrics needed to evaluate if recommendations are facilitating activity aligned with those values.

\textbf{Let users shape the recommendations.}
A theme in participant feedback was requests for more or less of a particular ``type'' of recommendation.
Explicit feedback enables users to shape the type of recommendations they receive, such as by changing the training data or specifying a specific model to use based on their motivation for viewing peer recommendations.
Prior work demonstrates that people will explore multiple recommender options if given the opportunity, and some people will even opt for non-personalized recommenders e.g. only sites with a specific health condition~\cite{ekstrand_letting_2015}.
Designers could add controls that highlight particular kinds of filtering, e.g. for sought characteristics like particular treatments or geographic proximity.
%salient in our study was spirituality e.g.\ more/less explicitly Christian and geographic proximity.
Based on a provided motivation, implicit feedback models could rely on different training signals and assumptions.
Our model assumed that interaction was the primary goal, but a model could be trained for sites that users want to read (trained on repeat visits), for sites that could benefit from additional comments (trained on first comments), or for sites on which a relationship is more likely to occur (trained on reciprocal interactions).
Users could select recommendations appropriate for their motivations, either as a support seeker or a support provider.

\textbf{Provide ``first-contact'' guidance.}
% mention community building here
The email recommendation interface we used in this feasibility assessment provided minimal guidance on intended usage.
We observed a variety of approaches to writing initial comments on a stranger's site, as well as multiple potential social norm violations. 
Prior work in CSCW has identified providing design support for navigating communicative norms as a priority, especially when engaging with potential peer supporters~\cite{andalibi_responding_2018}.
In other online communities, a visitor's first interaction is an important time to intervene, potentially by demonstrating appropriate norms~\cite{chandrasekharan_internets_2018}.
%When leaving an initial reaction, the interface might suggest reactions received on that site or similar sites in the past.
When composing an initial comment, the interface might highlight comments written on previous Journal updates or provide some other form of writing guidance such as a prompt~\cite{levonian_patterns_2021}.
%To facilitate supportive interaction, 
A recommendation interface can lead to the initial visit, but designing to reduce the friction of providing context-appropriate interaction will make it easier to provide meaningful support and could increase the likelihood of forming deep relationships.
Due to the siloed nature of CaringBridge sites, we observed minimal cross-site community building during this study, but on more collective OHCs the stakes for encouraging appropriate initial participation might be higher.
Recommendations alone are evidently not sufficient to cultivate relationships or broader community, but future qualitative and design work could focus on the flow from recommendation to first contact to meaningful and supportive relationships.

\subsection{Limitations \& Future Work}

Our feasibility assessment of peer recommendations has significant limitations, although we are excited by the prospect of additional research developing the systems and interventions we considered.
We developed and evaluated only a single interface and a single model. Our interface followed the existing design of CaringBridge notification emails, but a more iterative design process could produce more useful representations of recommendations~\cite{hartzler_leveraging_2016}.
The deep learning model we use is not ``state-of-the-art'', but represents a reasonable modern approach. Future studies should conduct more extensive offline and online model comparisons~\cite{beel_comparison_2015}---including with models that are focused on specifically facilitating reciprocal matching~\cite{palomares_reciprocal_2021}.
%Recommendation model development and design process could both be deeper, but they are good enough for accomplishing our research objectives.

We looked at the potential impact of the recommendations for 12 weeks during the field study and 13 weeks post-study.
We saw no influence of participants' visits on the journaling behavior of recommended sites in the 13 weeks post-study, but our dataset prevents us from analyzing any longer-term influence on the authors of recommended sites.
In particular, authors that perceive a change in their audience after visits or interactions from peers might change the \textit{content} of their Journal updates.
Writing updates involves self-disclosure, and self-disclosure can lead to vulnerability and potential disappointment in the absence of reciprocation~\cite{stepanova_strategies_2022}.
For an audience of peers, authors may be less likely to disclose negative information \cite{lepore_comparing_2014}, which could decrease the potential benefits of expressive writing and self-reflection~\cite{ma_write_2017} and, unintentionally, lead to posts that are less useful for both peer and non-peer readers.
Future studies should collect explicit feedback from authors to understand authors' perception of peer visitors.
%The relationship between perceived audience and self-disclosure behavior remains an important open question to consider before deployment of peer recommendation systems that may change that audience.
%Peers vs non-peers plausibly different here, so both potential opportunities but also greater potential harms if peer interaction is emphasized.

As an intervention, the peer recommender system we evaluated was intended to target behaviors correlated with social support benefits.
We proposed an RCT focused on quantifying the increase in reading and interaction as a result of exposure to peer recommendations.
However, due to the gap between received and perceived support~\cite{rains_coping_2018}, an increase in these behaviors may not produce corresponding increases in perceptions of social support.
%supportive comments may not result in perceptions of greater support.
Future experimental designs could include pre-post self-report measures, either for social support directly (e.g. using the social connectedness scale~\cite{lee_measuring_1995}) or for desired impact on downstream health (e.g. using the perceived stress scale~\cite{lee_review_2012} or measures for health-related quality of life~\cite{centers_for_disease_control_and_prevention_measuring_2000}).
\rev{Ultimately, the purpose of this feasibility study is to support the future execution of an RCT that moves beyond engagement metrics. An RCT for a peer recommendation intervention should evaluate not only that perceived social support increases for impacted users, but also that the increases in social support induced by peer recommendation are positively associated with more general self-reported health measures.}

We designed and evaluated the peer recommender system only with users of a single OHC.
We argue these results generalize most to OHCs with similar affordances to CaringBridge.
Rains argues that communication technologies for social connection have four primary affordances: visibility, availability, control, and reach~\cite{rains_coping_2018}. 
Control, for example, is the potential to manage interactions, while availability is the potential to interact at particular times or places, so other text-based asynchronous communities provide similar potential.
Visibility is the potential to make one's self known to others or to observe others' behavior; on CaringBridge, visibility is linked to specific blogs and communities that form around those blogs.
Forums or listservs offer very different trade-offs around visibility of user-generated content, as do OHCs with rich user profiles~\cite{hartzler_leveraging_2016,gatos_how_2021}.
Reach is the potential to contact specific individuals, groups, or communities: CaringBridge provides reach only to specific individuals known to the user by name~\cite{levonian_patterns_2021}. Peer recommendation might function differently in an ecosystem where other social discovery features are already present~\cite{yang_computational_2019}.
Future work should explore peer recommendation in OHCs with diverging affordances from those provided by CaringBridge---such as Q\&A communities.

We primarily collected quantitative feedback in this study, and the limited text feedback we did collect raised questions that would best be answered with future qualitative work.
Qualitative interviews might identify authors' \textit{motivations for seeking peer recommendations}, authors' \textit{perceptions of their ``imagined audience''}~\cite{litt_imagined_2016}, and the \textit{impact of peer visits} on those perceptions.
In-lab usability studies---representing a midpoint between online and offline evaluations~\cite{zangerle_evaluating_2022,beel_comparison_2015}---might identify what makes a peer recommendation \textit{good}. 
In particular, participants disliked some of our recommendations: in-depth qualitative elicitation methods could identify why these recommendations were ineffective and what kinds of recommendations would meet their peer connection needs.

\section{Conclusion}

This paper conceptualized peer recommendations as an intervention to increase reading and interaction behavior in online health communities.
We implemented and evaluated a peer recommendation system, demonstrating both the feasibility of and challenges associated with running a larger experimental study.
%We demonstrated the feasibility of peer recommendation as an intervention, discussing the design and evaluation of 
%This paper established the feasibility of peer recommendation interventions for online health communities.
%We explored peer recommendation as an approach to make finding peers easier, conceptualizing it as an intervention to increase peer connection behavior. 
Our system meets a demand for learning from the experiences of others and demonstrates the practicality of content-based models implemented in an email interface.
During a 12-week field study, peer recommendation emails successfully led participants to read about and interact with peers---with no evidence of harmful second-order effects.
We hope that our specific implementation serves to promote future investigation of peer recommendation: including models, interfaces, and outcomes for both individuals and communities.

%%%%%%%%%%%%%%%%%%%%%%%%%%%%%%%%%%%%%%%%%%%%%%%%%%%%%%%%%%%%%%%%% END OF DRAFT

%%
%% The acknowledgments section is defined using the "acks" environment
%% (and NOT an unnumbered section). This ensures the proper
%% identification of the section in the article metadata, and the
%% consistent spelling of the heading.
\begin{acks}
We would like to thank Daniel Kluver, Juan F. Maestre, Stevie Chancellor, Martin Michalowski, Haiwei Ma, and C. Estelle Smith for their feedback.
Special thanks to our collaborators at CaringBridge, especially Dennis Still, Brigid Bonner, Patricia McMorrow, Dale Durham, and Jon Linn.
The Minnesota Supercomputing Institute at the University of Minnesota provided computing support.
This research was funded in part by the University of Minnesota Doctoral Dissertation Fellowship.
\end{acks}

%%
%% The next two lines define the bibliography style to be used, and
%% the bibliography file.
\bibliographystyle{ACM-Reference-Format}
\bibliography{umn-caringbridge}

%%
%% If your work has an appendix, this is the place to put it.
\appendix
% Appendix (do not use /appendix here)
\section{Peer characteristics important for peer matching}
\label{app:sec:prior_work_characteristics}

Existing literature suggests a wide range of potential characteristics to incorporate in a peer recommender system.
Table \ref{tab:prior_work_characteristics} lists peer characteristics identified in prior work as salient or important for effective peer matching.  We distinguish these characteristics as either \textit{proposed} as an implication of a particular study, \textit{used} in practice to match peers in a study or support program, or \textit{expressed} by participants as preferences for or barriers to effective peer support. There are characteristics we don't represent in the table, such as abilities/skills~\cite{macleod_be_2017}, specific needs~\cite{eschler_im_2017}, interaction medium preferences (e.g.~email)~\cite{civan_locating_2009}, existing social connections~\cite{civan_locating_2009}, etc.

In addition to the comparative works identified in sec.~\ref{sec:related_work_peer_matching}, Boyes surveyed cancer patients about the importance of specific shared characteristics such as gender, age, and cancer type, although this data is currently unpublished~\cite{boyes_preferences_2018}.
While less relevant to our work with the CaringBridge OHC, Saksono et al.\ surveyed single mothers to compare the importance of demographic, life, and lifestyle characteristics for encouraging physical activity~\cite{saksono_evaluating_2023}.

\begin{table}[]
\caption{Peer characteristics identified in prior work as important for effective peer matching. We included the starred characteristics (*) in our preference survey.}
\label{tab:prior_work_characteristics}
\begin{tabular}{@{}llll@{}}
\toprule
Peer Characteristics & Proposed & Used & \begin{tabular}[c]{@{}l@{}}Expressed\\ Preference\end{tabular} \\ \midrule
\textbf{Health} \\
\quad Diagnosis* & \cite{levonian_patterns_2021,boyes_preferences_2018,anderson_peer_2021,eschler_im_2017,andalibi_considerations_2021,macleod_be_2017,tixier_counting_2016} & \cite{moulton_woman_2013,moulton_woman_2016,moran_theres_2021,walshe_peer_2020,frost_social_2008} & \cite{hartzler_leveraging_2016,embuldeniya_experience_2013,taylor_peer_2016,civan_locating_2009} \\  % cut haldar_patient_2020 from Proposed for space, should also list ceron-guzman_its_2025 in Expressed Preference
\quad Treatment* & \cite{boyes_preferences_2018,eschler_im_2017} & \cite{ieropoli_what_2011,frost_social_2008} & \cite{hartzler_leveraging_2016,taylor_peer_2016,civan_locating_2009} \\
\quad Symptoms* & \cite{oleary_design_2017} & & \cite{civan_locating_2009} \\
\quad Severity & \cite{eschler_im_2017,tixier_counting_2016} & & \cite{sandhu_peer--peer_2013} \\
\quad Timeline & \cite{levonian_bridging_2020} & & \cite{hartzler_leveraging_2016,civan_locating_2009,barta_similar_2023} \\
\quad Health role & \cite{levonian_patterns_2021} & & \cite{hartzler_leveraging_2016} \\
\quad Relevant knowledge & \cite{macleod_be_2017} & \cite{hughes_exploring_2009} & \cite{hartzler_leveraging_2016,heyer_opportunities_2020,barta_similar_2023} \\
\textbf{Demographics} \\
\quad Age & \cite{boyes_preferences_2018,anderson_peer_2021,eschler_im_2017,tixier_counting_2016} & \cite{hartzler_leveraging_2016,moulton_woman_2013,moulton_woman_2016,moran_theres_2021,ieropoli_what_2011,andalibi_considerations_2021,hughes_exploring_2009}  & \cite{hartzler_leveraging_2016,civan_locating_2009,barta_similar_2023} \\
\quad Gender & \cite{boyes_preferences_2018,anderson_peer_2021,eschler_im_2017,andalibi_social_2018} & \cite{lavender_exploring_2021,ieropoli_what_2011,walshe_peer_2020,hartzler_leveraging_2016,hughes_exploring_2009,saksono_evaluating_2023} & \cite{hartzler_leveraging_2016,sandhu_peer--peer_2013} \\
\quad Ethnicity & & \cite{dennis_effect_2009,hughes_exploring_2009,saksono_evaluating_2023} & \cite{butow_what_2007,barta_similar_2023} \\
\quad Sexual orientation & & & \cite{sandhu_peer--peer_2013} \\
\quad Nationality & & & \cite{hartzler_leveraging_2016} \\
\textbf{Life} \\
\quad Geography/location* & \cite{levonian_patterns_2021,anderson_peer_2021,tixier_counting_2016} & \cite{dennis_effect_2009,lavender_exploring_2021,walshe_peer_2020,frost_social_2008,saksono_evaluating_2023} & \cite{hartzler_leveraging_2016,ceron-guzman_its_2025} \\
\quad Cultural values/background* & \cite{boyes_preferences_2018,rehberg_facilitators_2021} & \cite{moulton_woman_2013,moulton_woman_2016}  & \cite{embuldeniya_experience_2013,butow_what_2007,civan_locating_2009} \\
\quad \quad Employment & & \cite{hughes_exploring_2009} & \cite{civan_locating_2009} \\
\quad \quad Religion & & \cite{moulton_woman_2013}  & \cite{barta_similar_2023} \\
\quad \quad Politics & & & \cite{sandhu_peer--peer_2013,barta_similar_2023} \\
\quad \quad Socio-economic status & & & \cite{taylor_peer_2016} \\
\quad \quad Education level & & & \cite{civan_locating_2009} \\

\quad Social role* & \cite{yang_seekers_2019,tixier_counting_2016,simoni_peer_2011} & \cite{moulton_woman_2016} & \cite{hartzler_leveraging_2016} \\
\quad \quad Marital status & \cite{eschler_im_2017} & \cite{ieropoli_what_2011,hughes_exploring_2009,saksono_evaluating_2023} & \cite{hartzler_leveraging_2016,barta_similar_2023} \\
\quad \quad Has children? & & \cite{moulton_woman_2013,moran_theres_2021,ieropoli_what_2011,saksono_evaluating_2023} & \cite{hartzler_leveraging_2016} \\ 
\quad Language & & \cite{moulton_woman_2016} & \\
\textbf{Other} \\
\quad Communication style* & \cite{newman_its_2011} & \cite{hartzler_leveraging_2016} & \cite{taylor_peer_2016} \\
\quad Lifestyle & & \cite{lavender_exploring_2021,saksono_evaluating_2023} & \cite{embuldeniya_experience_2013,civan_locating_2009,ceron-guzman_its_2025} \\
\quad Interests & & \cite{andalibi_considerations_2021,frost_social_2008} & \cite{civan_locating_2009} \\
\quad Personality & & \cite{lavender_exploring_2021} & \cite{heyer_opportunities_2020,taylor_peer_2016} \\
\quad \begin{tabular}[c]{@{}l@{}}Commitment to supp-\\ \quad ort \& recovery\end{tabular} & \cite{oleary_design_2017} & & \cite{embuldeniya_experience_2013,heyer_opportunities_2020} \\
%\quad Existing social connection & & & \cite{civan_locating_2009} \\
\bottomrule
\end{tabular}
\end{table}

\subsection{Alternatives to recommendation}
\label{app:sec:recommendation_alternatives}
Recommendation is not the only available mechanism for facilitating online peer matching. Two notable alternatives are improving search and filter tools and designing enriched profile pages to make it easier to represent one's diagnosis, expertise, and support needs~\cite{gatos_how_2021}.
Search is challenging in situations where a user's needs are known only implicitly to the user or are challenging to express in terms the system will understand~\cite{belkin_anomalous_1980}.
We suggest that peer support finding is an \textit{exploratory}~\cite{aula_query_2003} search task (e.g. see Pretorius et al.'s discussion of person-centered help-seekers~\cite{pretorius_searching_2020}).
Even with rich peer profiles available, it is challenging for searchers to formulate a query that captures their needs and intent~\cite{viswanathan_what_2022,wacholder_interactive_2011}. 
Other search systems for finding people---such as expertise-finding systems---were created based on interfaces designed to capture users' needs in a domain-specific query~\cite{husain_expert_2019}.
\textit{Mindsets} is a recent example of the design work needed to capture domain-specific intents during query formulation~\cite{viswanathan_what_2022}; additional research is needed in the peer support context to capture support seekers' and providers' intents.
In contrast to search, recommendation offers opportunities to engage with potential peers without explicitly articulating a person's current needs.
%Further, a search interface for peers would represent a larger change from the baseline CaringBridge experience than a recommendation email, as the existing search feature on CaringBridge is almost exclusively used to find specific others (see App. \ref{app:sec:search}).

\subsection{Peer matching without deep learning}

Social networking sites have used approaches based on neighborhoods and similarity of interactions to connect with strangers specifically.
Twitter's ``who to follow'' recommendations used an alternative approach similar to PageRank that uses only the existing follow network to make recommendations~\cite{gupta_wtf_2013}.
Guy et al.\ explicitly attempted to recommend strangers in an enterprise setting based on number of shared interests and memberships~\cite{guy_you_2011}.
The modeling problem closest to peer recommendation may be romantic relationship recommendation, a context that aims to encourage interaction between users and values reciprocity~\cite{tay_couplenet_2018,pizzato_recon_2010}.

\section{Use of search feature on CaringBridge}
\label{app:sec:search}

Do CaringBridge users use search to attempt to find information or supporters?
To address this question, we collected a dataset of user-initiated search queries on CaringBridge.  These queries were extracted from internal logs collected between July 4, 2021 and July 10, 2021. % note: UTC
Our dataset contained 103,830 searches comprising 32,722 unique query strings. We preprocessed the query strings by splitting them into tokens based on whitespace.

Based on a random sample of 100 queries with 1, 2, or 3 tokens and visual inspection of additional samples, all queries corresponded to either person names or existing site URL strings.  Thus, we conclude that a very high percentage of searches are for a specific CaringBridge site.  Queries with 4 or more tokens comprise $<1\%$ of the query strings.  Visual inspection suggests that the majority of queries with 4+ tokens are help requests or open-domain queries including spam. We could identify no instances of users searching for e.g.\ a particular health condition, treatment, or symptom; if search is used in this way, it occurs at a low prevalence.
Results were identical analyzing queries made specifically by author users.

\section{Associational evidence of peer interactions on author behavior}
\label{app:sec:assocational_peer_behavior_change}

\begin{table}[]
\caption{Associational differences between sites that receive interactions (ints) from at least one peer author and sites that only receive interactions from visitors. Site tenure is the number of months between the first and last Journal update on a site. \# updates is the number of Journal updates published more than 30 days after the first update. The common language effect size (CLES) is reported where appropriate. All comparisons are significant at the 99.5\% significance level.}
\label{tab:prior_use}
\begin{tabular}{@{}lllll@{}}
\toprule
 & 1+ peer int within 30 days & Non-peer ints only  & Difference & CLES \\ \midrule
Site Count & 92,352 sites & 100,424 sites & -8,072 sites & - \\
%Early Interactions (M; SD) & 375.5 (677.8) & 98.6 (178.5) & 276.9 & 21.3\% \\
%Early Journals (M; SD) & 14.0 (11.4) & 9.8 (9.9) & 4.2 & 64.2\% \\
%Site Tenure (M; SD) & 9.6 (14.3) & 5.7 (11.2) & +3.9mos & 62.4\% \\
Site Tenure (Median) & 3.9 months & 0.8 months & +3.2 months & 62.4\% \\
\# updates (M; SD) & 20.5 (47.6) & 12.4 (35.7) & +8.1 updates & 62.1\% \\
\# updates (Median) & 8 updates & 2 updates & +6 updates & - \\
\% sites with 2+ updates & 78.1\% & 59.5\% & +18.7pp & - \\
\bottomrule
\end{tabular}
\end{table}

Based on sec. \ref{sec:benefits}, we would expect that peer interaction is associated with behavior change for authors.
Prior work demonstrates that receiving interactions from visitors---peers and non-peers---is associated with retention on CaringBridge~\cite{wan_how_2020}, but are peer interactions more impactful than non-peer interactions?
%In addition to retention, we consider two additional behaviors: number of published Journal updates and number of future interactions with fellow authors.
Table \ref{tab:prior_use} shows associational differences between CaringBridge sites based on the interactions (reactions, comments, and guestbooks) received from visitors within 30 days of a site's first published Journal update.
Sites for which at least one visitor interaction is left by a peer author will publish on CaringBridge for a median of 3.2 additional months (with 6 additional updates) compared to sites that receive only interactions from visitors.
This analysis was conducted on 192,776 sites created between January 1, 2014 and September 1, 2021 (the start of the study) that received at least 1 visitor interaction in the first 30 days. These general results hold when adjusting for publishing rate, number of interactions received, number of visitors, and number of peer visitors. We omit full modeling tables (Poisson for \# updates and Cox's proportional hazards model for site tenure).
%In general, these correlations are hopelessly confounded; a positive correlation is a necessary but not sufficient condition for the influence of the intervention on retention (a secondary outcome discussed in sec.~\ref{sec:retention_outcomes}).}

%This positive correlation suggests that peer interaction could change author behavior in ways that are themselves correlated with author benefits~\cite{ma_write_2017}.
%A peer recommendation intervention intervenes in this existing ecosystem of peer interaction: we propose a design appropriate for the CaringBridge context in the next section.

\section{Modeling and optimization details}
\label{app:sec:hyperparameter_search}

This section provides additional details beyond those provided in sec. \ref{sec:model_development}.
Modeling and analysis code make primary use of Python's scikit-learn~\cite{pedregosa_scikit-learn_2011}, statsmodels~\cite{seabold_statsmodels_2010}, transformers~\cite{wolf_transformers_2020}, NumPy~\cite{harris_array_2020}, pandas~\cite{mckinney_data_2010}, and Matplotlib \cite{hunter_matplotlib_2007} packages.
The recommendation model was trained using PyTorch~\cite{paszke_pytorch_2019}.

\subsection{Implicit feedback}
\label{sec:implicit_feedback}

We processed all historical CaringBridge author interactions since January 1st, 2010.
We train our recommendation models on \textit{implicit feedback}---the behavioral signals that indicate an author is interested in reading a site or interacting with the author of that site~\cite{rendle_bpr_2009}. 
Three potential sources of implicit feedback are available: (1) \textit{first visits}---when an author visits another author's site for the first time, (2) \textit{initiations}---when an author interacts with a site for the first time, or (3) \textit{reciprocations}---when an author initiates with a site and an author of that site subsequently interacts on the initiator's site.\footnote{A fourth option is \textit{follows}. Twitter, for example, used reciprocal follows as an implicit signal for some user recommendations~\cite{su_experimental_2020}. Due to the unusual nature of Follows on CaringBridge (see sec. \ref{sec:prior_use}) and lack of clarity around this feature's usage, we omit it from consideration.}
Choice of implicit feedback signal can have a significant impact both on what recommendations are shown to users and how those users engage with those recommendations~\cite{nazari_choice_2022}.
We choose initiations as our implicit feedback signal, as it is less noisy than site visits (i.e. a visit may not indicate interest) while more plentiful than reciprocations (i.e. reciprocations are less common than unreciprocated initiations~\cite{levonian_patterns_2021}, which means less data available for model training).
By selecting initiations, we assume that leaving a reaction, comment, or guestbook on another author's site indicates a preference for reading that site and interacting with that author relative to other sites.
Optimizing for recommendations that increase initiations is likely to increase actual initiations~\cite{zhao_explicit_2018}, but we acknowledge a semantic gap between the implicit feedback metric and metrics of interest; a construct like ``perceived social support'' may not increase despite receiving more peer interaction~\cite{lakey_social_1996,del-pino-casado_social_2018}.

For each initiation between a source author and a target site, we generate training data.  
Using initiations presents three complexities: (1) a user may initiate with a site before they become an author, (2) authors may write multiple blogs, and (3) authors may co-write with other authors on a single blog.
We address these complexities by ranking \textit{author/site pairs} rather than sites alone. %by modeling the recommendation problem as ... given a rec-seeking author/site pair.  
For each initiation, we generate one positive sample for each author on the target site (target author/site pairs) \textit{and} each site written by the source author (source author/site pairs).
%Thus, for each initiation, we generate one positive sample for each author on the target site and each site written by the source author.
%If an initiating author does not yet have a site, then no samples will be produced.
An author/site pair is \textit{eligible} if the author has published at least three Journal updates on that site---a minimum activity threshold that ensures sufficient data is available during feature extraction.
%For each initiation, we generate one positive sample for each eligible author/site pair on the target site and each eligible author/site pair containing the source author.
%and the training data contain only initiations where both the source and target pairs are eligible.\footnote{If an initiating author does not yet have a site, then no samples will be produced.}
We include in the training data only initiations between eligible author/site pairs.\footnote{Eligibility is required at the time of the initiation; if an initiating author has not (yet) published three Journal updates on a site, no training samples will be produced for that initiation.}
At prediction time, given a recommendation-seeking source author, we score all \textit{candidate} author/site pairs: eligible author/site pairs that have been active on CaringBridge in the last week and have not previously been interacted with by the source author.\footnote{An active author is one that has created a Journal update, comment, guestbook, or reaction on any site within the last week. By focusing only on recommending active authors, we ensure that a recommendation-seeking author could receive a response from a site's author if they leave a comment.}
%for each site on which the source author has written at least three updates. 
As we ultimately recommend sites to authors, we convert the ranking of candidate author/site pairs into a ranking of sites.
If a site appears among candidate pairs twice (because that site has two authors), we remove all but the highest score from that site.  If a recommendation-seeking author is an author of multiple sites, we merge the scores for each source author/site pair by averaging the scores for each candidate pair.\footnote{These merging strategies reflect the intuition that a good match with any one author of a site makes the site relevant. In practice, these strategies have minimal impact on rankings.}

For each positive sample generated by initiation extraction, we sample an assumed-negative author/site pair to add to the training data.\footnote{While more complex approaches to negative sampling have been recently proposed~\cite{ding_simplify_2020}, we adopt uniform sampling of a single negative as a widely-used baseline.}
The negative author/site pair is randomly selected from among the candidates.
As a history of previous initiations are not required for eligibility, we avoid selection bias issues that arise in models that require user feedback on negative samples~\cite{yang_mixed_2020}.

\begin{figure}
\centering
\includegraphics[width=\textwidth]{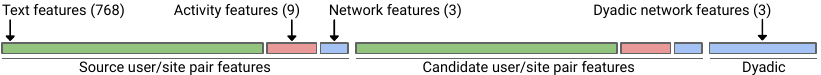}
\caption{Features available for recommendation, including 780 features for the recommendation-seeking \textit{source} author/site pair and the same for each \textit{candidate} author/site pair. Three dyadic features capture the relationship between the source and candidate within the initiation network. Total features: 1563}
\label{fig:feature_summary}
\Description{Visual depiction of the 768 text features, 9 activity features, and 3 network features that compose both the source and candidate user/site pair features, plus 3 dyadic network features.}
\end{figure}

\subsection{Model features}
\label{sec:model_features}

Given two author/site pairs, we use the recent behavior of the authors on CaringBridge to construct a feature representation of the two, summarized in Figure \ref{fig:feature_summary}.
Different models might use subset of these features, but we describe here the full set included in the model deployed during the field study.

\begin{itemize}
    \item \textbf{Text} ($768\times2$ features): We incorporate context from the three most recent Journal updates written on a site.  For each Journal update, we use the pre-trained RoBERTa~\cite{liu_roberta_2019} model available in the HuggingFace Transformers package~\cite{wolf_transformers_2020} to compute size-768 contextualized word embeddings. Then, we mean pool the token embeddings and then the update embeddings to produce a single vector representation, an effective general approach to using word embeddings~\cite{lin_pretrained_2020,reimers_sentence-bert_2019,levonian_trade-offs_2022}.\footnote{Analysis during model development suggested minimal impact of the pooling strategy (mean vs max vs concatenation) on model performance.}
    \item \textbf{Activity} ($9\times2$ features): For each of Journal updates, reactions, comments, and guestbooks, we include the count of that action within the last week and the time elapsed to the most recent action (in hours). In addition, we include the author's current tenure (time elapsed to the author's first published journal update).
    \item \textbf{Network} ($6\times2 + 3$ features): During initiation extraction, we maintain the interaction network between authors as described by Levonian et al., in which new edges are created between authors when an initiation occurs~\cite{levonian_patterns_2021}. For each author/site pair, we include: indegree, outdegree, and weakly-connected component size. In addition, we include three dyadic features: whether the two authors are weakly connected, whether the candidate author is the friend-of-a-friend of the source author (i.e. this initiation would create triadic closure), and whether the candidate author has previously initiated with the source author (i.e.\ this initiation would be a reciprocation to a prior initiation).
\end{itemize}

%We consider the specific utility of these features for recommendation in sec.~\ref{sec:feature_ablations}.

\subsection{Offline evaluation}\label{sec:offline_evaluation_methods}

\begin{figure}
\centering
\includegraphics[width=\textwidth]{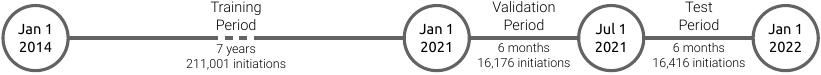}
\caption{Dataset splits and associated author initiation totals. All reported metrics are from the test period.}
\label{fig:evaluation_timeline}
\Description{The training period lasts from January 1, 2014 to January 1, 2021 and contains 211,001 initiations. The validation period lasts from January 1, 2021 to July 1, 2021 and contains 16,176 initiations. The test period lasts from July 1, 2021 to January 1, 2022 and contains 16,416 initiations.}
\end{figure}

Given an initiation, the goal for our recommender system is to rank the site that was actually initiated with as high as possible: ideally in rank 1, above other candidates.
%We select as candidates all authors who have published a Journal update in the last 10 days, who have published at least three Journal updates on a single site, and who the rec-seeking source author has not previously interacted with.
We evaluate various modeling approaches by splitting initiations chronologically into a training, validation, and test set---see Figure \ref{fig:evaluation_timeline} for initiation counts.
%We conducted an offline evaluation to compare models and get a sense for what was possible. 
Offline evaluations can diverge substantially from the usage preferences expressed by recommendation consumers~\cite{beel_comparison_2015}; we conduct one here to compare recommendation algorithms in terms of accuracy, coverage, and diversity of predictions on historical initiations.
%and identify patterns in the prediction of historical initiations.

During evaluation of each initiation, features are captured at the millisecond before the initiation actually occurred: for the source author/site pair (who did the initiation), for the target (who is being initiated with), and for the candidates (who are eligible, active author/site pairs that the source had not previously initiated with).
This evaluation approach intends to reward models that rank a site well if the source author/site pair actually initiated with that site at the time the recommendation was made.
The number of eligible, active candidates varies according to the time of the initiation; the median number of eligible, active authors during the test period was 13,252.  % was this actually the number of authors, or was it the number of author/site pairs?

We used two primary evaluation metrics: mean reciprocal rank (MRR) and hit rate (HR).  Mean reciprocal rank is computed based on the rank $r$ assigned to the target site, where 1 is the best rank given to the highest score.\footnote{Note that we punish ties by using competition ranking e.g. two sites scored the same both get assigned rank 2 and no site is assigned rank 1. At prediction time, we break ties randomly.}
MRR is the mean of the reciprocal ranks ($1/r$) for every test initiation.
Hit rate is the proportion of the time that $r$ is less than some threshold. As we provide 5 recommendations in a Site Suggestion email, we report HR@5 (how often would the model recommend the target site in a five-site set) and HR@1 (how often would the model rank the target site first). When we compare multiple hyperparameter configurations for a model, we select the best on the basis of validation MRR and report test metrics as the median from 3 random seeds.

Beyond accuracy-style metrics, we consider \textit{coverage} as an important secondary goal~\cite{ge_beyond_2010}. 
Coverage has two aspects: (1) how many users can receive recommendations or have their sites recommended and (2) the diversity of sites that are recommended in practice.
The first aspect of coverage is model agnostic and heavily affected by our decision to require three Journal updates to be eligible.  Of 124,051 new authors in 2020, only 51.4\% will publish three updates. Further, of those authors that do publish 3 updates, recommendations cannot be delivered until the publication of the 3rd update---a median wait of 2.5 days, but at least 72.4 days for the slowest-publishing 10\% of authors.
Beyond authors, a hypothetical system that served recommendations to any visitor with an interaction could reach an additional 1.3M new visitors who first interacted in 2020.

The second aspect of coverage---the diversity of recommended sites---is model dependent.
To evaluate the coverage of recommended sites, we randomly sampled 1000 eligible, active authors at the end of the training period\footnote{Specifically, recommendations are generated at the end of the training period + 12 hours. This approach captures a situation where a model is trained nightly and recommendations are generated the next day, as in the field study we conducted.} and produced recommendation sets with the 5 highest-scoring sites for each author. We then compared the sites that were actually recommended ($\text{R}$) to the sites that were not recommended ($\text{N}$), among the 12,432 candidate sites available at noon UTC on January 1, 2021.  % 12,493 authors for 12,432 sites
We consider three plausible goals for diverse recommendations, with an associated metric for each goal.
%The first goal is that a large number of unique sites are recommended rather than a few sites recommended many times; we compute the percentage of candidate sites that are in R ($|\text{R}| / |\text{R} \cup \text{N}|$).
The first goal is that a large number of unique sites are recommended rather than a few sites recommended many times; we compute the percentage of unique recommendations given ($|\text{R}| / 5000$).
The second goal is that newer sites are recommended, a period during which support may be particularly impactful~\cite{guy_you_2009}; we compute the minimum site tenure for each recommendation set and report the mean.
The third goal is that authors without previous interactions from other authors are recommended; we compute the proportion of recommended sites that have no prior connections ($|\text{S}_\text{R}| / |\text{R}|$ for ``siloed'' sites $\text{S}_\text{R}$) and report the ratio to the proportion for non-recommended sites ($|\text{S}_\text{N}| / |\text{N}|$).
If the ratio is less than 1, that model recommends fewer siloed sites than would be expected by chance.\footnote{Note that existing initiations are with siloed sites only 2.8\% of the time during the test period.}

\subsection{Model comparison}
\label{sec:offline_evaluation}

\begin{table}[]
\caption{Offline test performance for various accuracy and coverage metrics. Coverage is reported in terms of number of unique recommended sites ($|\text{R}|$), the percentage of unique recommendations made (out of 5000 total recommendations), and the mean of the minimum site tenure in each recommendation set (MMST). The final column displays the percentage of recced sites that have no prior connections (i.e. are ``siloed'', $|\text{S}_\text{R}| / |\text{R}|$), the same percentage for non-recced sites, and the ratio of those percentages. The bolded model was used during the field study.}
% if space needed, cut the |R| column; not sure it adds anything.
\label{tab:model_comparison}
\begin{tabular}{@{}lrrrrrrr@{}}
\toprule
Model & MRR & HR@1 & HR@5 & $|\text{R}|$ & \%Unique & MMST &  \multicolumn{1}{c}{$\frac{|\text{S}_\text{R}| / |\text{R}|}{|\text{S}_\text{U}| / |\text{U}|}$} \\ \midrule
$\text{MLP}_\text{Tuned}$ & 0.173 & 13.37\% & 20.27\% & 665 & 13.3\% & 6.2 weeks & 10.8\% / 24.6\% = 0.44 \\
\textbf{$\text{MLP}_\text{Study}$} & 0.163 & 12.76\% & 19.00\% & 735 & 14.7\% & 5.6 weeks & 10.6\% / 24.7\% = 0.43 \\
PeopleYouKnow & 0.102 & 7.86\% & 13.07\% & 3972 & 79.44\% & 29.2 weeks & 14.9\% / 28.1\% = 0.53 \\
CosSim & 0.002 & 0.05\% & 0.25\% & 3403 & 68.1\% & 27.1 weeks & 25.2\% / 23.3\% = 1.08 \\
MF & 0.002 & 3.05\% & 15.23\% & 13 & 0.26\% & 40.3 weeks & 0.0\% / 23.9\% = 0.00 \\
RecentInits & 0.036 & 1.27\% & 4.81\% & 6 & 0.12\% & 0.4 weeks & 0.0\% / 23.9\% = 0.00 \\
MostInits & 0.035 & 1.32\% & 4.90\% & 12 & 0.24\% & 112.9 weeks & 0.0\% / 23.9\% = 0.00 \\
NewestAuthor & 0.025 & 0.97\% & 3.19\% & 6 & 0.12\% & 0.1 weeks & 83.3\% / 23.8\% = 3.50 \\
RecentJournals & 0.016 & 0.19\% & 1.58\% & 6 & 0.12\% & 3.1 weeks & 16.7\% / 23.9\% = 0.70 \\
MostInteractive & 0.007 & 0.17\% & 0.68\% & 5 & 0.10\% & 0.9 weeks & 0.0\% / 23.9\% = 0.00 \\
MostJournals & 0.004 & 0.13\% & 0.38\% & 5 & 0.10\% & 0.4 weeks & 60.0\% / 23.8\% = 2.52 \\
Random & 0.001 & <0.0\% & 0.04\% & 4159 & 83.18\% & 17.6 weeks & 23.7\% / 23.9\% = 0.99 \\
\bottomrule
\end{tabular}
\end{table}

We describe the recommendation models we implemented and compared, starting with the model we used during the field study.
Table \ref{tab:model_comparison} presents the performance of these models in the order we describe them.

\textbf{MLP.} We used a multi-layer perceptron (MLP) with two hidden layers as a parsimonious yet effective deep recommender model.  All source, candidate, and dyadic features are concatenated into a single input vector. The output layer uses a sigmoid activation to score the inputs and we optimize the standard pointwise binary cross-entropy loss~\cite{xu_deep_2020}. 
We trained the model for 1000 epochs over the full training data, holding out a random 1\% of the training initiations to compute hold-out loss. 
We report results for two MLP models: $\text{MLP}_\text{Study}$ and $\text{MLP}_\text{Tuned}$.
$\text{MLP}_\text{Study}$ was the result of manual tuning before the study, and best reflects the model configuration used during the field study. $\text{MLP}_\text{Tuned}$ was the result of more systematic hyperparameter tuning (App. \ref{app:sec:mlp_model_details}) and adds additional hidden units, weight decay, and dropout.

We compare the MLP model to two personalized baselines. \textbf{PeopleYouKnow} uses only the dyadic network features, scoring highest the sites on which authors have already initiated with the source, then sites that are friends-of-friends with the source, then sites that are in the same network component as the source, and finally all other sites.
\textbf{CosSim} scores author/site pairs by computing the cosine similarity between source and candidate. Cosine similarity was used by Hartzler et al.\ to match peers by health interest~\cite{hartzler_leveraging_2016}.

We focus on two non-personalized baselines. \textbf{MostInits} scores each site by the number of initiations received by that site within the last week.\footnote{Following the recommendation of Ji et al., we use a popularity baseline that takes into account the time point when a user interacts with the system~\cite{ji_re-visit_2020}.}
In other recommendation contexts, popularity baselines like MostInits are strong contenders; not so on CaringBridge, where popularity follows a flatter distribution.  MostInits has the best MRR of several other plausible non-personalized activity baselines (defined in App.~\ref{app:sec:hyperparameter_search}).
\textbf{Random} ranks sites randomly, and is included as a useful point of comparison.

Table \ref{tab:model_comparison} compares the models by accuracy and coverage metrics. 
The best model ($\text{MLP}_\text{Tuned}$) would recommend the site that was actually initiated with more than 20\% of the time; the study model ($\text{MLP}_\text{Study}$) performs slightly worse.
Both MLP models are more likely to recommend newer sites than the baseline models, but less likely to recommend a variety of sites and sites that had not previously received an initiation.
PeopleYouKnow achieves impressive MRR and HR metrics, demonstrating the strong utility of the dyadic features; however, when the source and target are not connected, it predicts randomly (hence the high percentage of unique recommendations).
CosSim recommends the largest number of siloed candidates, although low MRR indicates that similarity alone is a poor predictor of historical initiation behavior.
%Based on an earlier form of this analysis, we selected the MLP model for use during the field study.

\subsection{Feature ablations}\label{sec:feature_ablations}

\begin{table}[]
\caption{Offline performance of an MLP model trained with combinations of the activity (A), network (N) and text (T) feature sets. The model using all feature sets (A+N+T) is $\text{MLP}_\text{Tuned}$ from Table \ref{tab:model_comparison}.}
\label{tab:feature_ablations}
\begin{tabular}{@{}lrrrrrrr@{}}
\toprule
Feature sets & MRR & HR@1 & HR@5 & $|\text{R}|$ & \% Unique & Site Age & \multicolumn{1}{c}{$\frac{|\text{S}_\text{R}| / |\text{R}|}{|\text{S}_\text{U}| / |\text{U}|}$} \\ \midrule
A+N & 0.204 & 16.12\% & 23.62\% & 598 & 12.0\% & 0.9 weeks & 12.0\% / 24.4\% = 0.49 \\
A+N+T & 0.173 & 13.37\% & 20.27\% & 665 & 13.3\% & 6.2 weeks & 10.8\% / 24.6\% = 0.44 \\
N+T & 0.144 & 11.75\% & 16.58\% & 803 & 16.1\% & 5.4 weeks & 12.5\% / 24.6\% = 0.51 \\
N (Network) & 0.136 & 11.70\% & 15.42\% & 842 & 16.8\% & 56.4 weeks & 11.6\% / 24.7\% = 0.47 \\
A (Activity) & 0.058 & 3.23\% & 6.99\% & 11 & 0.2\% & 0.8 weeks & 9.1\% / 23.9\% = 0.38 \\
A+T & 0.043 & 1.73\% & 5.25\% & 181 & 3.6\% & 0.8 weeks & 16.0\% / 24.0\% = 0.67 \\
T (Text) & 0.017 & 0.58\% & 1.85\% & 98 & 2.0\% & 0.6 weeks & 13.3\% / 23.9\% = 0.55 \\
\bottomrule
\end{tabular}
\end{table}

Providing personalized recommendations requires some form of sensitive data collection from users~\cite{terveen_social_2005}, but privacy and other ethical concerns necessitate collecting as little sensitive data as is required to produce useful recommendations (or in some cases opting not to provide recommendations at all).
Pervasive data collection contributes to perceptions of recommendation services as invasive ``little brothers''~\cite{goldschmitt_shaping_2019}, changing behavior and undermining the potential benefits of recommendation.  
Therefore, we compared the relative importance of the different feature sets in order to understand the utility of collecting particular types of sensitive data---results are shown in Table \ref{tab:feature_ablations}.
%, comparing MLP model performance with different combinations of the feature subsets described in sec.~\ref{sec:model_features}.
Surprisingly, the model trained without text features performs better on all accuracy metrics and similarly in terms of coverage, and it is also more likely to recommend newer sites.
This analysis suggests that the collection of textual data may not be necessary and peer recommendation may still be practical if interaction network data is available: reasonable recommendations could be generated purely based on usage metadata.
We still chose to include text features in the model we evaluated during the field study, as otherwise new authors with minimal activity and no peer interactions would all receive the same recommendations, conflicting with our design goal of personalization even in the cold-start setting.
We discuss these results and this decision further in Appendix \ref{app:sec:feature_ablations}.

\subsection{Model deployment \& required resources}\label{sec:required_resources}

\begin{figure}
\centering
\includegraphics[width=\textwidth]{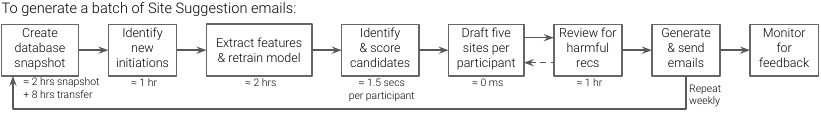}
\caption{Site Suggestion email generation during the field study. Recommendations were typically sent around noon, 36 hours after the database snapshot was taken.}
\label{fig:rec_generation_flow}
\Description{To generate a batch of Site Suggestion emails (flowchart): Create database snapshot (2 hours snapshot + 8 hours transfer), Identify new initiations (1 hour), Extract features and retrain model (2 hours), Identify and score candidates (1.5 seconds per participant), Draft five sites per participant (0 milliseconds), Review for harmful recs (1 hour), Generate and send emails (repeated weekly, looping back to Create database snapshot), and Monitor for feedback.}
\end{figure}

Figure \ref{fig:rec_generation_flow} shows the steps required to generate a batch of Site Suggestion emails.  
The most time-intensive step was the anonymization and manual transfer of a nightly database snapshot, which led to a 36-hour lag time between the snapshot and the associated email batch.\footnote{This 36-hour lag time could mean that the Journal update previewed in the email was no longer the most recent when the emails were sent, but we deemed this lag acceptable as it gave us time to do weekly robustness checks on the trained model and the identified recommendations.}
%, a process that required manual robustness checking during the study. 
%Because we wanted to send the updates at noon U.S. Central Time and database snapshots were created during nightly backup procedures, there was a 36 hour lag time between the database snapshot and the associated emails. 
%As batches were sent out manually, exact delivery time varied from week to week but generally occurred around noon U.S. Central Time on Wednesdays.
%To address an ethical concern that an influx of visitors to any single site could create a negative experience for that site's authors,
We created the Site Suggestion emails by retraining the recommender model on the most recent snapshot and scoring all eligible, active candidate sites.
Rather than recommending the top 5 sites by score, we decided to limit the total number of times a site was recommended in a single batch to at most 10 times.
We made this decision for two reasons: first, as the perception of stranger visits by site authors is unknown, a large influx of strangers might create a negative experience for the recommended site's authors; second, this restriction ensures that a diversity of sites are recommended each week.
To achieve this limit, we randomly drafted sites by score until each participant had five recommendations.\footnote{In the first batch, we applied the limiting only to the bottom quarter of recommended sites by total visit count. See details of the drafting process and an analysis of the small impact on recommendation quality in Appendix \ref{app:sec:site_rank_analysis}.}
For ethical reasons, the first author manually inspected all recommended sites, manually excluding sites that would be inappropriate to send to participants. Two sites were removed this way: one for spam and one for COVID-19-related health misinformation.
This small percentage of potentially harmful recommendations suggests that after some initial validation of quality, individual recommendations need not be inspected in practice.

For future studies, the actual resource requirements could be reduced by computing recommendations in-house.
Disk usage for the weekly snapshot and the feature databases was approximately 96GB. 
The most RAM-intensive processing is maintaining the author interaction network during feature extraction; all other processing tasks are parallelizable and generally IO-bound.
The CPU-based training and inference of the recommendation model could be accelerated with the use of GPUs, although at 1.5s per participant we found inference to be fast enough to serve even a much larger participant population.

\subsection{MLP model details}
\label{app:sec:mlp_model_details}

\begin{figure}
\centering
\includegraphics{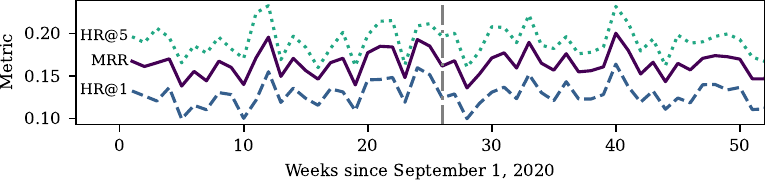}
\caption{$\text{MLP}_\text{Study}$ offline performance throughout the validation and test period. No evidence of a non-zero relationship between time since model training and offline performance metrics. (All significance tests have $p$>0.05.) This flat relationship suggests that (a) the offline analysis is minimally impacted by training set leakage and (b) frequent model retrains may not be necessary in practice, although we retrained the model weekly during the field study.}
\label{app:fig:mlpstudy_metrics_over_time}
\Description{A line chart shows essentially random-looking variation for the metrics HR@5, HR@1 and MRR.}
\end{figure}

Both $\text{MLP}_\text{Study}$ and $\text{MLP}_\text{Tuned}$ use two hidden layers, ReLU activation functions, and dropout.
Optimization occurs over 1000 epochs with a one-cycle learning rate scheduler proposed by Smith and Topin~\cite{smith_super-convergence_2018} and the Adam optimizer with default hyperparameters.
Following best practices to prevent overfitting, we used for prediction the weights from the epoch with the lowest hold-out loss.
Hold-out loss was computed over a random 1\% of training initiations.
We conducted random search experiments (not reported) using the learning rate and Adam hyperparameter distributions given by Sivaprasad et al., finding minimal differences~\cite{sivaprasad_optimizer_2020}.
$\text{MLP}_\text{Study}$ uses 100 hidden units per layer with a dropout $p$ of 0.1.
$\text{MLP}_\text{Tuned}$ uses 300 hidden units per layer with a dropout $p$ of 0.5 and adds a weight decay of 0.0001.
Hyperparameter tuning occurred using grid search to set hidden units $\in \{100, 300, 500\}$, dropout $p \in \{0.1, 0.5, 0.9\}$, weight decay $\in \{0, 0.0001, 0.01\}$, and maximum learning rate $\in [0.008, 0.016]$. We fit models from 3 random seeds at each hyperparameter combination and selected parameters based on the median of the 3 models' MRR.
We did not explore additional model architectures in depth, and offline metrics could be improved through the use of a model closer to the state-of-the-art or through a larger hyperparameter sweep.

One downside to our chronological validation and testing sets is that this setup deviates from the common modeling practice of retraining on regular intervals.
Fortunately, Figure \ref{app:fig:mlpstudy_metrics_over_time} indicates that this affect has a minimal impact on our offline evaluation.

\subsection{Other Baselines}
\label{app:sec:other_baselines}

This section defines less relevant baselines and provides more details on the baselines shown in Table \ref{tab:model_comparison}.
We explored CosSim with other feature sets; they all perform worse in terms of MRR and Coverage metrics than using all features.
\textbf{MF} is the conventional matrix factorization approach to collaborative filtering, using the dot product to compute similarity and selecting hyperparameters as described by Rendle et al.~\cite{rendle_neural_2020};\footnote{All matrix factorization models were trained for 100 epochs with the same optimizer as the MLP model; the best model (by validation MRR) used embeddings of size 128 and weight decay of 0.0001. Embeddings were trained for the 41,567 authors and 79,588 sites that appear at least twice in training-period initiations; a single embedding each was reserved for previously unseen authors and sites.} note the high hit-rate but low MRR, indicating a strong popularity bias.
The new non-personalized baselines are \textbf{RecentInits}, which ranks sites by the amount of time since the last initiation with that site, \textbf{MostJournals}/\textbf{RecentJournals}, which mirror MostInits and RecentInits but count published Journal updates rather than initiations, \textbf{NewestAuthor}, which ranks newest sites first, and \textbf{MostInteractive}, which ranks sites by the number of recent interactions made by authors on that site.

\subsection{Feature ablation discussion}
\label{app:sec:feature_ablations}

%new initiators
%303 1.0 9.26 82.84\%  
%MRR 0.190482 for 12212 test initiations
%previously initiated
%342 1.0 6.42 79.24\%
%MRR 0.147087 for 20380 test initiations

%new initiators    cov user total = 561
%previous initiators     cov user total = 439

The results presented in Table~\ref{tab:feature_ablations} suggest that RoBERTa text features are not important for peer recommendation, and that recommendations based solely on text data would require a different approach than the one we use here.
We suspect that the relative unimportance of text data reflects a bias in our implicit feedback signal and offline evaluation metrics toward authors already known to the source.
Chen et al.\ observed that social network information was more effective for discovering known contacts, while content similarity was more effective for discovering new connections~\cite{chen_make_2009}---a dynamic that may be recurring here, as most historical initiations are between known contacts~\cite{levonian_patterns_2021}.
Even if text-based recommendations are less useful, users may perceive them to be less invasive than network-based recommendations (or vice-versa).
Careful assessment of the perceived acceptability of this data collection for peer recommendation is necessary~\cite{fiesler_participant_2018}.

We chose to deploy the model that included the text data ($\text{MLP}_\text{Study}$) in part because it will still provide personalized recommendations, even to authors who have not yet interacted with peers.
We can observe this effect in the coverage predictions: while recommended site diversity is higher for authors who had previously initiated (342 unique sites for 439 source users, each site recommended on average 6.4 times) compared to authors who had no previous interactions (303 unique sites for 561 source users, each site recommended on average 9.3 times), these differences are relatively small.
Further, these recommendations are of a similar quality: MRR on test initiations is actually higher for people with no previous initiations than previous initiators (0.190 vs 0.147).
Thus, we were satisfied that deploying $\text{MLP}_\text{Study}$ satisfied our cold-start design requirement.
%$\text{MLP}_\text{Study}$ produces 0.78 unique site recommendations for authors who had previously initiated, compared to 0.54 unique site recommendations

%However, users may perceive the acceptability of network-based recommendations differently than text-based recommendations, necessitating further assessment~\cite{fiesler_participant_2018}.
%Careful assessment of the perceived acceptability of this data usage is necessary; users may perceive network-based recommendations to be more or less acceptable than text-based recommendations~\cite{fiesler_participant_2018}.
%As text data is seemingly unimportant in our results, recommendations based solely on text data would require a different approach than the one we use here.
%Given the implicit feedback signal used for training, the offline metrics reflect a bias toward authors already known to the source.

%Why is text data seemingly unimportant?  
%Further, new authors with minimal activity and no peer interactions will all receive the same recommendations, conflicting with our design goal of personalization even in the cold-start setting.
%Thus, we still chose to include all feature sets in the model we evaluated during the field study, which better reflects the desired conditions for peer connection.

\section{Survey Materials}
\label{app:survey}

All questions other than the eligibility, consent, and email questions are optional. All surveys were hosted on the Qualtrics platform.

\subsection{Preference Survey}
\label{app:sec:preference_survey}

All registered CaringBridge authors (18+ years old) are invited to take this 1-minute survey, which is part of a research collaboration by CaringBridge and a team of technology researchers at the University of Minnesota.

The purpose of CaringBridge is to help people get the support they need during health journeys—and support comes from many places! Many authors choose to make their CaringBridge sites open for anyone to read. \textbf{We want to send you emails with links to CaringBridge sites that we think you'll want to follow.}

If you are interested in participating in our study, we'll send you personalized emails with links to CaringBridge sites.  (As always, [privacy comes first]\footnote{Link to \url{https://www.caringbridge.org/what-we-offer/the-privacy-you-choose}} on CaringBridge: no one can read your site unless you want them to.) Opting in and sharing your opinion will help CaringBridge and the research community design features that make giving support easy and make a difference in the lives of the patients and caregivers you care about. Complete information and an FAQ for this study are available by clicking here [link].

\textit{\textbf{Click the arrow in the lower right corner of your screen to take the survey.}}

\begin{center}---Page Break---\end{center}

\subsubsection{Is Adult}
I am 18 years or older.
\begin{itemize}
    \item Yes, I am 18 years or older.
    \item No, I am less than 18 years old.
\end{itemize}

\subsubsection{Is Registered}
Do you have a registered CaringBridge account?
\begin{itemize}
    \item Yes, I have a CaringBridge account.
    \item No, I don't have a CaringBridge account.
\end{itemize}

\subsubsection{Is Author}
Are you can author or co-author of a CaringBridge site? You're an author or co-author if you've ever written a [Journal]\footnote{Link to \url{https://www.caringbridge.org/what-we-offer/a-journal-for-your-journey}} update on a CaringBridge site.
\begin{itemize}
    \item Yes, I am an author or co-author on a CaringBridge site.
    \item No, I am a visitor to CaringBridge and I haven't helped author a site.
\end{itemize}

\subsubsection{Opt-in}
Do you agree to receive personalized follow-up emails with links to CaringBridge sites that we suggest?
\begin{itemize}
    \item Yes, I want to participate in the study and receive follow-up emails with site suggestions.
    \item No, I am not interested in participating in the study.
\end{itemize}

\begin{center}---Page Break---\end{center}

\subsubsection{Email}
\textbf{We just need the email address that you use with CaringBridge so that we can send you personal site suggestions.}

We won't share this email address with anyone outside of CaringBridge or the research team; we'll only use this email address to contact you and to connect to your CaringBridge user account.
\begin{itemize}
    \item The email address I use with CaringBridge is: [Free Response]
    \item I'm not sure which email address is associated with my CaringBridge account, but the email I use most is: [Free Response]
    \item I changed my mind about participating in the study.
\end{itemize}

\subsubsection{Wants study result follow-up}
\textbf{Are you interested in reading about the results of this study when they become available in the future?} If you select yes, we will send you an email update with information about our results.
\begin{itemize}
    \item Yes, please email me with the results of this study.
    \item No, thanks.
\end{itemize}

\begin{center}---Page Break---\end{center}

Thanks for providing your info, you'll get emails with site suggestions from us soon. The following completely optional questions will help us create better site suggestions for you.

\subsubsection{Previous visit to stranger site}
(Optional) Have you ever visited the CaringBridge site of an author who you did not know personally?
\begin{itemize}
    \item Yes, I have visited the CaringBridge site of someone I didn't know.
    \item No, I have never visited a stranger's CaringBridge site.
\end{itemize}

\subsubsection{Motivations}
(Optional) Which of the following might motivate you to visit a fellow author's CaringBridge site, even if you didn't personally know them? Check all that apply.
\begin{itemize}
    \item To learn from the journeys of other CaringBridge authors.
    \item To help mentor or support newer CaringBridge authors.
    \item To receive advice or support from more experienced authors.
    \item To communicate with a peer who understands.
    \item I'm not interested in visiting other authors' CaringBridge sites right now, but I might want to in the future.
    \item I'm not interested in visiting other authors' CaringBridge sites right now, but I would have wanted to in the past.
    \item I'm never interested in visiting other authors' CaringBridge sites.
    \item Something else: [Free Response]
\end{itemize}

\subsubsection{Characteristics}
(Optional) What characteristics of an author or their site would make you want to read \& engage with that person's CaringBridge site? Check all that apply.
\begin{itemize}
    \item High-quality writing or photos
    \item Similar diagnosis or symptoms to you or the loved one you care for
    \item Similar treatments to you or the loved one you care for
    \item Lives near me
    \item Similar cultural background to you or the loved one you care for
    \item For caregivers: Sharing the same relationship (e.g. spouse, child) to the person they care for
    \item Something else: [Free response]
\end{itemize}

\subsubsection{Free Response -- General}
(Optional) Anything else you want to share with us about visiting the CaringBridge site of a fellow author? [Free Response]

\subsection{Feedback Survey}
\label{app:sec:feedback_survey}

Thank you for providing feedback.  All questions on this page are optional: submit your feedback by clicking the right arrow at the very bottom of this survey form.

Not sure why you received this email \& survey?  You agreed to receive site suggestion emails in a survey you took in August. See the FAQ [link]. If you don't want to receive these emails anymore, you can unsubscribe [link].

\subsubsection{Overall Interest}
Overall, did the suggested sites seem interesting to you?
\begin{itemize}
    \item Yes, the sites were generally interesting to me.
    \item Unsure or neutral.
    \item No, the sites were generally uninteresting to me.
\end{itemize}

\subsubsection{Free Response -- Interesting}
Briefly, what seemed interesting to you about the suggested sites? [Free Response]

\subsubsection{Free Response -- Uninteresting}
Briefly, what seemed uninteresting to you about the suggested sites? [Free Response]

\subsubsection{Specific Recommendation Relevance}
Specifically, how relevant did you find each of the suggested sites? [5-item response matrix with levels: Very Relevant; Somewhat Relevant; Unsure or Neutral; Somewhat Irrelevant; Very Irrelevant, Offensive, or Spam]
\begin{itemize}
    \item 1st suggested site
    \item 2nd suggested site
    \item 3rd suggested site
    \item 4th suggested site
    \item 5th suggested site
\end{itemize}

\subsubsection{Free Response -- General}
Any other thoughts you want to share about these site suggestions or your experience so far in this study? [Free Response]

\subsection{Unsubscribe Survey}
\label{app:sec:unsubscribe_survey}

\subsubsection{Page 1}

To unsubscribe: just enter your email address on the line below and click the right arrow.

Or, maybe you're looking for a feedback form [link] or the study FAQ [link] (including contact details for study coordinators) instead.

\subsubsection{Email}
\begin{itemize}
\item Unsubscribe me from additional emails from cb-suggestions@umn.edu. My email address is: [Free response]
\end{itemize}

\begin{center}---Page Break---\end{center}

\subsubsection{Free Response -- Unsubscribe}
\begin{itemize}
\item Tell us why you unsubscribed? [Free response]
\item Anything else you'd like to share with us? Thank you for participating in our study! [Free response]
\end{itemize}

\section{Click data}
\label{app:sec:click_data}

We identify recommendation clicks in the Site Suggestion emails via three data sources: (1) Google Analytics counts, (2) CloudFront logs, and (3) logged-in user visits.  Links to the recommended sites contain UTM tracking information, including the site, the batch ID, and a unique participant ID.

Each data source has limitations. The Google Analytics summary can only provide a total count and won't be incremented if JavaScript is disabled. An event won't be captured in the CloudFront log if UTM tracking tags are stripped and
in unknown other cases.\footnote{``We recommend that you use the logs to understand the nature of the requests for your content, not as a complete accounting of all requests. CloudFront delivers access logs on a best-effort basis. The log entry for a particular request might be delivered long after the request was actually processed and, in rare cases, a log entry might not be delivered at all.'' \url{https://web.archive.org/web/20211206215502/https://docs.aws.amazon.com/AmazonCloudFront/latest/DeveloperGuide/AccessLogs.html}}
Logged-in visits will only be recorded at or near the time of a recommendation click if the participant is already logged in or logs in while on the site.

Google Analytics reports 270 total recommendation clicks, while we have timestamped CloudFront log entries for only 232 clicks, suggesting that we are missing timestamps, participant, and site information for 14.1\% of clicks.  Of the 232 CloudFront clicks, 198 correspond to unique participant/site clicks.  % slight-of-hand, here: this is based on an earlier snapshot (I don't recall exactly when); in reality, we have 243 CF clicks
We can recover 22 clicks missing from the CloudFront request logs via the logged-in user visits, for a total of 220 total unique participant/site clicks. For subsequent statistical analyses, we assume that any additional clicks are missing at random, although in practice we are more likely to be missing clicks from participants who do not log in. If we assume that the Google Analytics count is authoritative and 14.1\% of clicks are missing at random from the CloudFront data, then the expected number of missing participant/site clicks is 10. Thus, we expect only a small impact on our statistical analyses.
Logged-in visit data are extracted from daily snapshots of the CaringBridge database. Thus, we only captured the most recent logged-in visit within a 24-hour period, sufficient to identify daily repeat visits but insufficient for fine-grained analysis of browsing behavior.

% \section{Engineering Miscellania}

% Currently, using this section to keep track of things related to the design and implementation of the system but aren't important enough to include in the main body of the paper. Leaving it in this document as a central place for this info, but moving it to a README file in the GitHub feels like the most likely future.

% \begin{itemize}
%     \item During the user study, a bug in the way activity ``time elapsed'' features was computed produced a bimodal distribution where no recent interaction of that type produced a large positive value.  This bug was corrected for the offline results reported in this paper.
%     \item During the user study, a timezone bug introduced during a system upgrade produced incorrect timestamps for specifically comments before March 28, 2018.  The data were reprocessed and the bug corrected for the offline results reported in this paper.
%     \item During the user study, a bug in the feature preprocessing code resulted in source USP activity and network features being omitted (specifically, replaced with duplicates of the candidate USP features) from training triples generated from historical data after July 1, 2020. This bug would likely have had a negative but minor impact on the model performance, roughly equivalent to dropping source USP activity and network features for about 12\% of the training data. % training data from Jan 2014 to current day
% \end{itemize}

\section{Participant Perspectives}
\label{app:sec:participant_quotes}

\subsection{Free-response: participant motivations}
\textit{``I would like to see how other CaringBridge authors articulate their reactions and feelings as they undergo the medical treatments for and rehabilitation from whatever medical conditions they experience.''}
Two free responses described motivations for not engaging with others' sites, both describing it as a distractor during a busy time. \textit{``it's hard to think about joining in someone else's journey.  I can focus on writing this blog discussing our journey because it is letting friends and family know what is happening so it is narrow enough that it doesn't take away from the other things needed to be accomplished during the rest of the day.''}
These responses indicate that demand for peer connection is goal-driven and subject to constraints. 

\subsection{Free response: peer characteristics}

Three participants specifically identified age of the author or patient as a relevant detail.
Two participants listed shared social context as important characteristics, such as already knowing the person or having ``common friendships''.
One participant said they sought Spanish-language updates, while another said they were looking for sites authored by healthcare professionals.
One participant wanted to see \textit{``multiple posts all the way through death. I wanted to see what I would probably be writing as time progressed. A glimpse into the future if you will.''}

\subsection{Free response: explicit recommendation feedback}

%``choppy, poor grammar and name calling''
\textbf{Negative feedback:} One participant requested the ability to filter sites \textit{``by the number of times Christ is mentioned''}; these objections to Christian religious content may reflect belief misalignment between reader and author~\cite{smith_what_2021}.
Less severe objections focused on aspects of the writing: ``boring'', ``wordy'', ``no reflection'', ``too much philosophizing'', ``bragging''. One participant desired sites that ``go beyond simply updating the reader'', describing their own efforts as an author to inspire readers and to include ``silver linings''. 

\textbf{Positive feedback:} Two participants valued the recency of the displayed updates.
\textit{``I’m finding myself following a lot of these sites. I enjoy the sites that update you often ... I feel like I’m abreast of what’s going on and take a personal interest into the person and their health battle.''}
%Participants valued concise descriptions of the author's emotional state.
Participants valued common ground---such as a patient being treated at the same hospital---and familiarity with issues faced. 
\textit{``I was more interested in the ones I was kind of familiar with.  Understood more.''}
One patient gave us a summary of how they engaged with recommendations:
\textit{``Even with all I’m going through ... I’ve come to care about these people who sites I am following. And I leave a comment every time I login on 95\% of them because I know what it’s like when nobody comments. So I’m really really grateful you guys have [sent me site recommendations] because it is just helps me personally to take my mind off of things when I can go and pray for some other poor souls problems. So good job!''}

\subsection{``Thank you'' email feedback}

The non-clicking participant who replied said: \textit{``CaringBridge filled its purpose and functioned as expected. ... I received a lot of care and support from friends.''} 
%We received three replies from participants who clicked at least once, with all responses emphasizing the importance of some kind of common ground.
The three participants who clicked at least once emphasized the importance of common ground.
One participant connected first with a site due to common ground---the patient and participant had previously lived in the same small suburb---and then kept following the site due to its use of poetry and song lyrics to process grief.
One participant described reading others' sites as ``helpful and informative'' due both to feeling comforted seeing a similar person navigate treatment and for learning strategies when writing their own Journal updates.
One participant decided not to follow some of the recommended sites because they had a lot of existing followers.
\textit{``Having a Caringbridge site I know what it’s like when you don’t have a lot of supporters. ... So I try to help by leaving comments 99\% of the time on the sites I follow to try and help the caregivers stay positive and give them kudos for all they are handling.''}

\section{Analysis of Journal update preview content}
\label{app:sec:thematic_analysis}

\begin{table}[] 
\centering
\caption{Prevalence of identified content categories in the Journal update previews. Only the presence of Expressive Writing was significantly associated with clicks (47.1\% of first-batch recommendations with Expressive Writing were clicked vs 28.7\% without, $p=0.017$).}
\label{tab:annotation_summary}
\begin{tabular}{lr}
\toprule
Preview category & First batch prevalence \\
\midrule
Reporting health status & 85.2\% \\
%\quad (No health status reporting) & 14.8\% \\
\quad Neutral disclosures & 31.5\% \\
\quad Positive disclosures only & 25.8\% \\
\quad Negative disclosures only & 22.2\% \\
\quad Positive \& negative disclosures & 5.8\% \\
Expressive Writing & 31.2\% \\
Managing Author/Audience Relationship & 17.5\% \\
Expressions of Appreciation & 5.8\% \\
\bottomrule
\end{tabular}
\end{table}

The primary information available to participants while they were deciding to click on a recommendation was the text preview of a recent Journal update. 
To capture the preview characteristics that our participants could see and respond to, we conducted a content analysis of these previews.
Two researchers generated open and axial codes independently, then used an affinity mapping process based on the Grounded Theory method to identify three high-level themes~\cite{charmaz_constructing_2006}.
To produce quantitative prevalence estimates, we integrated our three themes with taxonomic descriptions from prior work to identify a set of four categories~\cite{ma_write_2017,smith_i_2020}.
We conducted a regression analysis to identify which categories were associated with recommendation clicks.

\begin{table}[h]
\caption{Summary of themes in recommended Journal update previews}
\label{tab:thematic_analysis}
\resizebox{\textwidth}{!}{\begin{tabular}{p{0.25\linewidth} p{0.40\linewidth} p{0.35\linewidth}}
\toprule 
\textbf{Theme} & \textbf{Description} & \textbf{Example(s)} \\ 
\midrule
\multicolumn{3}{l}{\textbf{Reporting patient/caregiver status}} \\
 Past vs Future News & An axis that differentiates past occurrence and future plans. & This last week was oh so busy with all of my tests and appointments. \\
 & & Friday is a huge milestone for me. \\
Events & Those describing events, specifically related to a singular point in time. & Bryan met with his neurosurgeon yesterday. \\
Patient Symptoms, Status, and Health Processes & Those detailing how the patient is doing, and health processes related to a prolonged interval of time. & Well, Sid is getting stronger by the day. \\
Beginnings, Endings, and Transitions & Those describing singular health events that transition the patient into or out of a health process. & Kristen is finally off the ventilator. \\
Status sentiment & An axis that defines the sentiment of reported news (as “good” or “bad”). & Today Ashley had her second day of PT and she did amazing in every way.  \\
 & & We waited all night for and unfortunately he'll have to go through another round of chemo. \\
Emotion-laden Reporting & Those explicitly stating the author's attitudes towards an update using emotive language. & Tears of joy as I write this. We are sooo happy to announce Andrea is out of the ICU. \\
Reflection & Those reflecting on the author's or patient's experiences. & It's hard to believe it was only a year ago today that Jen started chemo. \\
\midrule
\multicolumn{3}{l}{\textbf{Managing author/audience relationship}} \\ 
Expressions of Gratitude & Those that positively acknowledge a specific type of support. & We are so thankful for your prayers. \\
Update Context & Those that explicitly provide contextual information about the post or its contents for the audience. & Today is going to be a little different. I don't have any medical updates, I just need to let some stuff out. \\
Comments on Update Frequency & Those acknowledging the time between multiple CaringBridge updates. & First, I need to apologize for how long it's been since our last update. \\
Reflection on Writing Process & Those reflecting on the author's experiences as a Caring Bridge author. & I've really struggled to put my feelings into words in these posts. \\ \bottomrule
\end{tabular}}
\end{table}

\subsection{Recommendation preview themes}
\label{sec:recommendation_characteristics}
\label{sec:thematic_analysis}

Our analysis of recommendation previews identified three high-level themes. We report the first-batch prevalence of the categories identified from these themes in Table~\ref{tab:annotation_summary}.
% (a) disclosing health status, symptoms, and treatment, (b) communicating emotions and reflection, and (c) managing author/audience relationship.

\textbf{Disclosing health status, symptoms, and treatment.} Previews that address the question, ``what or how is the patient doing?''
Authors report on both recent heath updates and future plans or expectations. \textit{``Friday is a huge milestone for me. My chemo port is going to be removed.''}
Disclosures vary from reporting on processes (\textit{``Sally is getting better day by day''}), to discrete events (\textit{``Bryan met with his neurosurgeon yesterday.''}), to transitions (\textit{``Kristen is finally off the ventilator.''}).

\textbf{Communicating emotion and reflection.} Previews that express the author's attitudes or reflect on author or patient experiences. 
Emotions are either linked to specific experiences during the health journey (\textit{``we are immensely happy to finally be home''}) or characterize the author's current mental state (\textit{``I wish you could experience this wonderful euphoria I am feeling''}).
Reflection is often used as the introduction to a narrative (\textit{``It's been 12 weeks since Beth’s accident. I cannot believe it's been that long...''}).

\textbf{Managing author/audience relationship.} Previews that engage with the reader. Includes requests, expressions of gratitude, and context about the update, author, or writing process. Requests acknowledge the reader as an active member in the patient's health journey. \textit{``Please send all your prayers as I start my journey through chemotherapy. I am nervous, but with your help I know I can get through this.''} Expressions of gratitude acknowledge received support from readers.
Previews that provide additional context make the author's writing work visible to the reader e.g.\ by discussing update frequency. \textit{``Sorry it’s been a while since I last posted an update. I've been finding it hard to work up the energy to write, but I’ll do my best to recap the last few weeks.''}

Beyond the three high-level themes identified in our content analysis, the full set of themes with descriptions and examples is shown in Table~\ref{tab:thematic_analysis}. From these themes, we identified four categories for quantitative analysis:

\begin{enumerate}
\item\emph{Reporting Health Status, Symptoms, and Treatment.} 
%Answering ``what or how is the patient doing?'' 
%Our thematic analysis found that most previews align with CaringBridge’s objective to facilitate keeping people connected on their health journeys, but this is not always the case. 
Includes previews that update on specific health events, symptoms, or emotions of the patient.
We include two sub-categories for \textit{positive} and \textit{negative} disclosures e.g. as used by Yang et al.~\cite{yang_seekers_2019}.
Previews can contain both positive and negative disclosures if both are provided or if the author qualifies the news (\textit{``in horrible pain, but finally a reason for hope''}).
%\item\emph{Positive/Negative Disclosures.} Sharing good or bad news about “what or how is the patient doing?” Yang et al. used self-disclosure language as a representative behavior to derive OHC roles \cite{yang_seekers_2019}. Previews in these categories address Reporting Health Status using positive/negative language or the content of the news is objectively good (finishing treatment, improving symptoms, etc) or bad (having an accident, diagnosis of illness, passing away, ect). Notably, some disclosures are neither positive nor negative when they do not meet this definition. Previews can be both positive and negative disclosures if any individual component of the post meets either definition or the author uses a qualifying override to describe good or bad news. For example, \textbf{“in horrible pain, but I’m feeling hopeful”} or \textbf{“Caleb is now able to rest in Heaven”}. 

\item\emph{Managing Author/Audience Relationship.} The author is visible through descriptions of their role as a writer or their relationship with their blog’s readers. See thematic description above. \textit{``writing these posts is somewhat cleansing.''}
%Evidently, previews can contain additional meta context on the reader’s behalf. Meta context surfaces through descriptions of the writing process. For example, \textbf{“writing these posts is somewhat cleansing”} and \textbf{“sorry it’s been a while since my last update”}. We also include previews containing requests directed at readers.

\item\emph{Expressions of Appreciation (EOA).} This theoretical construct was previously used by Smith et al., defined as explicit expressions of thanks, gratitude, blessing, or happiness that recognize the support received by the author or patient \cite{smith_i_2020}.
%We include expressions of appreciation as a category present in our thematic analysis (Communicating emotion) and align with prior research on expressions of appreciation as a useful concept in online health communities. 

\item\emph{Expressive Writing and Reflection.} We adopt the definition used by Ma et al.: Expressive Writing is ``completed using functionalities afforded by online communities to disclose users’ thoughts and feelings about personal experiences''~\cite{ma_write_2017}.
We adapt this definition to the context of Journal previews by including personal health-related reflection as a component of Expressive Writing.
%Based on our thematic analysis, we adapt this definition to the context of Journal previews by including personal reflection as a component of expressive writing, consistent with prior work [TODO cite me]. We narrow inclusion to previews containing author disclosures about health-related experiences(/events?), consistent with prior work [cite]. 
\end{enumerate}

Using these categories, two researchers separately annotated 658 update previews in three rounds of coding.
The researchers met between each round to discuss disagreements and update the codebook. 
We report pre-discussion IRR scores in Table \ref{tab:click_irr}.

\subsection{Quantitative content analysis methods}
\label{app:sec:thematic_analysis_methods}

\begin{table}[]
\centering
\caption{Inter-rater reliability as Cohen's $\kappa$ and percent agreement (\%A) between the two annotators in the final two coding rounds. Post-discussion codebook updates resulted in greater agreement during round 3.}
\label{tab:click_irr}
\begin{tabular}{@{}lllll@{}}
\toprule
  & \multicolumn{2}{l}{Round 2 ($n$=143)} & \multicolumn{2}{l}{Round 3 ($n$=376)} \\
Preview Category  & $\kappa$             & \%A            & $\kappa$             & \%A  \\ \midrule
Reporting Health & 0.56 & 83.9\% & 0.62 & 87.8\% \\
\quad Positive Disclosures & 0.72 & 87.4\% & 0.78 & 89.4\% \\
\quad Negative Disclosures & 0.61 & 88.1\% & 0.76 & 92.8\% \\
Managing Audience Relationship & 0.66 & 90.2\% & 0.74 & 91.0\% \\
Expression of Appreciation & 0.78 & 96.5\% & 0.90 & 98.1\% \\
Expressive Writing & 0.38 & 69.2\% & 0.48 & 73.7\% \\
All & 0.33 & 37.8\% & 0.45 & 48.9\% \\\bottomrule
\end{tabular}
\end{table}

We used the resulting annotations to model the relationship between these categories and clicks.
This problem is the ``post-presentation reward prediction'' problem~\cite{garcia-gathright_mixed_2018}.
Unfortunately, clicks are noisy and it is not possible to directly assess if the presence of a particular category is associated with a greater propensity to click.
This impossibility results from confounding: the rank of the recommendation in question, the other recommendations in the same email, when the email was generated and opened, and the specific participant are all confounding factors.
Instead, we build multiple models with varied sets of assumptions and look for convergence.
Specifically, we assume that clicks are independent---a common assumption in click models~\cite{chuklin_click_2015}---and report analysis for three subsets of recommendations:
(a) \textit{Clicked Batches Only} includes only recommendations in emails for which some but not all of the recommendations were clicked, so aims to acquire the clearest view of \textit{choice} among competing options. 
(b) \textit{B1 Only} includes only recommendations sent in the first batch of emails, eliminating temporal confounding and reducing bias introduced by varying levels of participant activity.
(c) \textit{Clickers Only} includes only recommendations sent to participants who clicked at least once during the study, reducing bias introduced by never-seen recommendations.
We fit logistic regression models to predict individual recommendation clicks. 
While we report only recommendation-level logistic regression models here, email-level models, multi-level models for participant and batch, models including subsets of the features, and feature transforms revealed similar patterns.

\subsection{Quantitative content analysis results}

% https://github.com/umncs-caringbridge/recsys-peer-match/blob/main/notebook/survey/ClickDataAnalysis.ipynb
\begin{table}[] 
\centering
\caption{Logistic regression models predicting recommendation clicks from preview content for the recommendation subsets defined in sec.~\ref{app:sec:thematic_analysis_methods}. Preview contents are poor predictors of clicks, as evidenced by near-chance ROC AUC scores (estimated using leave-one-out cross-validation). Note: $^{*}$p$<$0.05; $^{**}$p$<$0.01; $^{***}$p$<$0.001}
\label{tab:click_models}
\begin{tabular}{lccc}
\toprule
 & \multicolumn{1}{c}{Clicked Batches Only} & \multicolumn{1}{c}{B1 Only} & \multicolumn{1}{c}{Clickers Only}  \\
\midrule
Intercept & -0.467$^{}$ & -1.943$^{***}$ & -1.714$^{***}$ \\
  & (0.325) & (0.488) & (0.206) \\
Rank within email & 0.096$^{}$ & -0.023$^{}$ & 0.034$^{}$ \\
  & (0.085) & (0.115) & (0.052) \\
EOA & -0.581$^{}$ & 0.156$^{}$ & -0.546$^{*}$ \\
\quad 5.8\% of B1 recommendations  & (0.396) & (0.621) & (0.277) \\
Expressive Writing & 0.638$^{*}$ & 0.762$^{*}$ & 0.050$^{}$ \\
\quad 31.2\% of B1 recommendations  & (0.250) & (0.353) & (0.152) \\
Managing Audience Relationship & -0.293$^{}$ & 0.239$^{}$ & -0.358$^{}$ \\
\quad 17.5\% of B1 recommendations & (0.328) & (0.409) & (0.192) \\
\multicolumn{4}{l}{Reporting health status (categorical)} \\
\quad Base level: None (14.8\% of B1) &  &  &  \\
\quad Neutral disclosure & -0.171$^{}$ & -0.235$^{}$ & -0.013$^{}$ \\
 \quad\quad 31.5\% of B1 recommendations & (0.338) & (0.498) & (0.211) \\
\quad Positive disclosure only & -0.615$^{}$ & 0.125$^{}$ & -0.078$^{}$ \\
\quad\quad 25.8\% of B1 recommendations & (0.340) & (0.494) & (0.205) \\
\quad Negative disclosure only & -0.395$^{}$ & -0.540$^{}$ & -0.091$^{}$ \\
\quad\quad 22.2\% of B1 recommendations & (0.396) & (0.546) & (0.247) \\
\quad Pos \& neg disclosures & -1.070$^{}$ & -0.541$^{}$ & -0.706$^{}$ \\
\quad\quad  5.8\% of B1 recommendations & (0.551) & (0.768) & (0.368) \\
\midrule
Observations & 310 & 365 & 1,590 \\
Clicks & 120 & 51 & 220 \\
Log-Likelihood & -199.69 & -142.94 & -632.92 \\
ROC AUC & 0.554 & 0.469 & 0.488 \\
\bottomrule
\end{tabular}
\end{table}

We summarize the relationship between all four categories and click behavior in Table~\ref{tab:click_models}.
None of the models with any subset of the variables outperforms the null model (F-test $p>0.05$)---even the rank-only model.\footnote{Recommendations listed first \textit{were} most likely to be clicked, but this difference was statistically insignificant.}
Expressive Writing was associated with a greater probability of being clicked among Clicked Batches (50\% of recommendations with Expressive Writing clicked vs 36\% of recommendations without) and in B1 (47\% of recommendations with Expressive Writing clicked vs 29\% of recommendations without), but not when considering all recommendations sent to clicking participants.
Expressions of Appreciation were associated with a decreased probability of being clicked among participants who clicked at least once (7\% of recommendations with EOA clicked vs 12\% of recommendations without) but not in the other subsets.
Both of these results should be taken with a large grain of salt, and in general these results suggest that participants were making clicking decisions based on factors we did not annotate.

\section{Site Recommendation Analysis}
\label{app:sec:site_rank_analysis}

In sec. \ref{sec:recommendation_characteristics},  we describe characteristics of the recommended set of sites. In this section, we provide additional pre-click details and use them to create a pseudo-control comparison set of non-recommended sites.

\begin{figure}[h]
    \centering
    \begin{subfigure}[t]{0.5\textwidth}
      \includegraphics[width=\textwidth]{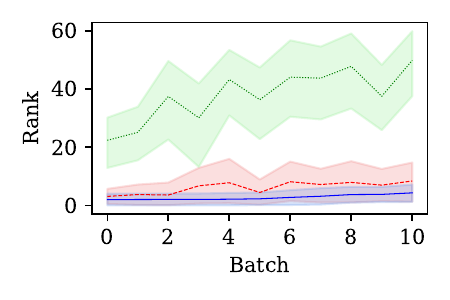}
    \end{subfigure}%
    \begin{subfigure}[t]{0.5\textwidth}
      \includegraphics[width=\textwidth]{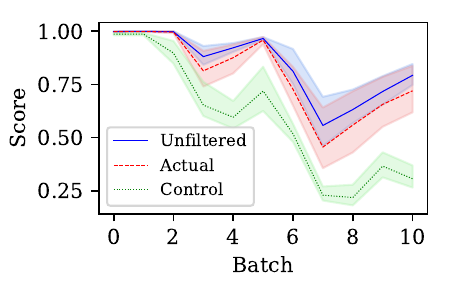}
    \end{subfigure}
    \caption{Distribution rank and the corresponding scores assigned by the recommendation model for unfiltered (blue, solid), actually recommended (red, dashed), and pseudo-control (green, dotted) sets by weekly batch. Across all batches, the unfiltered group contains $n$=397 unique sites, the actual contains $n$=526 unique sites, and the pseudo-control contains $n$=511 unique sites. Each line shows the mean rank and score respectively, while the shaded region indicates the max and min value within each set.}
    \label{fig:sub:score_rank_by_batch}
    \Description{Rank and score are negatively correlated in the expected ways. Control sites receive lower recommendations than Unfiltered and Actual sites.}
\end{figure}

\subsection{Site filtering}

In sec.~\ref{sec:required_resources} we discuss a decision limiting the total number of times a site was recommended in a single batch to at most 10 times. 
Specifically, we conducted a five-round draft.
In each round, a random participant ordering is chosen and participants take turns selecting their highest-scored site that has been picked fewer than 10 times.
We analyze the impact of this decision in Figure~\ref{fig:sub:score_rank_by_batch} by plotting the mean of the rank and score distributions for hypothetical recommendation sets generated without filtering. % to evaluate the trade-off between recommendation diversity and average score/rank. 
We observe that the maximum model score varies between batches as the model was retrained weekly, and that the filtered sites were similar in score to the sites that would have been recommended without filtering---all filtered sites were still in the top 0.1\% of the model's scores.
% observe trends in the average rank distribution of the recommendation algorithm by batch. We find earlier batches are generally associated with smaller differences between actual and unfiltered rank/batch. 
%Still, differences at batch 10 are small which presents a effective trade-off given that filtering results in 32.5\% more unique sites after 11 batches of recommendation.
The use of filtering resulted in a 32.5\% increase in the number of unique recommended sites, a fair trade-off for the modest decrease in score and rank induced by filtering.

\subsection{Psuedo-control set of non-recommended sites}
\label{app:sec:pseudocontrol}

The pseudo-control set is composed of the top five sites for each participant within a batch of recommendations that were never recommended during the duration of the study.
We visualize this set's ranks in Figure~\ref{fig:sub:score_rank_by_batch} along with the resulting effect on the corresponding score assigned to each set by the recommendation algorithm. 
In Table~\ref{tab:rp_prestudy_difference}, we see that the set of recommended author/site pairs were significantly different from the group of sites that could have been recommended (pseudo-control). Values were calculated at the time a recommendation was clicked; in the case of non-clicked and pseudo-control sites, a click time was randomly chosen from the set of actual click times from the same recommendation batch.

In Tables~\ref{tab:cn_prestudy_difference} and~\ref{tab:cp_prestudy_difference}, we extend this analysis to include the subset of clicked recommended author/site pairs. Again, we see many significant differences between the set of clicked recommended author/site pairs and the pseudo-control set. However, we find little significant differences between clicked and non-clicked recommended author/site pairs.

\begin{table}[]
    \caption{Observed recommended pseudo-control USP site behavior from the 35 day window before the point they were clicked (or could have been clicked). Site tenure is in days, Total \# of authors is since the sites creation, \# of authors and \# days visiting peers is for the entire window period, while all other rows are the number of weekly actions.}
    \label{tab:rp_prestudy_difference}
    \begin{tabular}{@{}lrlrlll@{}}
    \toprule
     & \multicolumn{2}{l}{Recommended} & \multicolumn{2}{l}{Pseudo-Control} & & \\
     & \multicolumn{2}{l}{($n_\text{P}$=4190)} & \multicolumn{2}{l}{($n_\text{C}$=4190)} & & \\
     & Med. & M (SD) & Med. & M (SD) & $\text{M}_\text{P}$ - $\text{M}_\text{C}$ & $U_\text{P} / (n_\text{P}n_\text{C})$     \\ \midrule
    %%%% copy cell output below
     Site tenure (days) & 113 & 227.4 (519.2) & 152 & 262.7 (425.2) & -35.3* & 42.4\%* \\
          Journal updates & 1 & 1.9 (2.0) & 1 & 1.6 (2.3) & 0.3* & 40.4\%* \\
            \# of authors & 1 & 1.3 (0.7) & 1 & 1.2 (0.5) & 0.2* & 44.9\%* \\
      Total \# of authors & 2 & 1.8 (1.0) & 2 & 1.7 (0.8) & 0.2* & 46.1\%* \\
              Peer visits & 0 & 0.3 (1.2) & 0 & 0.1 (0.2) & 0.2* & 26.6\%* \\
       Repeat user visits & 0 & 2.1 (6.3) & 0 & 0.9 (1.5) & 1.2* & 49.5\% \\
         Peer initiations & 0 & 0.8 (1.7) & 0 & 0.4 (0.5) & 0.4* & 47.2\%* \\
        Peer interactions & 0 & 3.6 (8.0) & 0 & 1.8 (3.9) & 1.8* & 47.3\%* \\
   \# days visiting peers & 7 & 11.6 (11.7) & 7 & 9.3 (9.0) & 2.3* & 47.4\%* \\
 Site author interactions & 0 & 0.0 (0.2) & 0 & 0.0 (0.2) & 0.0 & 50.0\% \\
  Site author initiations & 0 & 0.0 (0.0) & 0 & 0.0 (0.0) & 0.0 & 50.0\% \\
Site author self interactions & 6 & 15.1 (24.9) & 3 & 8.4 (15.4) & 6.8* & 40.8\%* \\
    %%%% copy cell output above
    \bottomrule
    \end{tabular}
\end{table}

\begin{table}[]
    \caption{Observed clicked and non-clicked recommended site behavior from the 35 day window before the point they were clicked (or could have been clicked). Site tenure is in days, Total \# of authors is since the site's creation, \# of authors and \# days visiting peers is for the entire window period, while all other rows are the number of weekly actions.}
    \label{tab:cn_prestudy_difference}
    \begin{tabular}{@{}lrlrlll@{}}
    \toprule
     & \multicolumn{2}{l}{Clicked} & \multicolumn{2}{l}{Non-Clicked} & & \\
     & \multicolumn{2}{l}{($n_\text{P}$=220)} & \multicolumn{2}{l}{($n_\text{C}$=3970)} & & \\
     & Med. & M (SD) & Med. & M (SD) & $\text{M}_\text{P}$ - $\text{M}_\text{C}$ & $U_\text{P} / (n_\text{P}n_\text{C})$     \\ \midrule
    %%%% copy cell output below
     Site tenure (days) & 107 & 267.4 (609.1) & 113 & 225.2 (513.8) & 42.2 & 49.5\% \\
          Journal updates & 1 & 2.1 (2.2) & 1 & 1.9 (2.0) & 0.2 & 46.6\% \\
            \# of authors & 1 & 1.4 (0.7) & 1 & 1.3 (0.7) & 0.1 & 45.7\%* \\
      Total \# of authors & 2 & 1.9 (1.0) & 2 & 1.8 (1.0) & 0.1 & 47.4\% \\
              Peer visits & 0 & 0.5 (1.7) & 0 & 0.3 (1.1) & 0.2 & 43.9\%* \\
       Repeat user visits & 1 & 3.1 (8.1) & 0 & 2.0 (6.2) & 1.1 & 43.6\%* \\
         Peer initiations & 0 & 1.1 (2.1) & 0 & 0.8 (1.7) & 0.3 & 45.2\% \\
        Peer interactions & 0 & 4.6 (9.5) & 0 & 3.6 (7.9) & 1.0 & 45.7\% \\
   \# days visiting peers & 11 & 13.0 (11.7) & 7 & 11.5 (11.7) & 1.5 & 45.0\% \\
 Site author interactions & 0 & 0.1 (0.4) & 0 & 0.0 (0.2) & 0.0 & 49.2\% \\
  Site author initiations & 0 & 0.0 (0.1) & 0 & 0.0 (0.0) & 0.0 & 49.2\% \\
Site author self interactions & 5 & 14.0 (25.8) & 6 & 15.2 (24.9) & -1.2 & 48.0\% \\
    %%%% copy cell output above
    \bottomrule
    \end{tabular}
\end{table}

\begin{table}[]
    \caption{Observed clicked and pseudo-control site behavior from the 35 day window before the point they were clicked (or could have been clicked). Site tenure is in days, Total \# of authors is since the site's creation, \# of authors and \# days visiting peers is for the entire window period, while all other rows are the number of weekly actions.}
    \label{tab:cp_prestudy_difference}
    \begin{tabular}{@{}lrlrlll@{}}
    \toprule
     & \multicolumn{2}{l}{Clicked} & \multicolumn{2}{l}{Pseudo-Control} & & \\
     & \multicolumn{2}{l}{($n_\text{P}$=220)} & \multicolumn{2}{l}{($n_\text{C}$=4190)} & & \\
     & Med. & M (SD) & Med. & M (SD) & $\text{M}_\text{P}$ - $\text{M}_\text{C}$ & $U_\text{P} / (n_\text{P}n_\text{C})$     \\ \midrule
    %%%% copy cell output below
     Site tenure (days) & 107 & 267.4 (609.1) & 152 & 262.7 (425.2) & 4.7 & 43.0\%* \\
          Journal updates & 1 & 2.1 (2.2) & 1 & 1.6 (2.3) & 0.5* & 37.3\%* \\
            \# of authors & 1 & 1.4 (0.7) & 1 & 1.2 (0.5) & 0.2* & 40.7\%* \\
      Total \# of authors & 2 & 1.9 (1.0) & 2 & 1.7 (0.8) & 0.2* & 43.4\%* \\
              Peer visits & 0 & 0.5 (1.7) & 0 & 0.1 (0.2) & 0.4* & 23.1\%* \\
       Repeat user visits & 1 & 3.1 (8.1) & 0 & 0.9 (1.5) & 2.3* & 43.4\%* \\
         Peer initiations & 0 & 1.1 (2.1) & 0 & 0.4 (0.5) & 0.6* & 47.2\% \\
        Peer interactions & 0 & 4.6 (9.5) & 0 & 1.8 (3.9) & 2.8* & 47.5\% \\
   \# days visiting peers & 11 & 13.0 (11.7) & 7 & 9.3 (9.0) & 3.7* & 42.3\%* \\
 Site author interactions & 0 & 0.0 (0.2) & 0 & 0.0 (0.1) & 0.0 & 49.8\% \\
  Site author initiations & 0 & 0.0 (0.0) & 0 & 0.0 (0.0) & 0.0 & 49.8\% \\
Site author self interactions & 5 & 14.0 (25.8) & 3 & 8.4 (15.4) & 5.6* & 42.5\%* \\
    %%%% copy cell output above
    \bottomrule
    \end{tabular}
\end{table}

\section{Estimates of study impact on behavior for authors and sites}
\label{app:sec:retention_outcomes}

\begin{table}[]
\caption{Activity variables included in the participant author model and the visited site model. We removed some variables to avoid collinearity, but otherwise opted to incorporate as many potentially relevant behaviors as possible~\cite{weisberg_applied_2013}.}
\label{app:tab:covariates}
\begin{tabular}{@{}ll@{}}
\toprule
Author model variables & Site model variables \\ \midrule
\# Journal updates                           & \# Journal updates \\
\# first visits to other authors' sites      & \# unique author visits (recent) \\
\# repeat visits to other authors' sites     & \# unique authors (all time) \\
\# unique days visiting other authors' sites & \# first visits to site from peers \\
\# interactions on other authors' sites      & \# repeat author visitors to site \\
\# interactions on their own sites           & \# unique days other authors visited \\
\# self-authored sites interacted with       & \# interactions from other authors \\
Author tenure (log) & \# initiations to site \\
& \# peer interactions by site authors \\
& \# self-site interactions by site authors \\
& \# initiations by site authors \\ 
& Site tenure (log) \\ \bottomrule
\end{tabular}
\end{table}

\begin{figure}
    \centering
    \includegraphics[width=\textwidth]{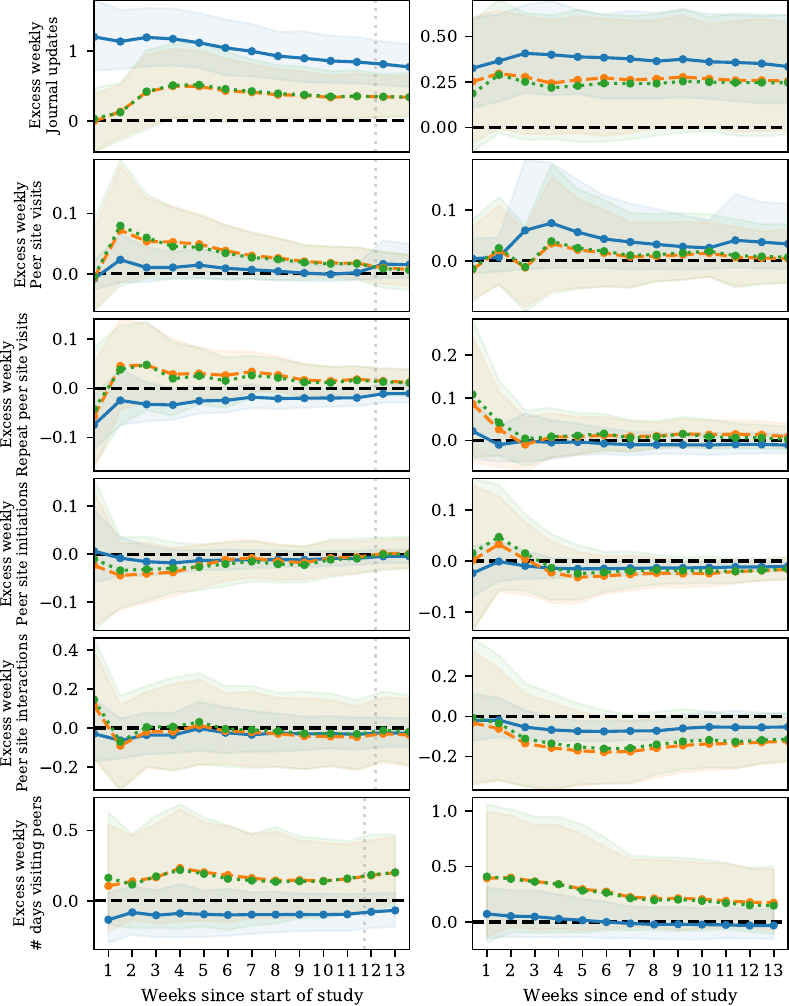}
    \caption{Estimated impacts of receiving Site Suggestion emails on author behavior, both during the study (left column, the vertical dotted line indicates the date of the last Site Suggestion email) and after the study (right column). The estimators shown are as described in sec. \ref{sec:retention_outcomes}: raw (solid, blue), OLS (dashed, orange), and DR (dotted, green). Figure~\ref{fig:participant_outcomes} captures the vertical slice at post-study week 13. 95\% confidence intervals are computed via bootstrapping (1000 iterations).}
    \label{app:fig:participant_outcomes_all}
    \Description{A 2x6 grid of Excess weekly actions relative to the start and end of the study. Most CIs are wide, but there are minimal week-to-week changes in the mean estimate.}
\end{figure}

In sec.~\ref{sec:retention_outcomes}, we estimated the impact of recommendation on second-order behaviors. We estimate this impact by comparison to a pseudo-control group of non-enrolled authors and non-visited sites (see section \ref{sec:participant_prior_use} and Appendix~\ref{app:sec:pseudocontrol} respectively). We provide additional method details and sensitivity analysis in this section.

In Figures~\ref{fig:participant_outcomes} and~\ref{fig:recommendation_outcomes}, we showed post-study outcomes by comparing 5 weeks (35 days) of pre-study behavior to 12 weeks (91 days) of post-study behavior.  We conducted a sensitivity analysis to determine the impact of two decisions: examining behavior during the post-study period (rather than the ``during study'' period) and choosing a time window of 12 weeks.
Figure~\ref{app:fig:participant_outcomes_all} is a high-detail summary of those decisions, demonstrating that differences are small depending on the selected time window: smaller post-study time windows are generally associated with higher-variance effect size estimates.
In pre/post panel data analyses, it is common to use the same time window pre- and post-intervention, which helps to control for longer-term  behavioral trends. We tried both approaches, and found no difference except increased variance, so we use data from the full pre-study time window (5 weeks) even if the post-study time window is less than 5 weeks.

\begin{figure}
    \centering
    \includegraphics[width=\textwidth]{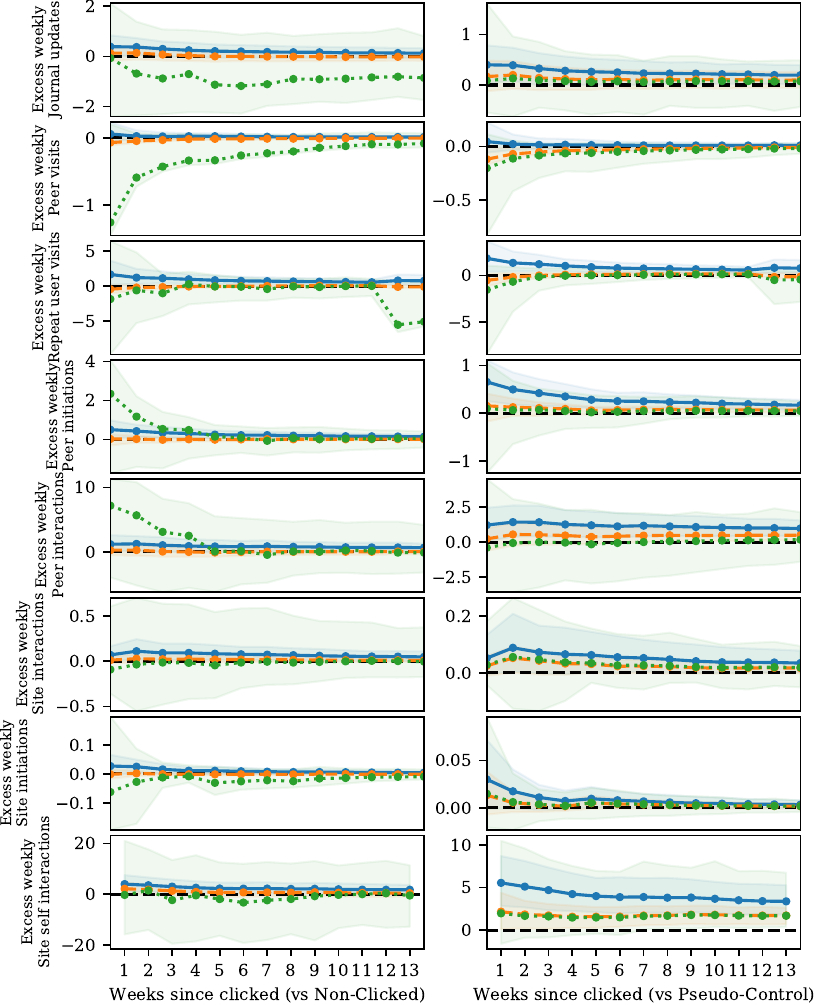}
    \caption{Estimated impacts of Site Suggestion emails on recommended site behavior post click, both using non-click sites (left column) and pseudo-control sites (right column). The estimators shown are as described in sec. \ref{sec:retention_outcomes}: raw (solid, blue), OLS (dashed, orange), and DR (dotted, green). Figure \ref{fig:recommendation_outcomes} captures the vertical slice at week 13 in the right column. 95\% confidence intervals are computed via bootstrapping (1000 iterations).}
    \label{app:fig:recommendation_outcomes_all}
    \Description{A 2x8 grid of Excess weekly actions relative to the click time for both he non-clicked and pseudo-control sites. Most CIs are wide.}
\end{figure}

Using a similar approach, we extend this analysis to investigate post-click outcomes on recommended site behavior.
In Figure \ref{fig:recommendation_outcomes}, we showed post-study outcomes by comparing 5 weeks (35 days) of pre-click behavior to 12 weeks (91 days) of post-click behavior. Similar to Appendix \ref{app:sec:site_rank_analysis}, values were calculated at the time a site was first clicked and in the case of non-clicked and pseudo-control sites, a click time was randomly chosen from the set of actual click times in the
first batch that site was/could of be recommended. We conducted a sensitivity analysis to examine the impact a participant visiting a recommended site had on the sites behavior during the post-click period. Figure~\ref{app:fig:recommendation_outcomes_all} is a high-detail summary demonstrating we were unable to capture statistically significant effects from participant clicks on recommended site behavior. Moreover, we find that effect estimates based on comparison to the non-clicked and pseudo-control groups are similar.

As noted by Hernán and Robins, successful causal inference predominantly rests on untestable assumptions \cite{hernan_causal_2020}. The doubly robust estimates we produce depend on five assumptions:
\begin{itemize}
    \item \textbf{Exchangeability} Exchangeability refers to the probability of being in the treatment group is independent of the causal outcome (depending only on $A$). Exchangeability is likely false: it is reasonable to assume that authors that opted to participate in a recommendation intervention are more likely to engage with a recommendation intervention than non-participating authors, independent of e.g. their tenure as an author or their prior peer interaction behavior. If participants are not very different from non-participating authors in terms of their responsiveness to treatment, then it may still be reasonable to assume exchangeability. The same reasoning holds for clicked and non-clicked sites.
    \item \textbf{Positivity} Positivity requires that the distributions of the covariates for the treated and pseudo-control groups to fully overlap. Positivity is approximately true in our data, i.e. all values of the covariates observed in the treated group are also observed in the pseudo-control group. The positivity assumption is why we omit causal estimates of the impact of deploying Site Suggestion emails to the whole author population in terms of e.g. click rates; we have no ability to estimate the behavior of the pseudo-control population when exposed to Site Suggestion emails.
    \item \textbf{Consistency} Consistency refers approximately to the assumption that the treatment varies only in ways unrelated to the outcomes. Recommendation varies, so participants will be exposed to different versions of the treatment; different participants are shown different recommendations, so clicked sites will be exposed to visits from different participants. Intuitively, these differences can be causally related to the outcomes: a participant with no prior interaction behavior, for example, may receive less relevant recommendations than a participant with a long history of peer interaction. A site clicked by a participant that visits and leaves a comment is different than a site clicked by a participant that only visits.
    Joachims et al. argue that recommender systems ought to be viewed as policies that select interventions in order to optimize a desired outcome~\cite{joachims_recommendations_2021}; it's a policy \textit{designed} to select variable interventions.
    By assuming consistency, we assume that the recommendations are of similar quality and that participant visits have similar effects on the visited sites. The degree to which this assumption is violated will bias the resulting estimates.
    \item \textbf{No measurement error} Measurement error in the observed behaviors is likely non-existent, given the use of a complete database snapshot.
    \item \textbf{No model misspecification} To adjust for confounding, the models we use need to include all relevant covariates and assume the correct functional form. By fitting linear models, we make a parsimonious but possibly false assumption of functional relationship between the covariates and the outcome of interest.
    More serious is the assumption that all relevant covariates are available; as discussed, we believe that unmeasured confounders such as an interest in peer connection affect response to treatment in a way that is independent of the measured covariates.
\end{itemize}

Given this discussion, it seems reasonable to object that three of the assumptions are very likely false. In practice, we are assuming that these assumptions are \textit{close enough} to true, such that we can still attain some insight from computing causal estimates based on these assumptions in order to compare to the observational estimates.

\section{Sample Size Calculations}
\label{app:sec:sample_size_calculations}

In sec. \ref{sec:effect_size_estimation}, we described a power analysis for an uncontrolled peer recommendation study. In this section, we provide additional details and add estimates for an RCT.
In Fig. \ref{fig:power_analysis}, we showed sample size estimates for two variations of a replication of our feasibility study by computing the effect of two author behaviors associated with receiving recommendations. Here, we extend this analysis to include 5 behaviors: unique repeat visits to recommended sites,  interactions with recommended sites, unique repeat visits to all sites,  interactions with all sites, and recommended site updates. We consider the 35 days before and after the first exposure to recommendation or first stranger visit for the one-time email and 35 days before and the study period plus 35 days after for the 12 week recurring email intervention.

\begin{table}[htb]
    \caption{Mean and standard deviation estimates for a one-time and recurring recommendation email intervention. Site descriptions in parenthesis for participant groups denote the target of the intended action.}
    \label{tab:sample_size_calculations}
    \begin{tabular}{@{}lllll@{}}
    \toprule
    & \multicolumn{2}{c}{Recurring 12-Week} & \multicolumn{2}{c}{One-Time} \\
     Behavior & $\text{M}_\text{P}$ ($\sigma^2$) & $\text{M}_\text{C}$ ($\sigma^2$) & $\text{M}_\text{P}$ ($\sigma^2$) & $\text{M}_\text{C}$ ($\sigma^2$)  \\ \midrule
    %%%% copy cell output below
\textbf{Participants} & \textbf{Study} &  & \textbf{Study} &  \\
\textbf{(Recommended Sites)} & \textbf{($n_\text{P}$=79)} &  & \textbf{($n_\text{P}$=73)} &  \\
\quad Second visits & 0.06 (0.06) &  & 0.04 (0.02) &  \\
\quad Interactions & 0.47 (16.36) &  & 0.07 (0.22) &  \\
\\
\textbf{Participants} & \textbf{Participants} & \textbf{Pseudo-control} & \textbf{Participants} & \textbf{Pseudo-control} \\
\textbf{(All Sites)} & \textbf{($n_\text{P}$=79)} & \textbf{($n_\text{C}$=1759)} & \textbf{($n_\text{P}$=73)} & \textbf{($n_\text{C}$=1759)} \\
\quad Second visits & -0.07 (0.07) & 0.02 (0.01) & -0.05 (0.03) & 0.02 (0.01) \\
\quad Interactions & -0.34 (14.81) & 0.11 (0.49) & -0.04 (0.72) & 0.09 (0.49) \\
\\
\textbf{Recommended Sites} & \textbf{Clicked} & \textbf{Non-clicked} & \textbf{Clicked} & \textbf{Non-clicked} \\
& \textbf{($n_\text{P}$=158)} & \textbf{($n_\text{C}$=368)} & \textbf{($n_\text{P}$=51)} & \textbf{($n_\text{C}$=55)} \\
\quad Updates & 0.024 (2.11) & 0.023 (1.80) & 0.02 (1.33) & 0.04 (3.45) \\
    %%%% copy cell output above
    \bottomrule
    \end{tabular}
\end{table}

\begin{table}[htb]
    \caption{Observed effect sizes and required sample size needed for an appropriately powered RCT for a one-time recommendation email and a recurring, weekly recommendation email. Site descriptions in parenthesis for participant groups denote the target of the intended action.}
    \label{tab:sample_size_estimates}
    \begin{tabular}{@{}lllll@{}}
    \toprule
    & \multicolumn{2}{c}{Recurring 12 Week} & \multicolumn{2}{c}{One-Time} \\
     Behavior & Effect Size & Sample Size & Effect Size & Sample Size \\ \midrule
    %%%% copy cell output below
\textbf{Participants (Rec)} \\
\quad Second visits & 0.28 & 75 & 0.26 & 81 \\
\quad Interactions & 0.12 & 444 & 0.14 & 314 \\
\\
\textbf{Participants (All)} \\
\quad Second visits & 0.85 & 5 & 0.67 & 10 \\
\quad Interactions & 0.43 & 29 & 0.18 & 189 \\
\\
\textbf{Recommended Sites} \\
\quad Updates & -0.14 & 299 & 0.23 & 110 \\
    %%%% copy cell output above
    \bottomrule
    \end{tabular}
\end{table}

Table \ref{tab:sample_size_calculations} outlines the mean, variance, and sample size for each group in terms of weekly actions used in the calculations.
The structure of our data let us consider two potential future interventions: a one-time recommendation email and a recurring, weekly recommendation email.
We estimate the effects of a one-time recommendation email by including only participant exposures and visits from the first batch of recommendation emails (B1).
We use the pseudo-control group (see sec. \ref{sec:participant_prior_use} and Table \ref{tab:prestudy_difference}) and non-clicked recommendations (see Appendix \ref{app:sec:site_rank_analysis}) to analyze the effect of receiving recommendations and stranger visits.
For participants (recommended sites) behaviors, we report weekly actions from a 5 week pre-inflection period to a post-inflection period: 12 weeks + 5 weeks and 5 weeks. 
For participants (all sites) and recommended sites behaviors, we report the difference in weekly actions from similar pre- and post-inflection periods. 
%Effect size calculations were made using 17 decimal places.

Using these statistics, Table \ref{tab:sample_size_estimates} estimates the sample size required for a replication (same as Fig. \ref{fig:power_analysis} and for an RCT, with 50\% not receiving recommendation emails.
For author behaviors targeted at recommended sites, we compute the effect as simply as \(d = M_P / SD\).
For all other behaviors, we control for prior behavior by computing the effect as \(d = (M^{\text{before}}_{C} - M^{\text{after}}_{C}) - (M^{\text{before}}_{P} - M^{\text{after}}_{P}) / SD_{\text{pooled}}\), where P is the group of participants or clicked recommended sites and C is the control group of pseudo-control participants or non-clicked recommended sites.
We present \(M^{\text{before}} - M^{\text{after}}\) because the expected churn of users dictates over time they will become less active. We find this is true in our sample; all users that published at least one journal update in July 2021 subsequently published, on average, 1.32 (5.44) less updates in the following month.
Here, a positive effect size means that being a participant results in less of a difference from before and after when compared to the control group.
Using G*Power 3.1.9.7, the estimated required sample sizes for an appropriately powered RCT were calculated based on observed effect sizes using a one tailed point biserial model at 80\% power with \(\alpha = 0.5\) \cite{faul_g*power_2007}.

Unsurprisingly, we see large differences in participant behavior towards recommended sites when compared to the pre-study period. 
More interesting is the fact that participants, on average, sustained an increase in weekly actions including non-recommended sites after receiving recommendations. 
While this translates to a seemingly large effect size during calculation, we attribute these differences primarily to the significant differences pre-study between both groups outlined in Table \ref{sec:participant_prior_use}.
Here, we believe any differences that exist between groups can largely be attributed to confounding variables (i.e.\ author/site tenure).

\end{document}